\documentclass[12pt]{article}
\usepackage{geometry}
\usepackage{bbold}
\usepackage{amsmath,amssymb,amsfonts}
\usepackage{graphicx,axodraw}
\usepackage[footnotesize]{caption2}
\usepackage{mathrsfs}
\usepackage{bbm}
\usepackage{braket}
\usepackage{cancel}
\usepackage{color}

\allowdisplaybreaks

\definecolor{darkblue}{rgb}{0,  0,  .5}
\definecolor{lightgray}{gray}{0.5}

\newcommand{\Mg}{M_{\textnormal{GUT}}}
\newcommand{\Mi}{M_{in}}
\newcommand{\Mw}{M_{\textnormal{EW}}}

\newcommand{\V}{V_{\textnormal{CKM}}}

\newcommand{\eq}[1]{Eq~\eqref{#1}}
\newcommand{\Figref}[1]{Fig~\ref{#1}}

\newcommand{\Secref}[1]{Sec~\ref{#1}}

\newcommand{\mfs}{m^2_{f}} %Changed this temporarly, as it seems this is the notation we are using
\newcommand{\mQs}{m^2_{ Q}}
\newcommand{\mds}{m^2_{D}}
\newcommand{\mus}{m^2_{U}}
\newcommand{\mls}{m^2_{L}}
\newcommand{\mes}{m^2_{E}}
\newcommand{\mQsSK}{\hat{m}^2_{ Q}}
\newcommand{\mdsSK}{\hat{m}^2_{D}}

\newcommand{\mfive}{m^2_{\bar{5}}}
\newcommand{\mten}{m^2_{\widetilde {10}}}

\newcommand{\skm}{\rm{SCKM}}

\newcommand{\aDH}{\hat{a}_{D}}

\newcommand{\aDHE}{V^{D\dagger}_L a_D V^{D}_R}
\newcommand{\BBsm}{\text{BR}(B_s\rightarrow\mu^+\mu^-)}
\newcommand{\BBsme}{\text{BR}(B_s\rightarrow\mu^+e^-)}
\newcommand{\BBdm}{\text{BR}(B_d\rightarrow\mu^+\mu^-)}
\newcommand{\Bsg}{\text{BR}(B\rightarrow X_s\gamma)}
\newcommand{\Bmueg}{\text{BR}(\mu\rightarrow e\gamma)}
\newcommand{\Btaueg}{\text{BR}(\tau\rightarrow e\gamma)}
\newcommand{\Btaumug}{\text{BR}(\tau\rightarrow \mu\gamma)}

\newcommand{\Btaunu}{\text{BR}(B \rightarrow  \tau \nu)}

\newcommand{\BRKLpinunu}{\text{BR}(K_L^0 \rightarrow \pi^0 \nu\nu)}
\newcommand{\BRKPpinunu}{\text{BR}(K^+ \rightarrow \pi^+ \nu\nu)}
\newcommand{\amuegG}{a_{\mu e \gamma}}

\newcommand{\amuegL}{a_{\mu e \gamma L}}
\newcommand{\amuegR}{a_{\mu e \gamma R}}

\newcommand{ \deltaLL   }{(\delta_{12}^E)_{LL}^{}}
\newcommand{ \deltaRR   }{(\delta_{12}^E)_{RR}^{}}

\newcommand{\deltaXY}{(\delta_{ij}^f)_{\{ XY \} }}

\newcommand{\nn}{\nonumber}

\def\beq{\begin{equation}}
\def\eeq{\end{equation}}
\def\bea{\begin{eqnarray} }
\def\eea{ \end{eqnarray} }

\def\mgut{M_{GUT}}
\def\calh{\mathcal{H}}

\def\mten{m^2_\mathbf{10}}

\DeclareMathOperator{\Tr}{Tr}
\def\nl{\hfill\nonumber\\&&}

\begin{document}

\begin{flushright}
{\tt KCL-PH-TH/2016-20}, {\tt LCTS/2016-13}, {\tt CERN-PH-TH/2016-095}  \\
{\tt UMN-TH-3526/16, FTPI-MINN-16/16} \\
\end{flushright}

\vspace{1.5cm}
\begin{center}
{\Large\bf 
Maximal Sfermion Flavor Violation in Super-GUTs\\
}
\end{center}

\vspace{0.5cm}
\begin{center}{\large
{\bf John~Ellis}$^{1}$,
{\bf Keith A. Olive}$^2$ and {\bf L. Velasco-Sevilla}$^{3}$\\
}
\end{center}

\begin{center}
{\em $^1$Theoretical Particle Physics and Cosmology Group, Department of
  Physics, King's~College~London, London WC2R 2LS, United Kingdom;\\
Theoretical Physics Department, CERN, CH-1211 Geneva 23,
  Switzerland}\\[0.2cm]
  {\em $^2$William I. Fine Theoretical Physics Institute, School of Physics and Astronomy,\\
University of Minnesota, Minneapolis, MN 55455, USA}\\[0.2cm]
{\em $^3$ University of Bergen, Department of Physics and Technology,\\
PO Box 7803, 5020 Bergen, Norway}\\
\end{center}

\vspace{1.5cm}
\begin{center}
{\bf Abstract}
\end{center}
We consider supersymmetric grand unified theories with soft supersymmetry-breaking scalar masses $m_0$ specified
above the GUT scale (super-GUTs) and patterns of Yukawa couplings motivated by upper limits on flavour-changing
interactions beyond the Standard Model. If the scalar masses are smaller than the gaugino masses $m_{1/2}$, as is expected in
no-scale models, the dominant effects of renormalization between the input scale and the GUT
scale are generally expected to be those due to the gauge couplings, which are proportional to $m_{1/2}$
and generation-independent. In this case, the input scalar masses $m_0$ may violate flavour maximally,
a scenario we call MaxSFV, and there is no supersymmetric flavour problem.
We illustrate this possibility within various specific super-GUT scenarios that are deformations of no-scale gravity.

\vspace{1.5cm}
\begin{flushleft}
%May
October 2016
\end{flushleft}

\newpage

\section{Introduction}

Ever since the earliest days of supersymmetric model-building, it has been emphasized that
data on flavour-changing processes suggest the existence of a `super-GIM' mechanism to ensure that
the effective electroweak-scale slepton and squark mass matrices are almost diagonal
with small generational mixing and eigenvalues that are almost degenerate~\cite{EN}. This constraint
on supersymmetric model-building has subsequently been dubbed the `supersymmetric flavour problem'.
Soon after~\cite{EN}, it was recognized that one possible scenario for solving this `problem' 
in the squark sector would be to
postulate that all supersymmetric flavour violation is proportional to the Cabibbo-Kobayashi-Maskawa
(CKM) mixing between quarks~\cite{BG}, a scenario that has come to be known as minimal flavour violation (MFV).
This approach left open the question how MFV came to be, one possible answer being provided by
gaugino mediation of supersymmetry breaking~\cite{gaugino}.

A suitable framework for studying the supersymmetric flavour problem is provided by a supersymmetric GUT such
as SU(5)~\cite{DG, Sakai} in which the soft supersymmetry-breaking scalar masses $m_0$, the trilinear soft
supersymmetry-breaking parameters $A_0$ and the gaugino masses $m_{1/2}$ are input at the
GUT scale $M_{GUT} \simeq 10^{16}$~GeV. Upper limits on the deviations from Standard Model
predictions for flavour-changing processes motivate the hypothesis that the $m_0$ parameters for
chiral supermultiplets with the same gauge quantum numbers are identical at this input scale~\cite{BG},
and the GUT symmetry requires them to be identical for all the sparticles in the same GUT
multiplet. Thus, in SU(5) all the $\mathbf{\bar 5}$ sfermions would have a common $m_0$,
and all the $\mathbf{10}$ sfermions would have another (potentially different) common $m_0$.
It is often assumed, with no clear phenomenological motivation apart from simplicity and
possible embedding in a larger supersymmetric GUT such as SO(10), that these
two $m_0$ parameters are identical at the GUT scale, a scenario called the constrained
minimal supersymmetric extension of the standard model or CMSSM \cite{cmssm}.

The question remains, however, what might be the origin of any such universality in the $m_0$
parameters. This would occur in minimal supergravity models with trivial, flat K\"ahler metrics~\cite{bfs},
but would not happen in more general supergravity models \cite{gauge}, as discussed recently in the context of
compactified M-theory~\cite{EK}. One interesting exception is no-scale
supergravity \cite{nosc1}, in which the input soft supersymmetry-breaking scalar masses $m_0$ vanish at
the input scale. In this case, the electroweak-scale soft supersymmetry-breaking scalar masses
are all generated by gauge interactions, and hence are identical for different sparticles with the
same gauge quantum numbers, in a manner reminiscent of gauge-mediated supersymmetry-breaking models~\cite{GMSB}.
The phenomenological constraints on sparticle masses exclude models with no-scale
boundary conditions at the supersymmetric GUT scale~\cite{eno5}, but no-scale
boundary conditions at higher input scales may be acceptable~\cite{eno5,emo2,ENO2013}.

The principal purpose of this paper is to study  the constraints on the flavour structure of the
soft supersymmetry-breaking scalar masses for models that are similar to such no-scale
models, with $m_0 \ll m_{1/2}$ at some input scale $M_{in} > M_{GUT}$, scenarios
we call super-GUTs. One may regard such scenarios as deformations of the simple
no-scale framework, as might occur in realistic string models via higher-order corrections
to the  (over-simplistic?) no-scale K\"ahler potential, cf, the studies in~\cite{EK}. Intuitively, it is
clear that the constraints on the non-universalities between the diagonal $m_0$ parameters
and on the ratios of off-diagonal to diagonal entries in the soft supersymmetry-breaking
scalar mass matrix must become progressively weaker as the no-scale limit: $m_0 \to 0$ is
approached. 

Indeed, close to this no-scale limit a completely anarchic $m_0$ matrix is allowed.
{In this sense, we consider possible anarchic structures which we term as
maximal flavour violation (MaxSFV).} Thus,
there is no `supersymmetric flavour problem' for super-GUTs with input boundary conditions
at some scale $M_{in} > M_{GUT}$ that are small deformations of the idealised no-scale limit.
A primary objective of this paper is to quantify this statement within illustrative super-GUT scenarios.

The effects of flavor-violating sfermion mass parameters on hadronic and leptonic flavor observables, as well as their
correlations,   have been 
 studied previously in the context of Grand Unified Theories. These
 studies typically
 assume the mass insertion approximation without a complete
 top-down running of the soft supersymmetry-breaking parameters (see, e.g., \cite{Ciuchini:2007ha}). We do not strive to
 study the generalities of such correlations, instead we consider a specific
 set up  where we establish limits on the maximal values of the off-diagonal entries
 in the $m_0$ matrix using a complete running of soft supersymmetry-breaking parameters. 
Also, we constrain the parameter space via EW observables as well as flavor-violating
effects. To our knowledge, this is the
first complete and realistic study that takes into account running from a scale
above the unification scale.

The layout of this paper is as follows. We begin in Section~2 by setting up 
our super-GUT model framework~\cite{eno5,sg,Ellis:2010jb,emo2,dmmo},
focusing in particular on its implementation in no-scale supergravity~\cite{emo2}. We use weak-scale
measurements to specify the gauge and Yukawa couplings, whereas the soft supersymmetry-breaking scalar masses,
trilinear and bilinear terms are specified at the input scale, $\Mi$. The matching conditions at $M_{GUT}$ are
discussed in Section~2.3.  We then specify  in Section~2.4 the illustrative flavour-mixing models
that we choose for further study. In Section~3 we analyse the case of pure no-scale boundary conditions,
in which all soft supersymmetry-breaking scalar masses,
trilinear and bilinear terms are set to zero at $\Mi$. We display the running of of these parameters
as well as the Yukawa couplings between the input and weak scales for our representative flavour-mixing scenarios.
Then, in Section~4 we analyse super-GUT scenarios in which $\Mi > M_{GUT}$, studying the upper bounds
on non-universality in $m_0/m_{1/2}$ that are permitted by the experimental upper limits on flavour-changing interactions as
functions of $\Mi$ in our illustrative flavour-mixing scenarios. Finally, Section~5 summarises our conclusions.

\section{Model Framework}

\subsection{No-scale SUGRA model}
We first consider a low-energy effective theory that is based on an $N = 1$ supergravity model 
with the simplest no-scale structure~\cite{nosc1} defined by a K\"ahler potential
\beq
\label{noscale}
K \; = \; - 3 {\rm ln} (T + T^\dagger - \sum_i |\Phi_i|^2/3)
\eeq
where $T$ is a modulus field and $\Phi$ represents matter fields present in the theory.
The no-scale form (\ref{noscale}) for $K$ ensures that all soft supersymmetry-breaking
scalar masses and bi- and trilinear terms vanish
at some input universality scale, $\Mi$. However, we recall that non-zero gaugino masses arise
independently from a non-trivial gauge kinetic function $f_{\alpha \beta}$ in the
effective supergravity theory. It is known that vanishing soft masses at the GUT scale $M_{GUT}$
in a theory with universal gaugino masses are in general phenomenologically disastrous~\cite{eno5}. However, 
this problem may be circumvented if the input universality 
scale is between the GUT scale and the Planck scale \cite{eno5,emo2,ENO2013}.

A K\"ahler potential of the form (\ref{noscale}) arises in generic manifold compactifications
of string theory, in which $T$ is identified as the manifold volume modulus. In such a scenario,
the $\Phi_i$ are identified as untwisted matter fields. In general there would, in addition, be
twisted matter fields $\varphi_a$ described by additional terms in the K\"ahler potential of the form
\beq
\Delta K \; = \; \sum_a \frac{|\varphi_a|^2}{(T + T^\dagger)^{n_a}} \, ,
\label{twistedK}
\eeq
where the parameters $n_a$ are model-dependent modular weights. These give rise to
$n_a$-dependent soft supersymmetry-breaking terms whose magnitudes and flavour
structure are also model-dependent and violate the MFV assumption,
in general. Here, we assume that the
MSSM matter fields are assigned to the untwisted sector.

The renormalization of MSSM parameters at scales above $M_{GUT}$ requires the inclusion
of new particles and parameters in addition to those in the generic MSSM, including GUT-scale Higgses, their 
self-couplings and couplings to matter. For simplicity, we assume here minimal SU(5),
in which one introduces a single SU(5) adjoint Higgs
multiplet $\hat{\Sigma}(\bf{24})$, and the two Higgs doublets of the MSSM, $\hat{H}_d$ and $\hat{H}_u$,
are extended to five-dimensional SU(5) representations $\hat{\calh}_1(\bf{\overline{5}})$ and 
$\hat{\calh}_2(\bf{5})$ respectively.
The minimal renormalizable superpotential for this model is~\cite{Polonsky:1994rz,Ellis:2010jb}
\bea
W_5 &=& \mu_\Sigma \Tr\hat{\Sigma}^2 + \frac{1}{6}\lambda'\Tr\hat{\Sigma}^3
 + \mu_H \hat{\calh}_{1\alpha} \hat{\calh}_2^{\alpha} 
 + \lambda \hat{\calh}_{1\alpha}\hat{\Sigma}^{\alpha}_{\beta} \hat{\calh}_2^{\beta} \nl
 +\frac{1}{4}({\bf h_{10}})_{ij} \epsilon_{\alpha\beta\gamma\delta\zeta}
   \hat{\psi}^{\alpha\beta}_i \hat{\psi}^{\gamma\delta}_j \hat{\calh}_2^{\zeta} 
 +\sqrt{2}({\bf h_{\overline{5}}})_{ij} \hat{\psi}^{\alpha\beta}_i \hat{\phi}_{j\alpha} \hat{\calh}_{1\beta} ,
\label{eq:sup_matt}
\eea
where Greek letters denote SU(5) indices, $i,j=1..3$ are generation indices and $\epsilon$ is
the totally antisymmetric tensor with $\epsilon_{12345}=1$.
The $\hat{D}^c_i$ and $\hat{L}_i$ superfields of the MSSM reside in the $\bf{\overline{5}}$
representations, $\hat{\phi}_i$, while the $\hat{Q}_i,\ \hat{U}^c_i$ and $\hat{E}^c_i$ 
superfields are in the $\bf{10}$ representations,
$\hat{\psi}_i$. The new dimensional parameters $\mu_H$ and $\mu_\Sigma$ are of $\mathcal{O}(\mgut)$.
The soft supersymmetry breaking part of the Lagrangian involving scalar components of
chiral superfields can then be written as 
\bea 
\mathcal{L}_{soft}(\psi,\phi)  &\ni& -m^2_{\bar{5}} |\phi|^2  - m^2_{10} \rm{Tr} \left[ \psi^\dagger \psi\right]\nonumber\\
 &+ & \left[ b_\Sigma \Tr\hat{\Sigma}^2 + \frac{1}{6} a' \Tr\hat{\Sigma}^3 + b_H \hat{\calh}_{1\alpha} \hat{\calh}_2^{\alpha} 
 + a \hat{\calh}_{1\alpha}\hat{\Sigma}^{\alpha}_{\beta} \hat{\calh}_2^{\beta} \right] \nonumber \\
 &+& \left[\frac{1}{4} \ a_{10}\  \epsilon_{ijklm} \psi^{ij}\psi^{kl} \  {\hat{\mathcal{H}}}_2^m 
 + \sqrt{2} \ a_{\bar 5} \ \psi^{ij}  \phi_i   \  {\hat{\mathcal{H}}}_{1j} + H.c. \right],
 \label{eq:lagr}
\eea
where the soft parameters $\{a, b \}$ are assumed to be of the same order as
$m^2_{\bar{5}}$ and $m^2_{10}$, and hence of $\mathcal{O}(M_{weak})$.

\subsection{Boundary conditions at $\Mi$ \label{subsec:BCMin}}

%\paragraph{Supersymmetry-breaking parameters}

The no-scale structure (\ref{noscale}) requires that 
all supersymmetry-breaking soft masses and bi- and tri-linear terms for the fields $\Phi_i$ vanish
at $\Mi$, so that
\bea
\label{eq:noscaleconds}
m_0=B_0=A_0=0,
\eea
where the bilinear couplings $b_\Sigma$ and $b_H$ in  (\ref{eq:lagr}) are related to the
corresponding superpotential terms by
\bea
\label{eq:condatMinBb}
b_\Sigma \; \equiv \; B_{\Sigma} \mu_{\Sigma},\quad
b_H \; \equiv \; B_{H} \mu_{H}.
\eea
and the trilinear couplings $a', a, a_{10}$ and $a_{\bar 5}$ in  \eq{eq:lagr} are related to the
corresponding Yukawa couplings by
\bea
\label{eq:condatMinAa}
a' \; \equiv \; A' \lambda',\quad
a \; \equiv \; A \lambda,\quad
(a_{10})_{ij}\; \equiv \; A_{10 ij} (h_{10})_{ij},\quad
(a_{\bar{5}})_{ij} \; \equiv \; A_{5 ij} (h_{\bar{5}})_{ij} ,
\eea
where no summation over repeated indices is implied.
Having specified the boundary conditions on scalar masses
as well as setting $A_0 = B_0 = 0$, we no longer have the freedom of choosing $\tan \beta$ as a free
parameter. Instead, the minimization of the Higgs potential provides the solutions for both the MSSM
Higgs mixing parameter $\mu$ and $\tan \beta$~\cite{vcmssm}.
Thus the theory is defined by 4 parameters: 
\beq
 m_{1/2},\ M_{in},\ \lambda,\ \lambda',
\eeq
where we will denote the gaugino mass above the GUT scale by $M_5$. In addition, the sign of the MSSM $\mu$ 
parameter must also be specified~\footnote{It is determined from $\mu_H$, $\lambda$, and the
vacuum expectation value of $\Sigma$ \cite{Borzumati}.}.

\subsection{Boundary conditions at $\Mg$}

In the previous subsection, we specified the boundary conditions on the soft supersymmetry-breaking 
parameters at the input universality scale $\Mi$.  Using the GUT RGEs, these are run down to $\mgut$ where 
they must be matched with their MSSM equivalents.
At the GUT scale, we have 
\bea
\label{eq:bcatMGUT}
\begin{array}{ll}
M_i=M_5, &  \\
%m^2_{D_{1, 2}}=m^2_{L_{1, 2}}=m^2_{\bar{5}_{1, 2}}, & a_t= {4} a_{10},\\
%m^2_{D_3}=m^2_{L_3}=m^2_{\bar{5},3}, & a_b=a_\tau=a_{\bar{5}}/{{\sqrt{2}}} ,\\
m^2_{D}=m^2_{L}=m^2_{\bar{5}}, & a_t= {4} a_{10},\\
%m^2_{Q_{1, 2}}=m^2_{U_{1, 2}}=m^2_{E_{1, 2}}=m^2_{10_{1, 2}}, & m^2_{H_d}=m^2_{\mathcal{H}_1},\\
%m^2_{Q_3}=m^2_{U_3}=m^2_{E_3}=m^2_{10,3},& m^2_{H_u}=m^2_{\mathcal{H}_2}.\\
%m^2_{Q}=m^2_{U}=m^2_{E}=m^2_{10}, & m^2_{H_d}=m^2_{\mathcal{H}_1},\\
m^2_{Q}=m^2_{U}=m^2_{E}=m^2_{10}, & a_b=a_\tau=a_{\bar{5}}/{{\sqrt{2}}} ,\\
m^2_{H_d}=m^2_{\mathcal{H}_1}, & m^2_{H_u}=m^2_{\mathcal{H}_2}.\\
\end{array}
\eea
We treat the gauge and Yukawa couplings differently,
inputting their values at the electroweak scale and matching 
to their SU(5) counterparts at $\Mg$. 
The minimal SU(5) relations 
\bea
\label{eq:minimal_su5}
h_E(\Mg)= {h_D(\Mg)}^T,
\eea
between the charged-lepton and for the down-type Yukawa couplings
are unrealistic since they do not produce the right values of  lepton
masses at $\Mw$, except possibly for the third generation.
We assume here that at $\Mg$
\bea
\label{eq:hdmatch_atMg}
h_D(\Mg)= {h_{\bar{5}}(\Mg)}/\sqrt{2},
\eea
but we do not match $h_E$ to  $h_5$ at $\Mg$. Instead we use the
values that $h_E$ should have at $\Mg$ in order to produce the
observed lepton masses. One way to justify this assumption would be
to allow the lepton sector to have additional, non-renormalizable couplings
besides the minimal renormalizable $\bar{5}$ couplings~\cite{EG}, so at $\Mg$ one has
\bea
h_E(\Mg)= {h_{\bar{5}} (\Mg)}^T/\sqrt{2} + \ \text{other interactions},
\eea
where these other interactions are too small to be important for the quark sector.
Thus for the gauge and Yukawa couplings, we determine their SU(5) counterparts as
\bea
%\begin{array}{ll}
\label{eq:YukMSSM_to_SU5}
g_i=g_5 \nonumber, \\ h_t=4 h_{10}, \nonumber \\
(h_D+ h_E) _{33}/2= {h_{\bar{5}}}_{33}/\sqrt{2}, \nonumber\\
{h_D(\Mg)}_{ij}= {h_{\bar{5}}(\Mg)}_{ij}/\sqrt{2}, \ \forall \ \{i,j\} \ \text{except}\
\{i,j\}={33}, 
%\end{array}
\eea
The corresponding experimental inputs for the Yukawa couplings at the weak scale are shown in Table~\ref{tab:fmassesMZ}.

\begin{table}[ht!]
\centering
\begin{tabular}{|c|c|c|}
\hline
\hline
&  Mass values [GeV] & $\begin{array}{c} m^{\overline{MS}}_f(M_Z)  \text{[GeV]}\end{array}$\\
\hline
%${m}_t$ &   $173.07\pm 0.52\pm 0.72 $   &    $171.31 \pm 0.98$  \\
${m}_t$ &   $173.21\pm 0.51\pm 0.71 $   &    $171.46 \pm 0.96$\\ %Updated Feb 8, 2016
%${m}_b^{\text{OS}}$ &   $4.66 \pm 0.03$   &  \\
$m_b$ &   $4.18\pm 0.03 $     &       $2.85 \pm 0.04 $  \\
$m_c$ &   $1.275\pm 0.025 $   &     $0.63 \pm 0.025 $  \\
$m_s$ &   $0.095\pm 0.005 $  &     $0.059 \pm 0.0033 $ \\   
$m_d$ &   $4.8^{+0.5}_{-0.3}\times 10^{-3}$  & $0.0028 \pm 0.0004 $ \\
%$m_u$ &    $(0.0018,0.003)$    &  $0.0013 \pm 0.0005$ \\
$m_u$ &    $2.3^{+0.7}_{-0.5}\times 10^{-3}$    &  $0.0013 \pm 0.0005$ \\
\hline
$m_e$ & $(0.51 \pm (1.1\times 10^{-8}  ) ) \times 10^{-3} $ & $(0.49\pm (4.2\times 10^{-8}))\times 10^{-3} $\\
$m_\mu$ & $(105.66 \pm (3.5 \times 10^{-6}))\times 10^{-3}$ & $(102.72\pm (9.2 \times 10^{-6}))\times 10^{-3} $\\
$m_\tau$& $1.78\pm (1.2\times 10^{-4})$ &$1.75\pm (2\times 10^{-4})$\\ 
\hline
\end{tabular}
\caption{\footnotesize{\it Values of the fermion masses, in GeV, as appear in current edition of the 
PDG review~\cite{rpp}. The quoted quark mass values at $M_Z$ were obtained with the program 
{\tt RunDec}~\cite{Chetyrkin:2000yt}. 
The values of the charged lepton masses were taken from~\cite{Xing:2007fb}.}}
\label{tab:fmassesMZ}
\end{table}

We use for our renormalization-group calculations the program {\tt SSARD}~\cite{ssard}, 
which computes the sparticle spectrum on the basis of
2-loop RGE evolution for the MSSM and 1-loop evolution
for minimal SU(5). 
We define $\mgut$ as the scale where $g_1=g_2$, so that
$\mgut \simeq 10^{16}$~GeV, with its exact value depending on the values of other parameters. 
The value of $g_3$ at $\mgut$ is within the threshold uncertainties in the GUT matching conditions.

\subsection{Non-zero off-diagonal Yukawa couplings}
 
It is well known that renormalisation interrelates the soft masses-squared, trilinear
and Yukawa couplings. In particular, a non-zero diagonal
soft mass-squared term, the  K\"ahler potential, the F terms and the
Yukawa couplings could be seeds for non-zero off-diagonal soft masses-squared and trilinear
terms. Alternatively, even if the Yukawa couplings were flavor-diagonal, there would be non-diagonal soft masses-squared and
trilinear terms if the K\"ahler potential~\cite{EK} or the F terms were flavor non-diagonal.

However, in the case of no-scale supergravity boundary conditions at $\Mi$ there is no source
of non-zero trilinear or soft masses-squared, apart from the running
induced by the renormalization-group $\beta$ functions.  
However, off-diagonal Yukawa couplings
at $\Mi$ would generate, via renormalization-group running, off-diagonal soft masses-squared and
trilinear terms. The ultimate goal of our study is to quantify how large
these parameters could be near $\Mi$ before they become problematic
at $\Mw$. 

In order to understand the effects of the Yukawa couplings on the evolution of the soft
supersymmetry-breaking parameters, we initialize our study by enforcing conditions
at $\Mw$ that reproduce the CKM matrix. 
In particular, we take the values of the quark masses at $\Mw$ to be those at $M_Z$ 
(see the second column of Table~\ref{tab:fmassesMZ}), and use 
$h_D(M_Z)=\sqrt{2}\, m_D(M_Z)/v/\cos\beta$ and $h_U(M_Z)=\sqrt{2}\,  m_U(M_Z)/v/\sin\beta$.
The diagonalization of the Yukawa couplings is defined by 
\bea
h_D=V^D_R\hat h_D V^{D\dagger}_L,\quad
h_E=V^E_R\hat h_E V^{E\dagger}_L,
\label{eq:VdandVe}
\eea
where $\hat h_D$ and $\hat h_E$  are diagonal matrices, and the unitary matrices
$V_{L, R}^{D, E}$ are such that the CKM matrix is $V^{U\dagger}_L V^D_L$.

Since the structure of the Yukawa couplings cannot be determined in a model-independent way,
we adopt the minimal assumption that the CKM matrix is the only source of flavor violation
in the Yukawa couplings in the $D$ sector, and study the differences
induced by different assumptions for the $E$ sector. Thus, we are
assuming that $h_U$ is diagonal at $\Mw$.
Using this condition, 1-loop running does not generate off-diagonal terms,
and the off-diagonal entries generated at  $\Mg$ at the 2-loop level are negligibly small.

We remind the reader that, in contrast to the Standard Model,  the
MSSM observables are sensitive to right-handed currents, which
have no Standard Model counterparts.  Therefore, it is not possible to make
predictions without additional assumptions~\footnote{Such assumptions can have important effects on the observables,
and can help to determine the choices of Yukawa structures compatible with a given supersymmetric model~\cite{Olive:2008vv}. } 
on the Yukawa couplings, and hence the $V_R^{D, E}$
as well as the $V_L^{D, E}$.

We therefore consider the following illustrative Ans\"atze that illustrate the range of possibilities:
\bea
A1. && V^{D*}_R=V^{D}_L=V_{\rm{CKM}}, \quad \quad \quad V^{E}_R=V^E_L=1, \label{eq:Az1}\\
A2. && V^D_L=V_{\rm{CKM}}, \quad V^D_R=1, \quad \ V^E_R=V^E_L=1, \label{eq:Az2}\\
A3. && V^D_R =V^{E}_L=1,\quad \quad \quad \quad \quad  \ V^D_L=V^{E*}_R=V_{\rm{CKM}}, \label{eq:Az3}\\
A4. && V^{D*}_R=V^D_L=V_{\rm{CKM}}, \quad  \quad \quad V^{E*}_R=V^E_L=V_{\rm{CKM}}.\label{eq:Az4}
\eea
The Ans\"atze A3 and A4 are compatible with the minimal
SU(5) conditions (\ref{eq:minimal_su5}) and (\ref{eq:VdandVe}). However, we will focus later on Ans\"atze A1 and A2
because, as we shall see,  Ans\"atze A3 and A4 give rise to unacceptably large
flavour-violating processes in the lepton sector.
We use examples A1 and A2 to illustrate the determination of the diagonal and off-diagonal 
soft masses-squared and trilinear terms that are generated below $\Mi$. These are constrained by flavour 
observables, and our goal is to determine how large the deviations from pure no-scale boundary conditions
can be, before they induce flavour violations in contradiction with experiment.

%\section{Flavor violating parameters}
\section{Running of Parameters}

In this Section, we restrict our attention to pure no-scale boundary conditions, and
assume the Ansatz A1 for the CKM mixing among fermions.  
As already mentioned, the boundary conditions for the soft supersymmetry-breaking parameters 
are fixed at $\Mi$, the gauge and Yukawa couplings are fixed at the weak scale,  and
all parameters are matched at $\mgut$ to allow for running above and below the GUT scale. 
Though the off-diagonal sfermion masses begin their RGE evolution with the no-scale 
boundary conditions, i.e., they vanish at $\Mi$, they contribute to
low-energy flavour observables in a model-dependent way after renormalisation,
as we now calculate.

To be concrete, we choose a limited set of 
benchmark points with different choices of  $\lambda$, $\lambda'$ and $\Mi$
(all masses are expressed in GeV units). 
The 4-dimensional parameter space of the super-GUT no-scale model was explored in \cite{emo2}.
It was found that unless $\lambda/\lambda' < 0$ with $|\lambda | < |\lambda' |$, both $m_{1/2}$ and $\Mi$
are pushed to relatively low values. However, the low values of $m_{1/2}$ are now in conflict with LHC searches
for supersymmetric particles \cite{lhc}. Therefore we restrict our analysis to an illustrative benchmark point {\bf B}
defined by
 \bea
\label{eq:benchyuklambd}
&&\mathbf{B}: \quad M_5=1500~{\rm GeV}, \quad \Mi=1\times 10^{18}~{\rm GeV},\quad  \lambda'=2,\quad \lambda=\ -0.1,
%&&\mathbf{B2},\quad  M_5=1000, \quad \Mi=1.2\times 10^{17},\quad  \lambda'=2,\quad \lambda=-0.1,\nn\\
%&&\mathbf{B3},\quad M_5=1050, \quad \Mi=1\times 10^{18},\quad \lambda'=2,\quad \lambda= - 0.1.
\eea
%In both examples, we choose the Higgs couplings $\lambda$ and $\lambda'$
suggested by a no-scale model \cite{ENO2013} in which the right-handed sneutrino 
is responsible for Starobinsky-like inflation. The value of 
$M_5$ is chosen so that we obtain the relic abundance corresponding to the cold
dark matter density determined by Planck and other experiments~\cite{planck}. As
noted earlier, after setting $A_0 = B_0 = 0$, we no longer have the freedom of choosing $\tan \beta$ as a free
parameter.  For benchmark {\bf B}, we find $\tan \beta \simeq 52$.
The value of $M_5$ is large enough to satisfy LHC bounds from 
supersymmetric particle searches and the lightest Higgs mass is $m_h = 125.0 \pm 3.1$ GeV
when calculated with the {\tt FeynHiggs} code~\cite{fh}, which is comfortably consistent with the
joint ATLAS and CMS measurement of $m_h$~\cite{LHCmh}.
It is important to compare correctly theoretical predictions of  $B_s \to \mu^+ \mu^-$ with
its experimental value, as was reviewed in~\cite{DeBruyn:2012wj}. In particular,  one should compare the
``untagged'' computed value, instead of the tagged one, to its experimental counterpart. The relevance of this comparison for some supersymmetric scenarios was studied in~\cite{Arbey:2012ax}, where it
was pointed out that this difference is important for evaluating the validity of some scenarios. Using the {\tt
  SUSY\_FLAVOR} code~\cite{susyflavor} and the latest hadronic observables, we find that for the benchmark 
  point under consideration the tagged value is $3.42 \times 10^{-9}$, whilst using a modified
version of the {\tt SUSY\_FLAVOR} code we find that the untagged value is $3.76 \times 10^{-9}$.
Because $\tan \beta$ is relatively high for this benchmark point,
the branching ratio of $B_s \to \mu^+ \mu^-$ is somewhat large,
but within the experimental 95\% CL upper limit \cite{bmm}.

\subsection{Runnings of SU(5) parameters}

\begin{figure}[!ht]
\centering
\includegraphics[width=8.1cm, height=5.5cm]{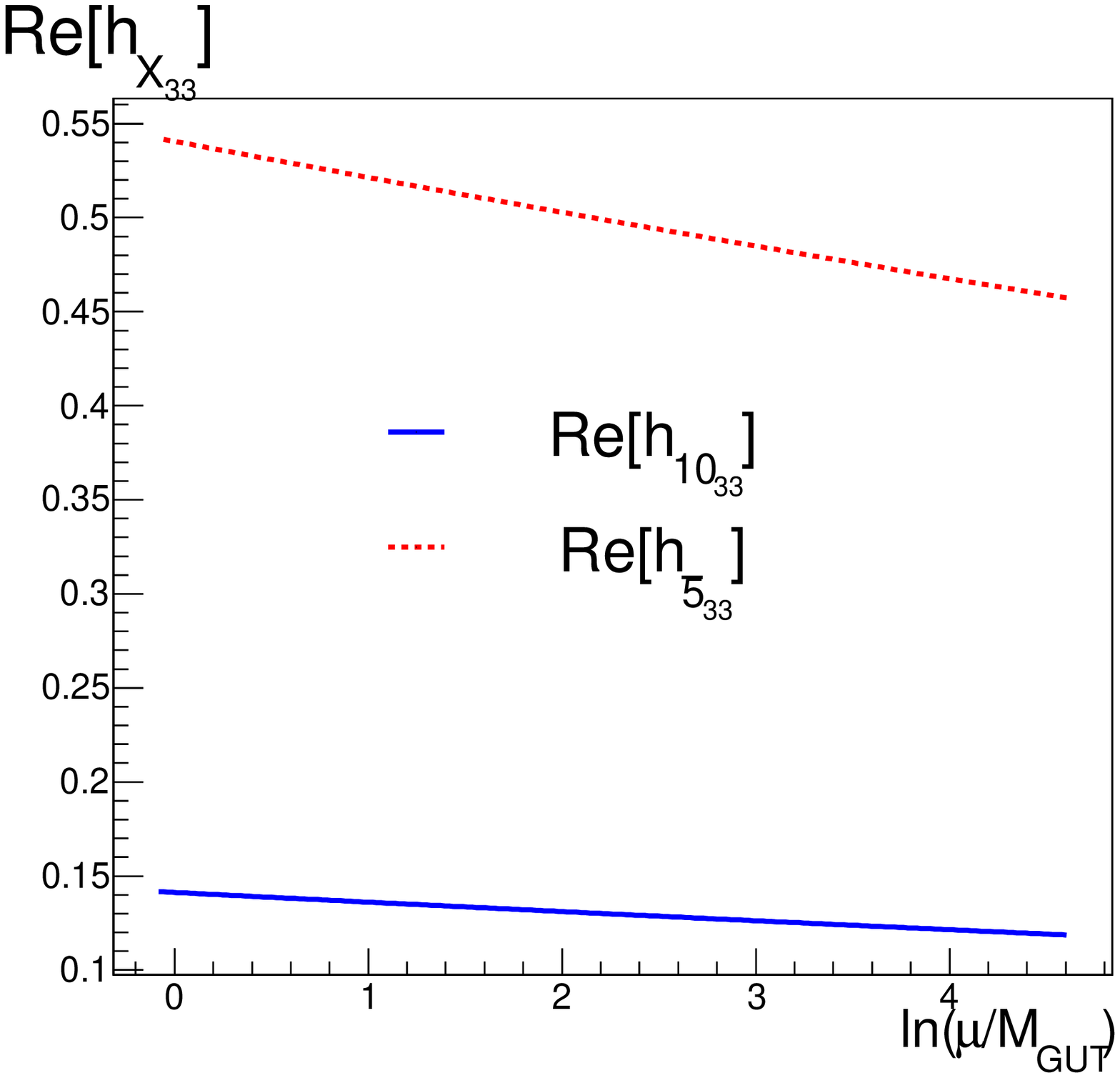}
\includegraphics[width=8.1cm, height=5.5cm]{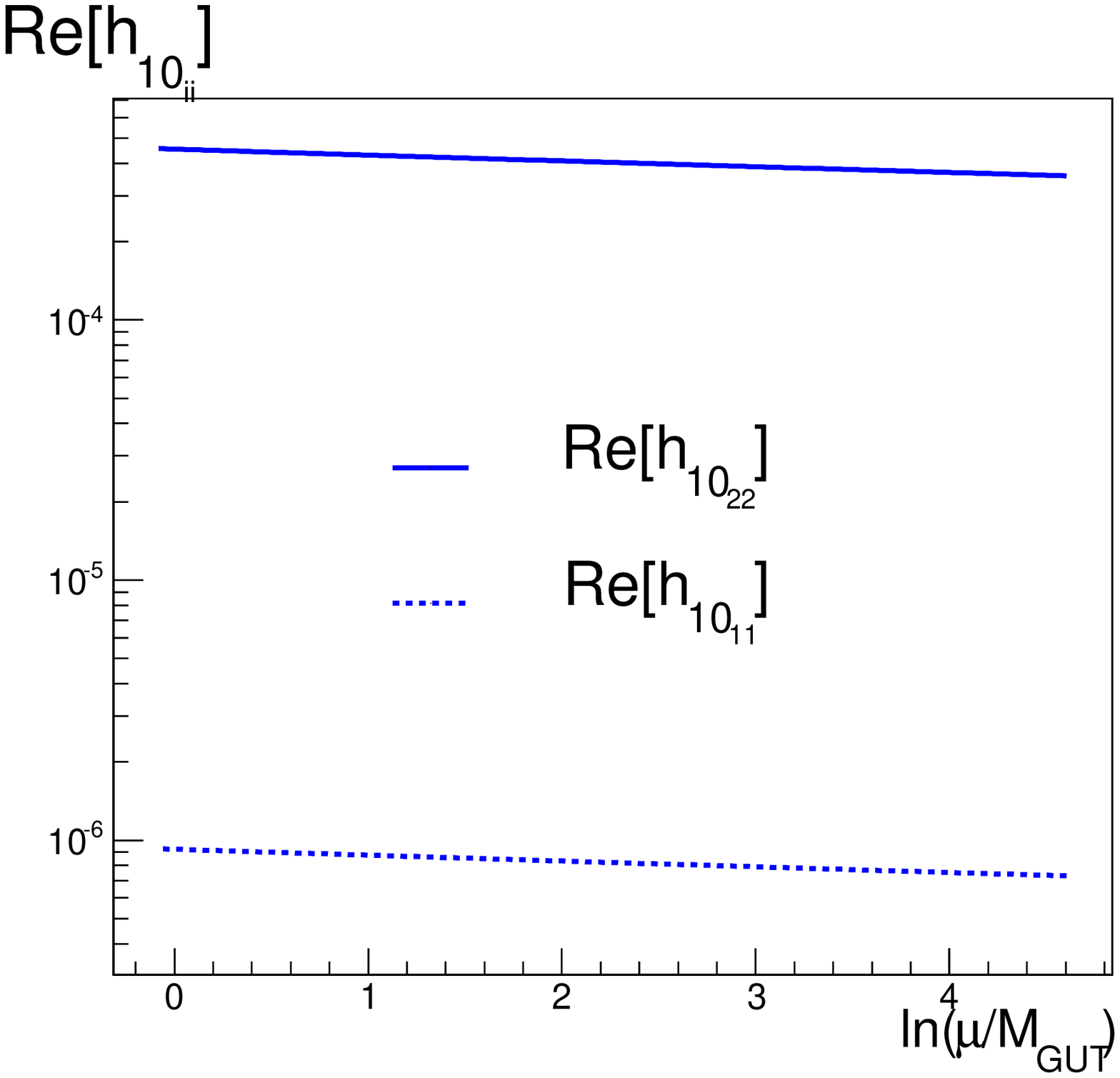}\\
\includegraphics[width=8.1cm, height=5.5cm]{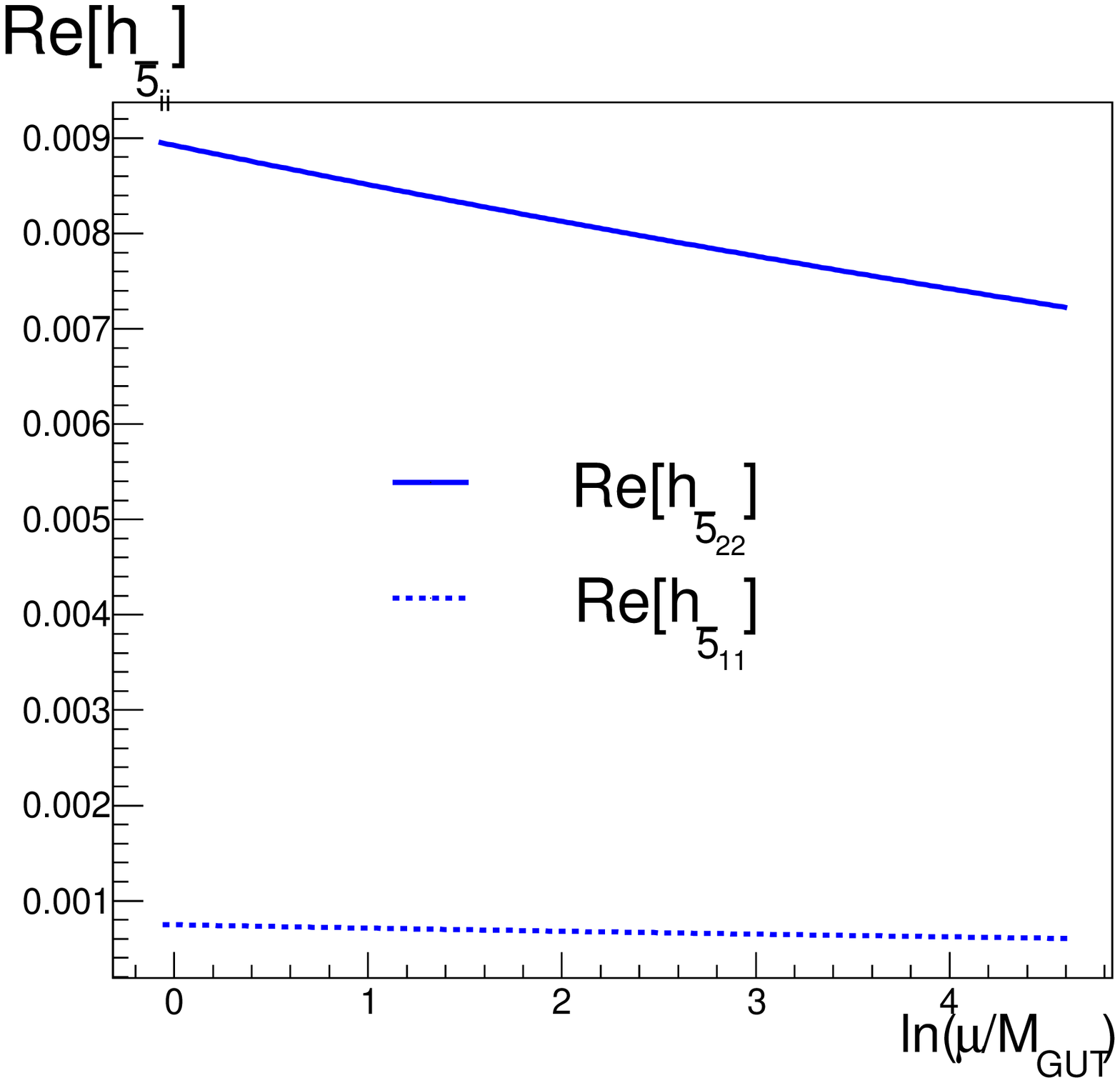}
\includegraphics[width=8.1cm, height=5.5cm]{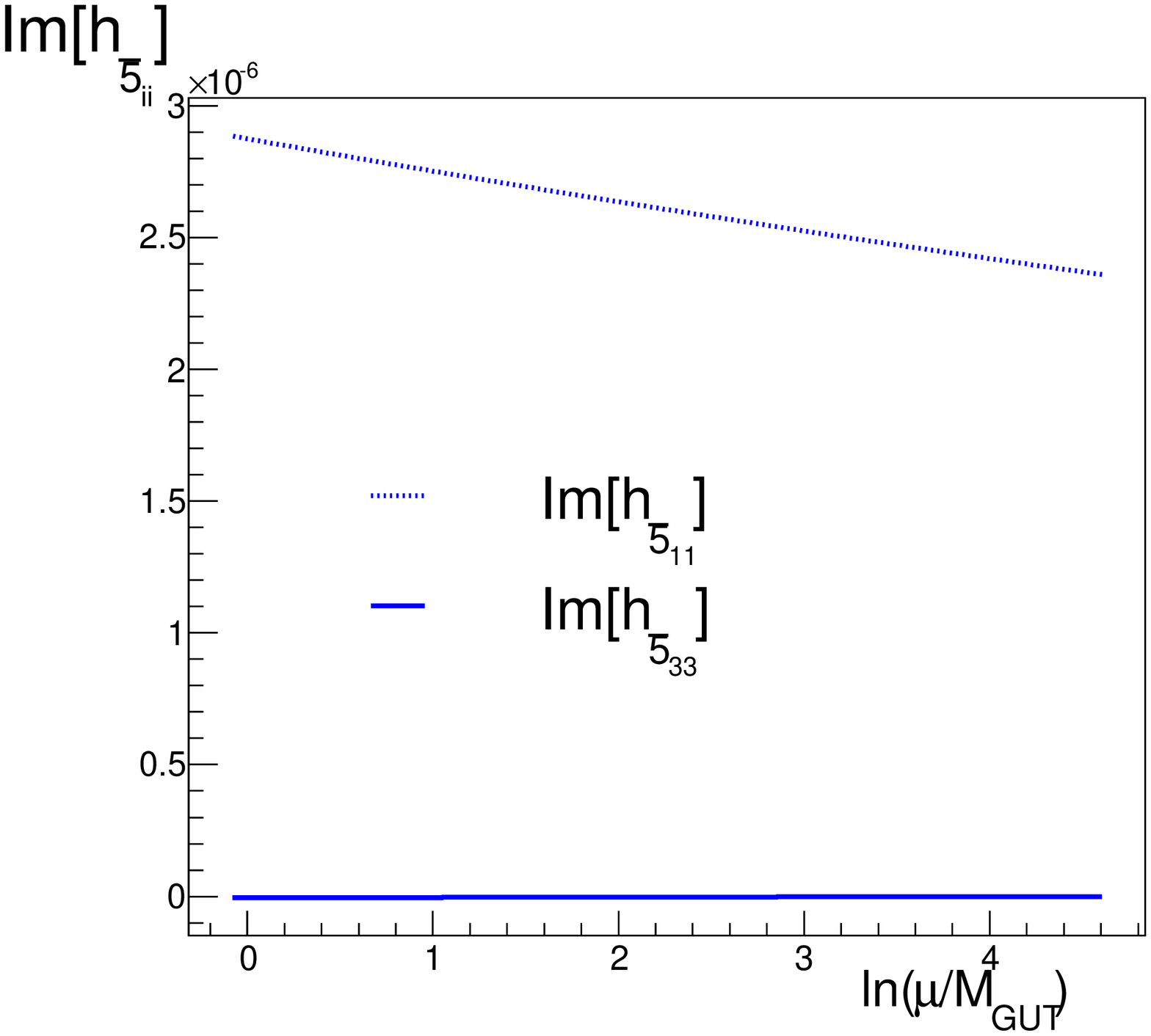}
\includegraphics[width=8.1cm, height=5.5cm]{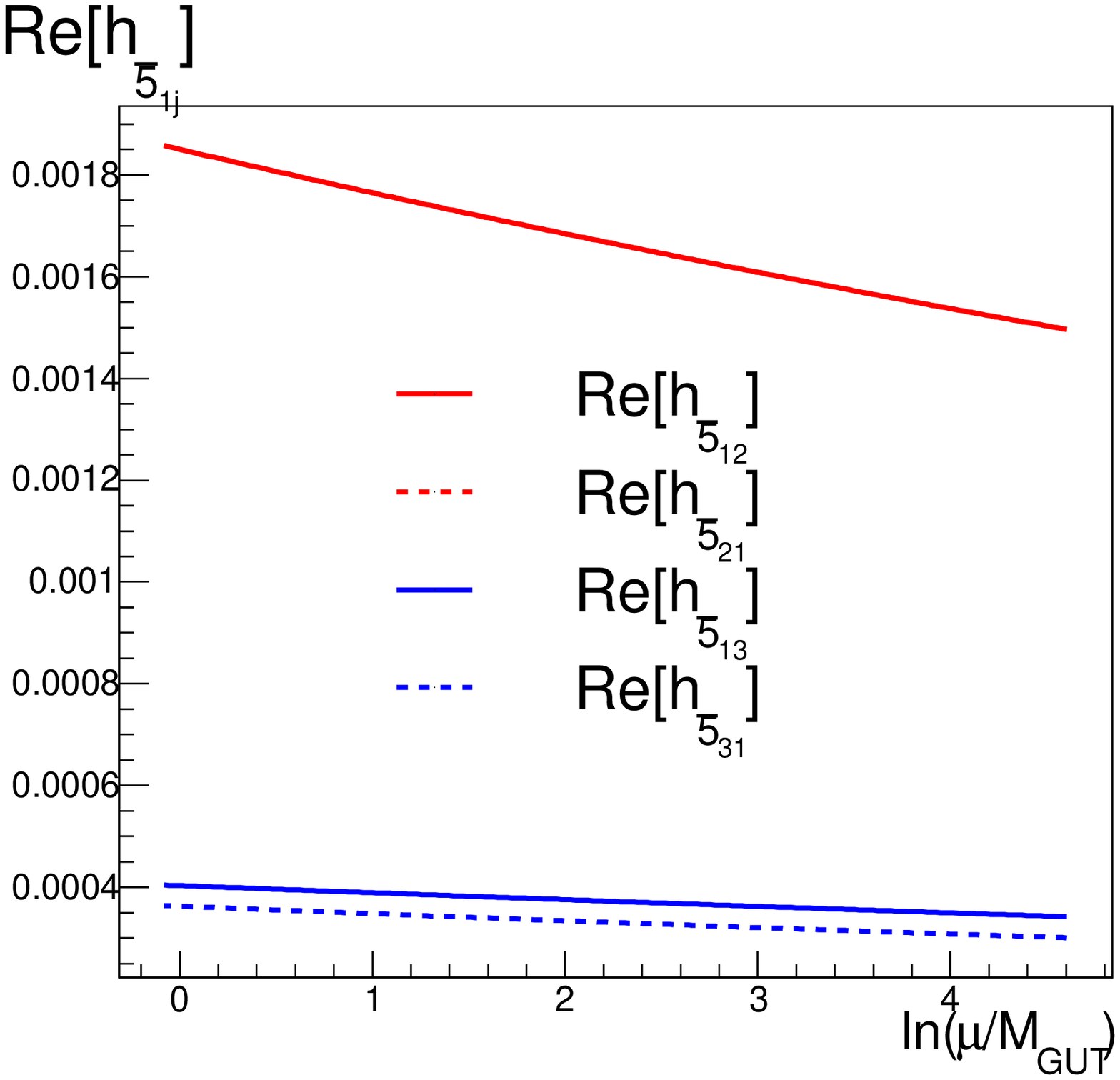}
\includegraphics[width=8.1cm, height=5.5cm]{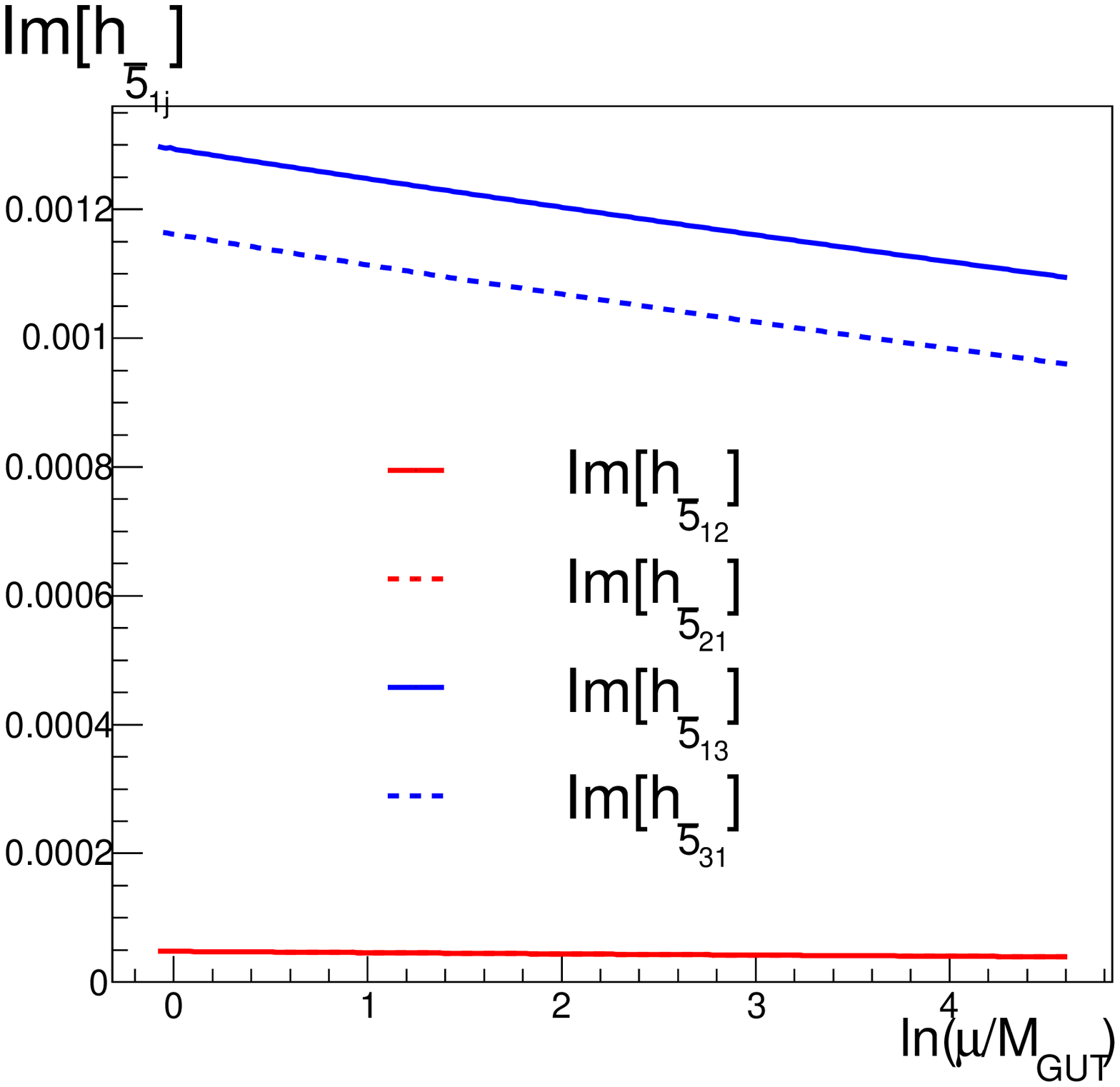}\\
\includegraphics[width=8.1cm, height=5.5cm]{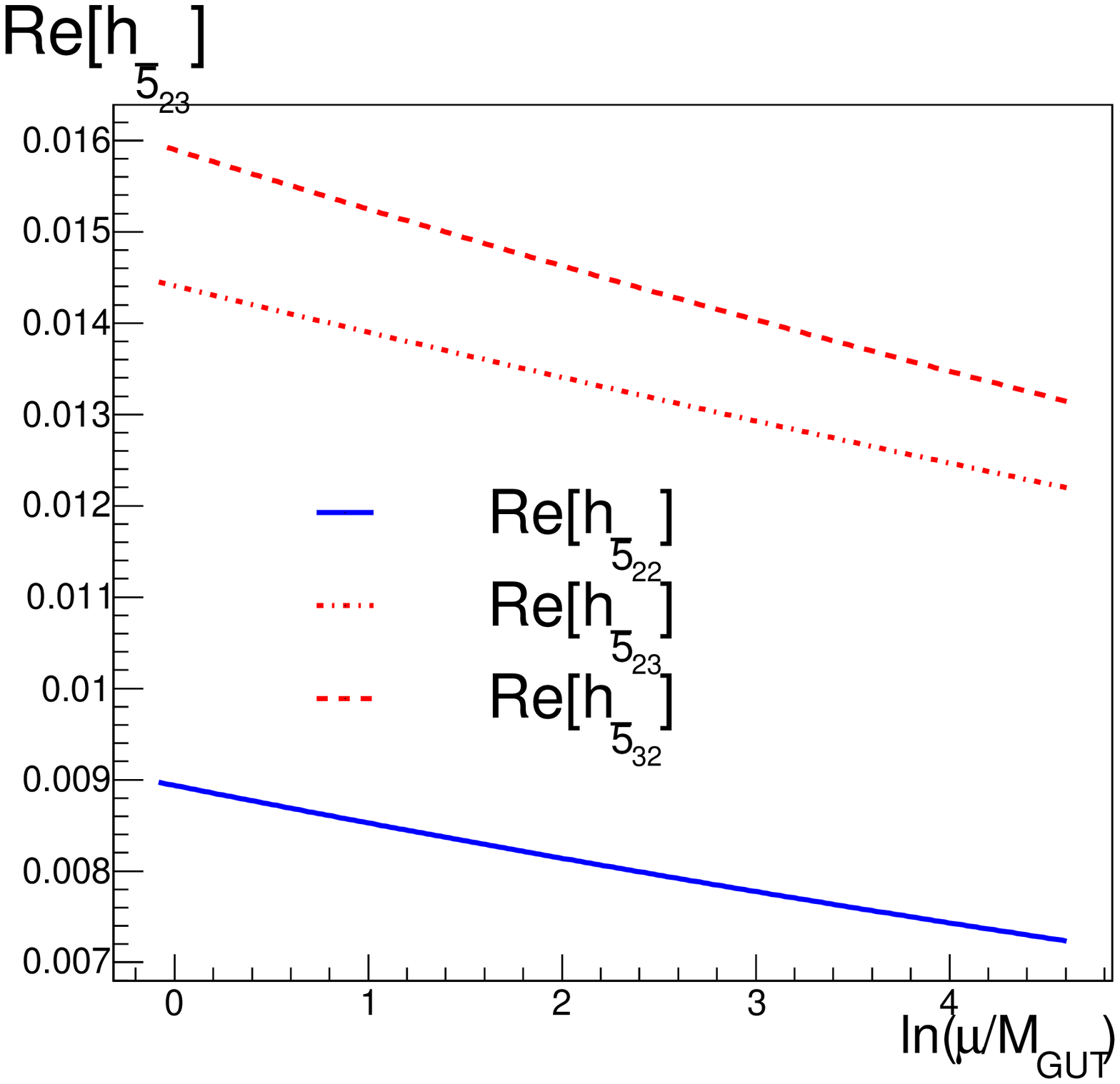}
\includegraphics[width=8.1cm, height=5.5cm]{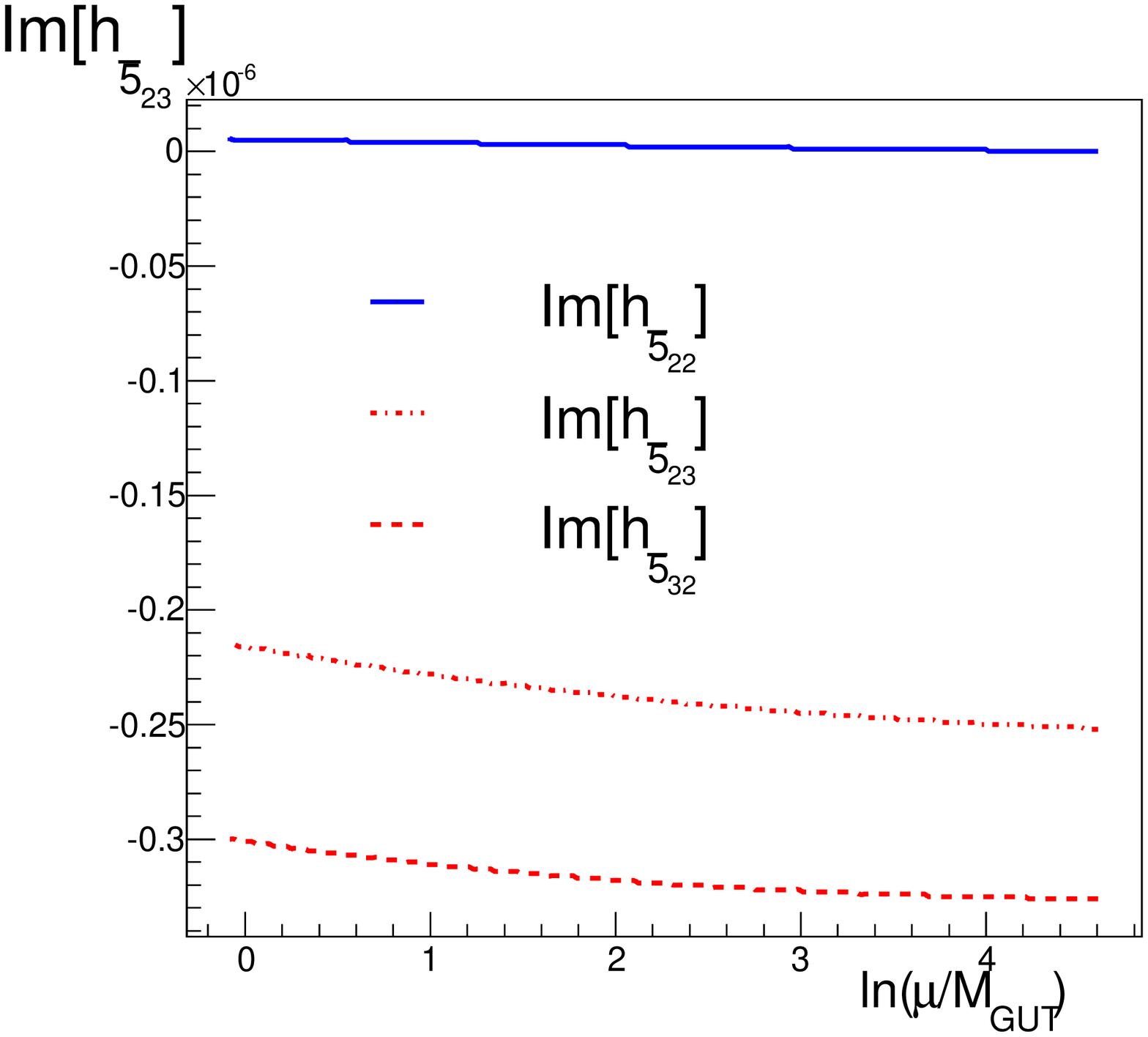}
\caption{\it{The runnings of the Yukawa couplings from $\Mi$ to $\mgut$ for the
benchmark point $\mathbf{B}$, using the patterns specified
via Ansatz A1. Note that the lines for $h_{\bar{5}_{12}}$ and $h_{\bar{5}_{21}}$ lie on top of each other.
%{\bf{$\rm{Im}[h_{\bar{5}_{3}}]$ is of $\cal{O}(10^{-9})$.}}
}
\label{fig:Trilinearssu5}}
\end{figure}

As already emphasised, the soft supersymmetry-breaking mass parameters are zero at the input scale in a no-scale model, 
but running between $\Mi$ and $\mgut$ leads in general to non-zero masses for both
diagonal and non-diagonal elements. In particular, the latter are induced by the non-diagonal Yukawa couplings
assumed in Ansatz A1. 
The runnings of the Yukawa couplings for benchmark {\bf B} are shown in Fig.~\ref{fig:Trilinearssu5}.
Recall that we have assumed diagonal Yukawa matrices for the up-quark sector and the {\bf 10} of SU(5), 
and therefore we show only the evolution of the real part of $h_{10 ii}$.  In contrast, for the ${\mathbf{\bar 5}}$ of 
SU(5), the Yukawa matrices are determined from (\ref{eq:VdandVe}) using Ansatz A1. Note that,
with this definition, these Yukawa matrices are in general not symmetric (though the runnings of $h_{5_{12}}$ and
$h_{5_{21}}$ are indistinguishable in the figure). 
The Figure shows the runnings of the Yukawa couplings from the input scale $\Mi$ ($\ln (\mu/M_{GUT}) \approx 4.6$ for
benchmark {\bf B}) down to the GUT scale ($\ln (\mu/M_{GUT}) = 0$).

\begin{figure}[!t]
\centering
\includegraphics[width=8.1cm, height=6cm]{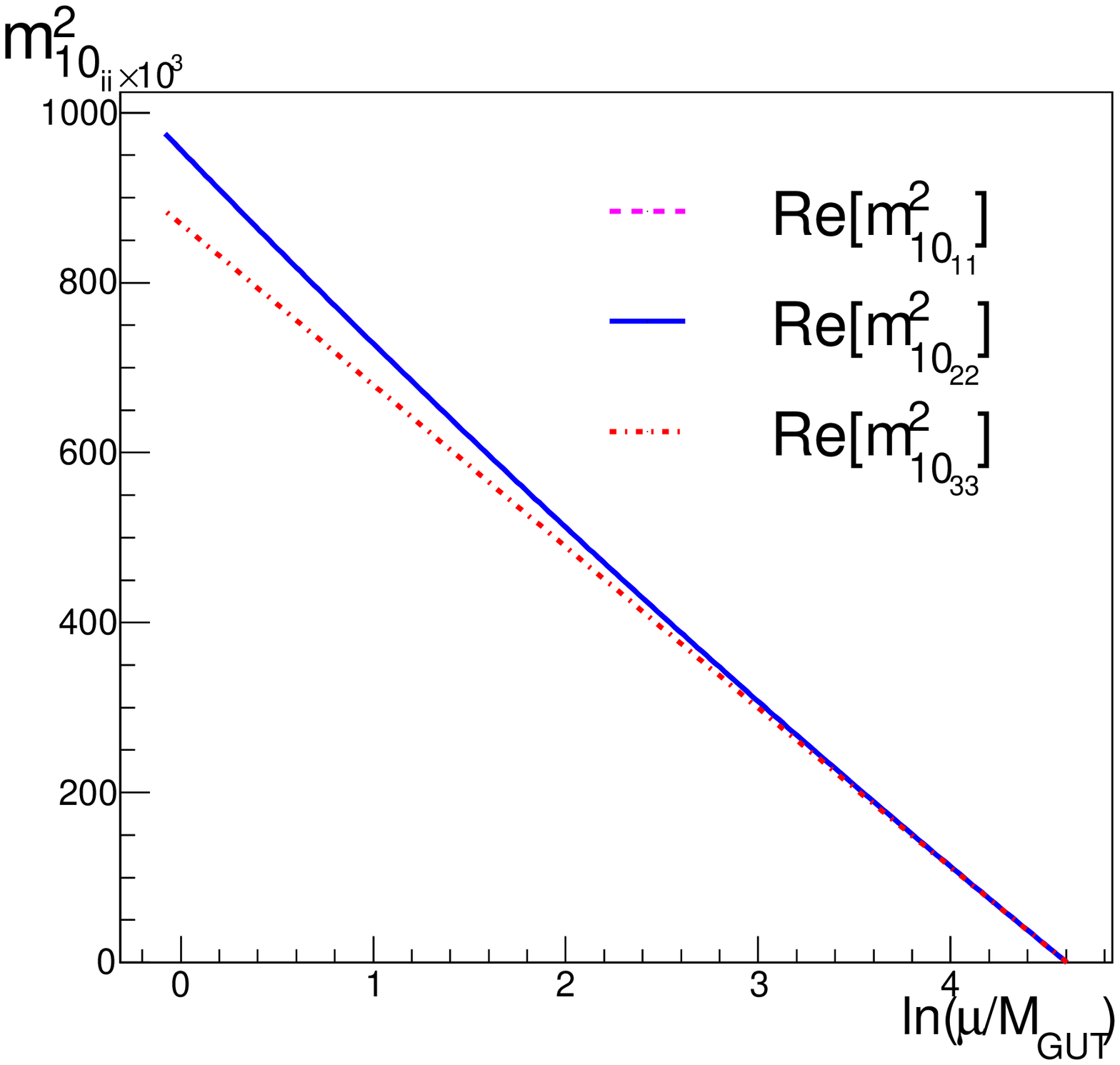}
\includegraphics[width=8.1cm, height=6cm]{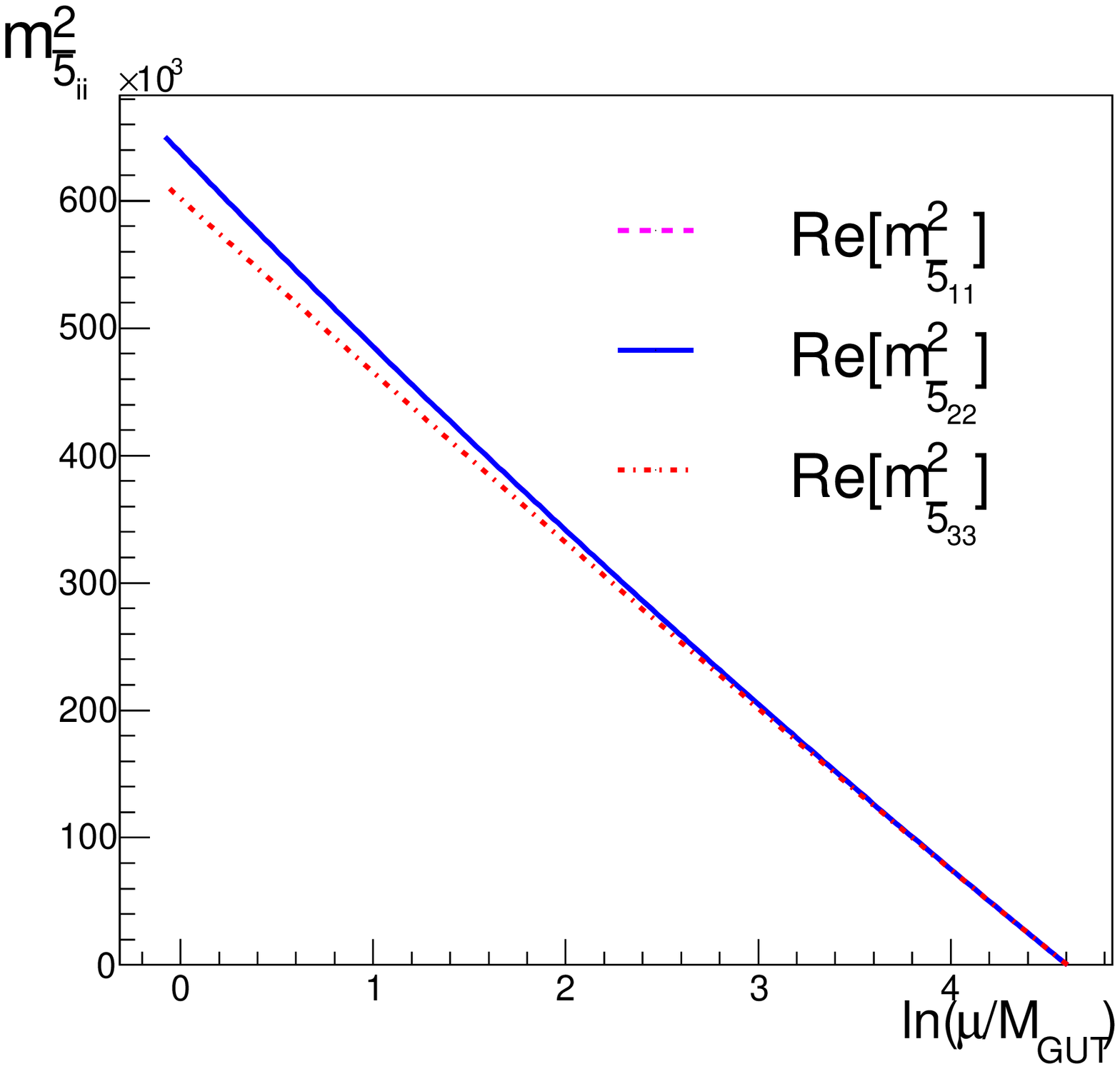}\\
\includegraphics[width=8.1cm, height=6cm]{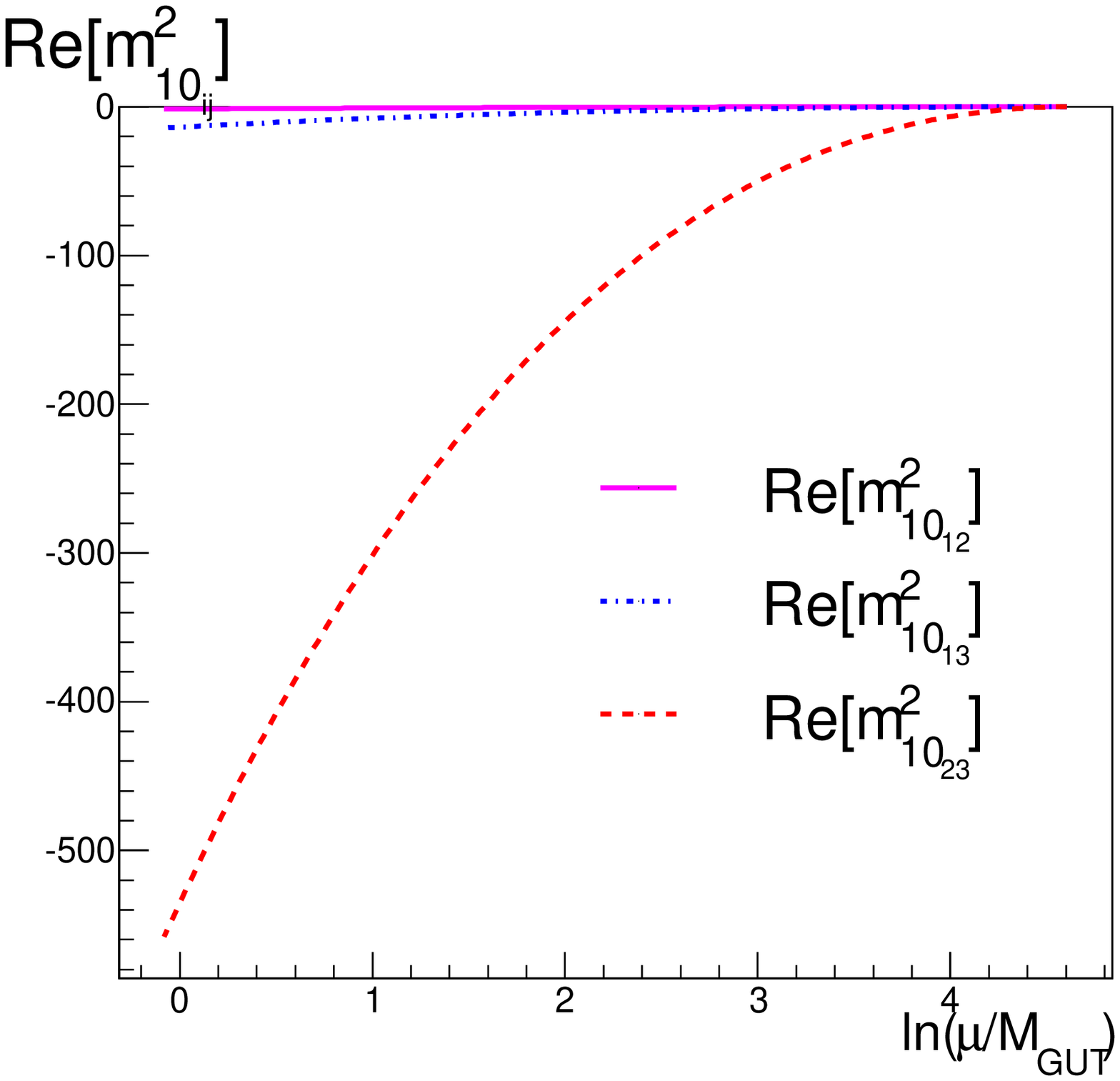}
\includegraphics[width=8.1cm, height=6cm]{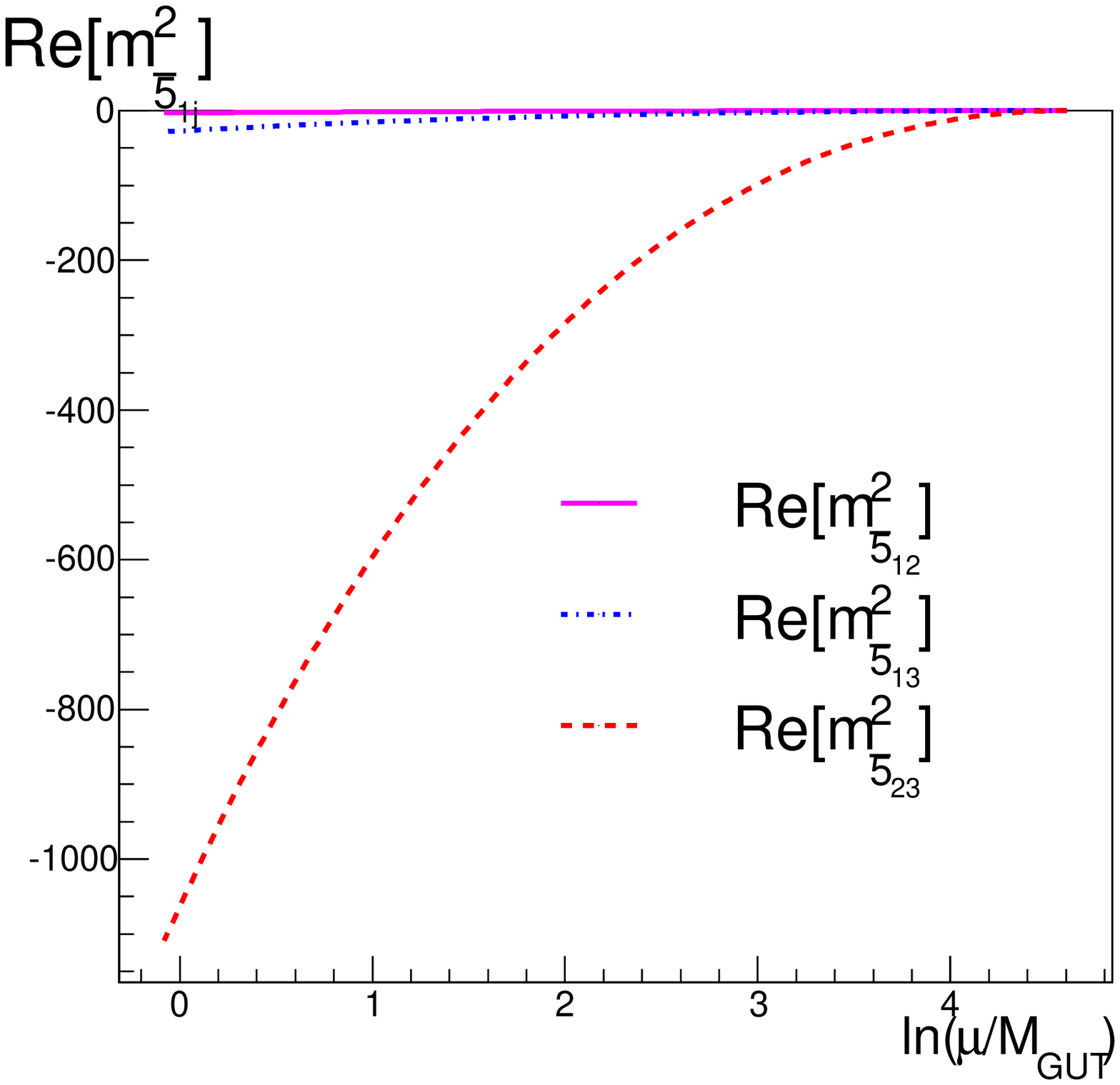}\\
\includegraphics[width=8.1cm, height=6cm]{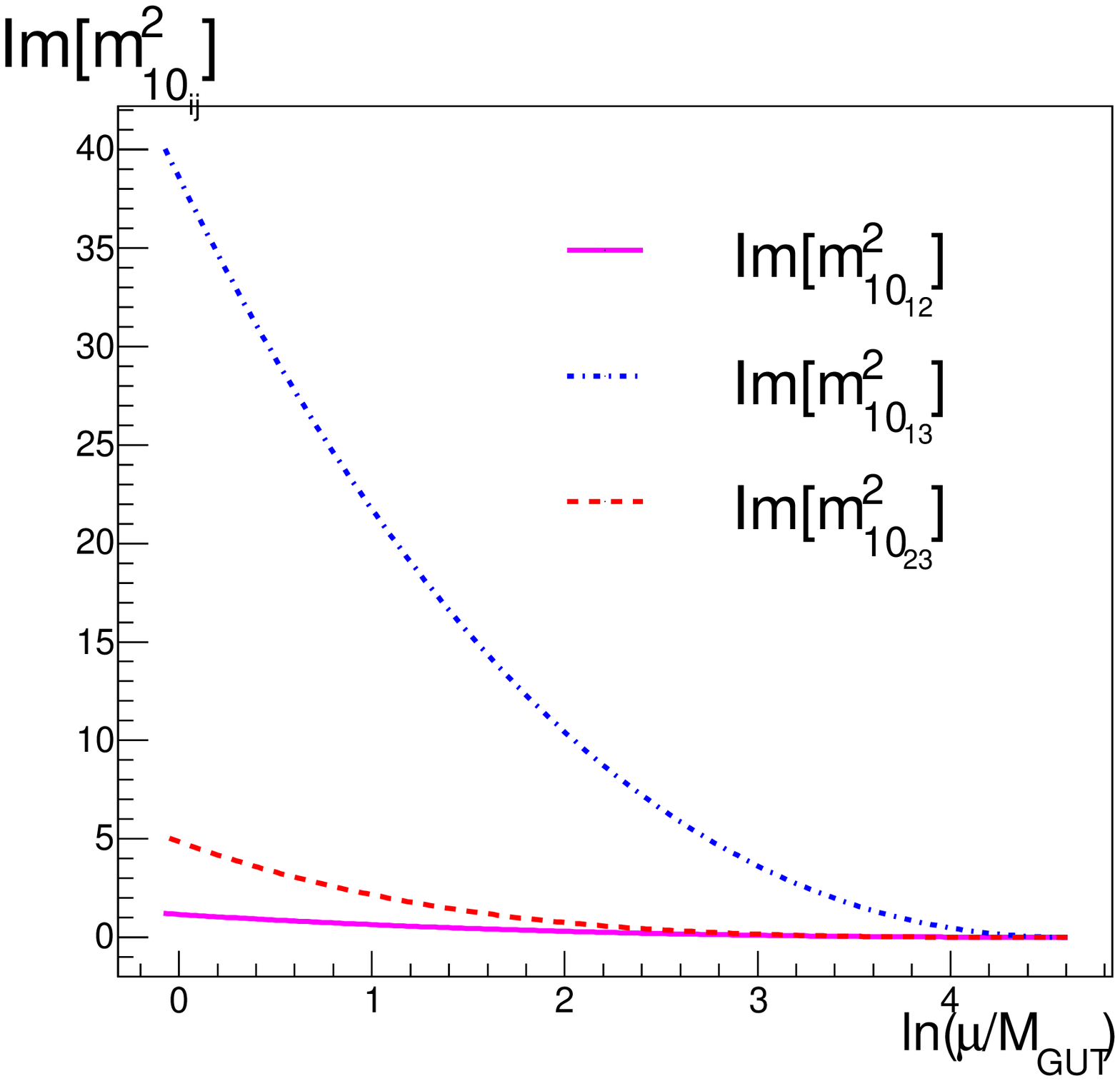}
\includegraphics[width=8.1cm, height=6cm]{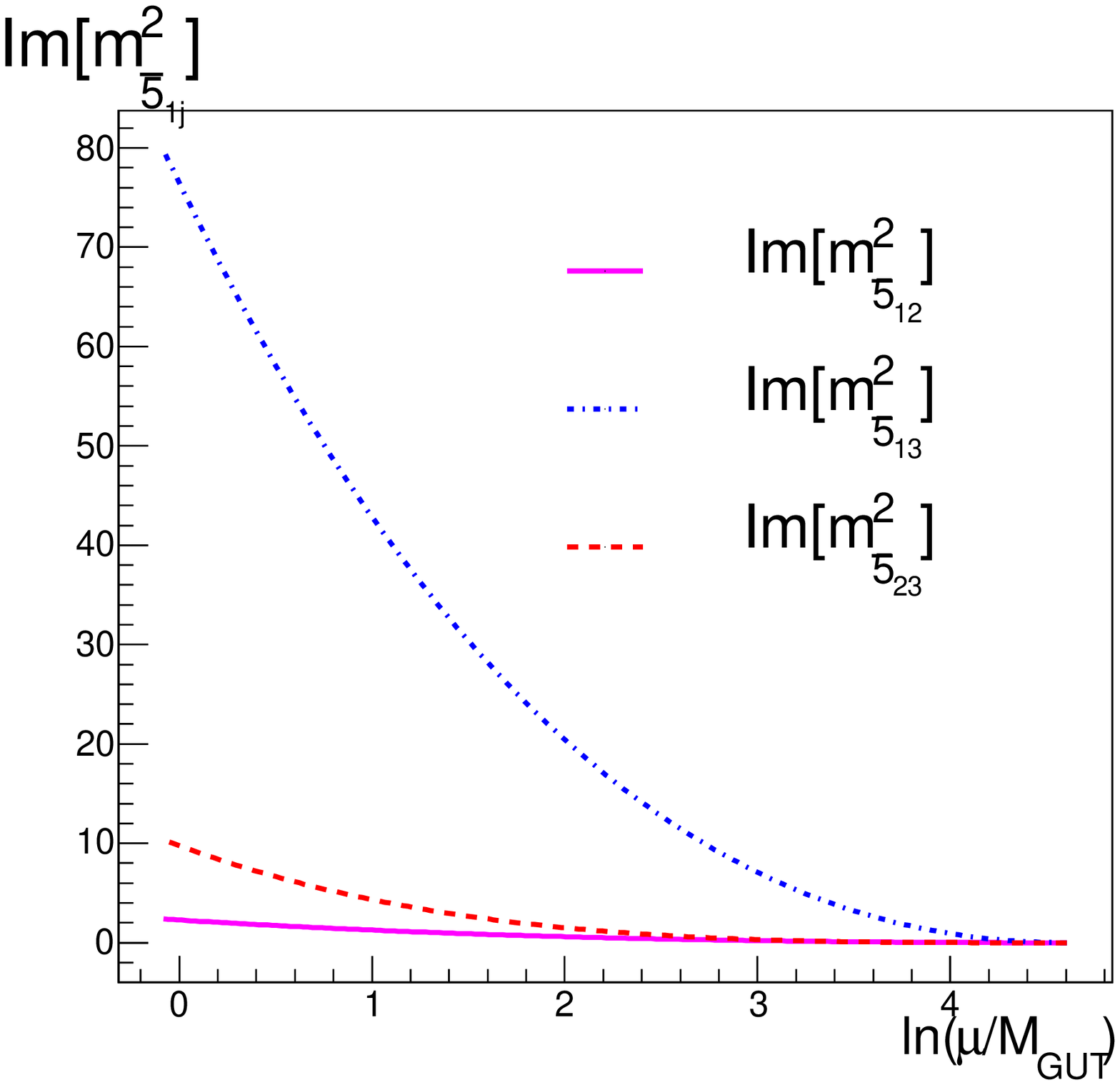}
\caption{\it{The runnings of  diagonal (left) and off-diagonal (right) soft masses-squared
associated with the {\bf 10} (left) and $\mathbf{\overline{5}}$ (right)
representations in the SU(5) model between $\Mi$ and $\mgut$ for the
benchmark point $\mathbf{B}$ (in units of GeV$^2$) using Ansatz A1 for the Yukawa couplings.  }
\label{fig:M2su5}}
%\vspace{-2cm}%Deleted space
\end{figure}

The running of the soft scalar masses is shown in \Figref{fig:M2su5},
where we have again assumed benchmark point {\bf B} and ansatz A1. 
The top panels show the running of the diagonal soft masses for the {\bf 10} and ${\mathbf{\bar 5}}$ of SU(5).
The middle panels show the real parts of the off-diagonal entries, and the bottom panels show
the imaginary parts of the same off-diagonal entries (as these are hermitian quantities, information
on the transposed entries are already given by the real and imaginary parts).  
All of the squark masses begin their evolution with $m^2 = 0$ at $\Mi$. 
The running of the diagonal components is driven by the value of the gaugino mass, here set at 
$M_{1/2}=1500$~GeV.  We clearly see from \Figref{fig:M2su5} % that  the soft mass
that the off-diagonal elements of the squark mass matrices induced by the non-diagonal
Yukawa matrices remain very small after we have imposed the no-scale boundary conditions:
we find $(\mfive)_{12},  (\mten)_{12}  \ll (\mfive)_{13} \approx (\mten)_{13}\approx  -20 
 \ \rm{GeV}^2 $, $(\mfive)_{23}\approx -1100  
\ \rm{GeV}^2, \quad  (\mten)_{23}\approx  -550
\ \rm{GeV}^2$, while  $(\mfive)_{ii}$ and $(\mten)_{ii}$ are of order $10^5$ and $10^6$ GeV$^2$, respectively. 
Later we will use this evolution to place constraints on the size of 
the possible sizes of the off-diagonal elements at $\Mi$.

\subsection{Running of MSSM parameters}
\begin{figure}[ht!]
\centering
\hspace{-0.5cm}
\includegraphics[width=8.1cm, height=5.5cm]{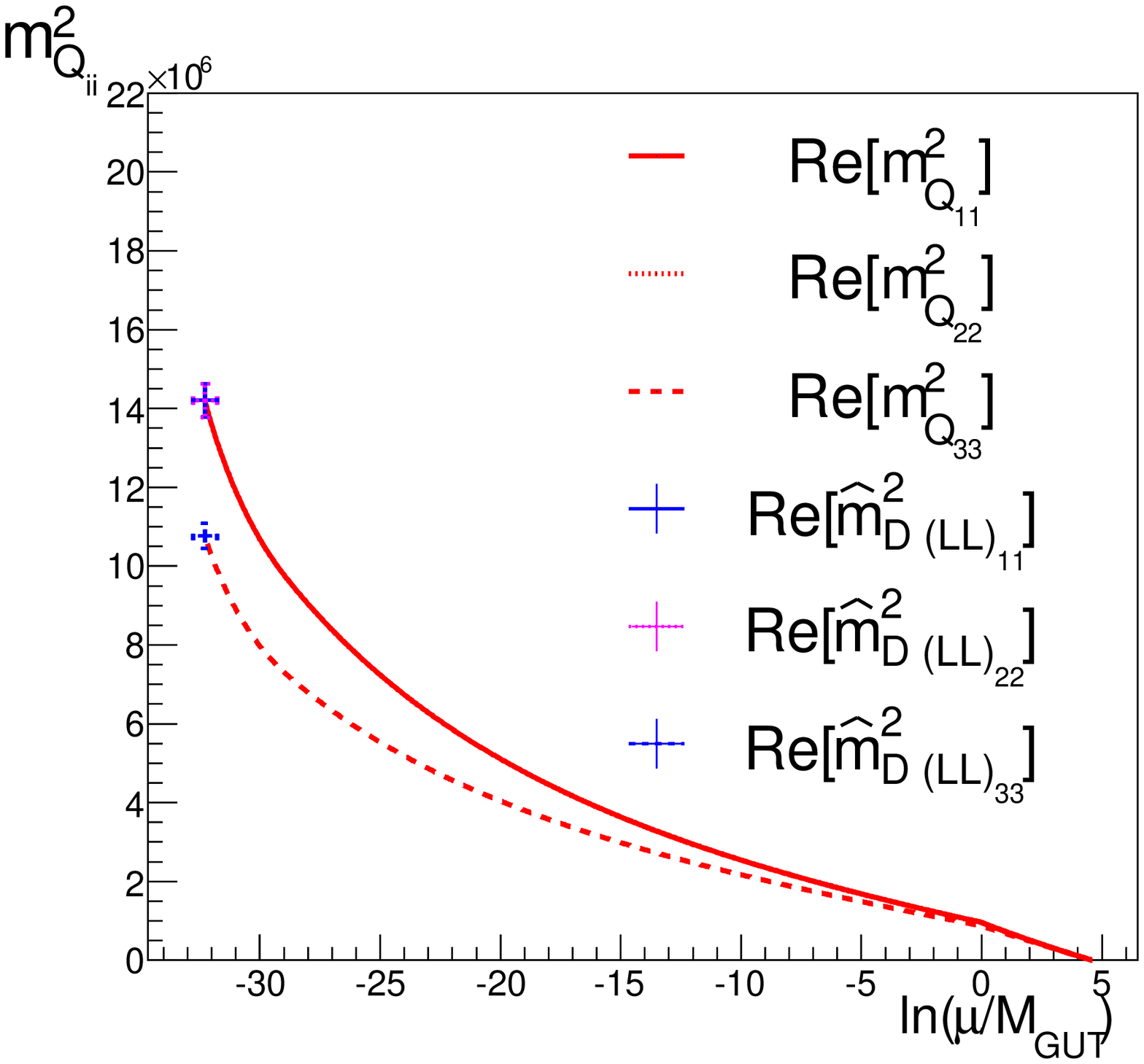}
\includegraphics[width=8.1cm, height=5.5cm]{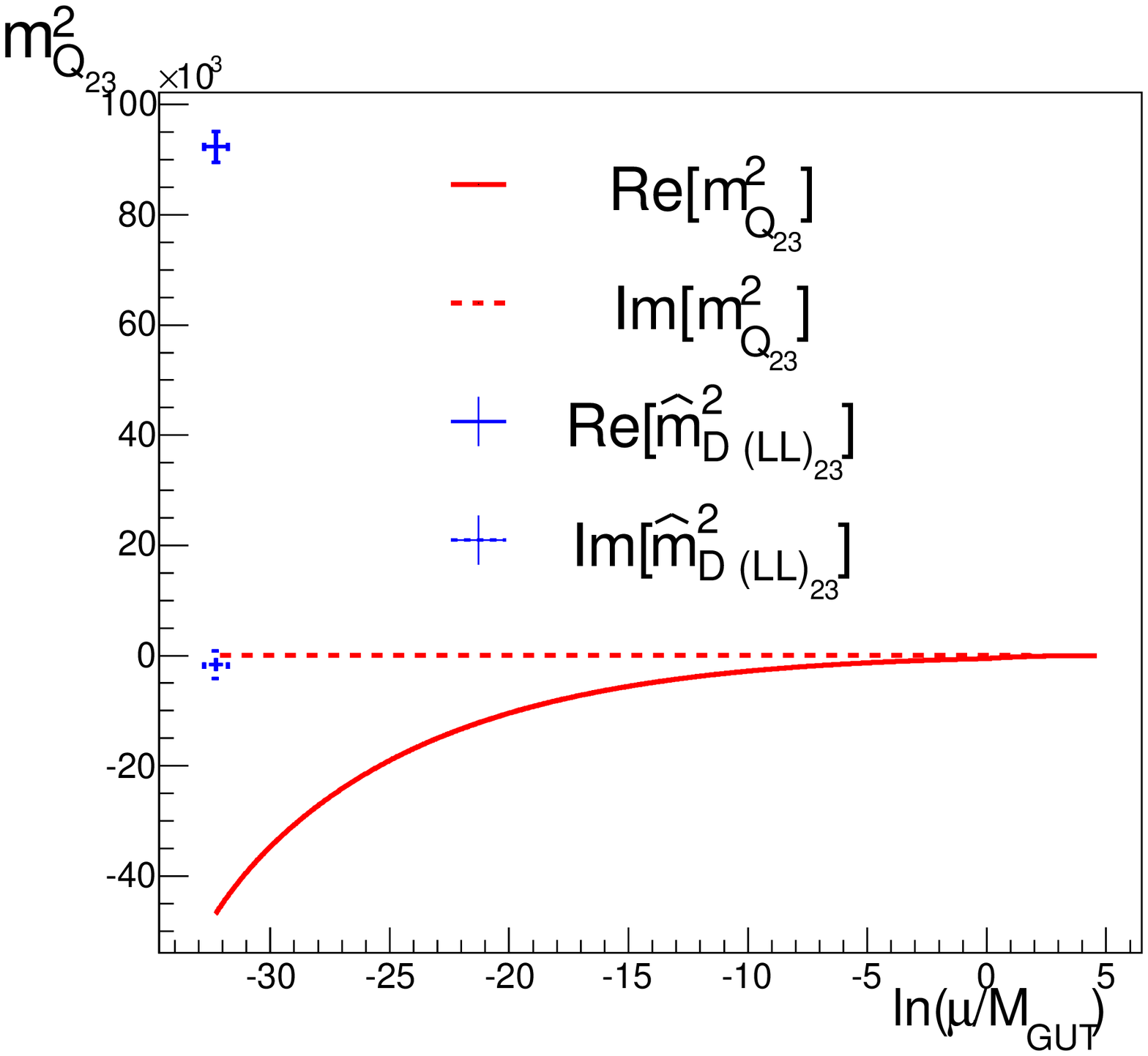}
\includegraphics[width=8.1cm, height=5.5cm]{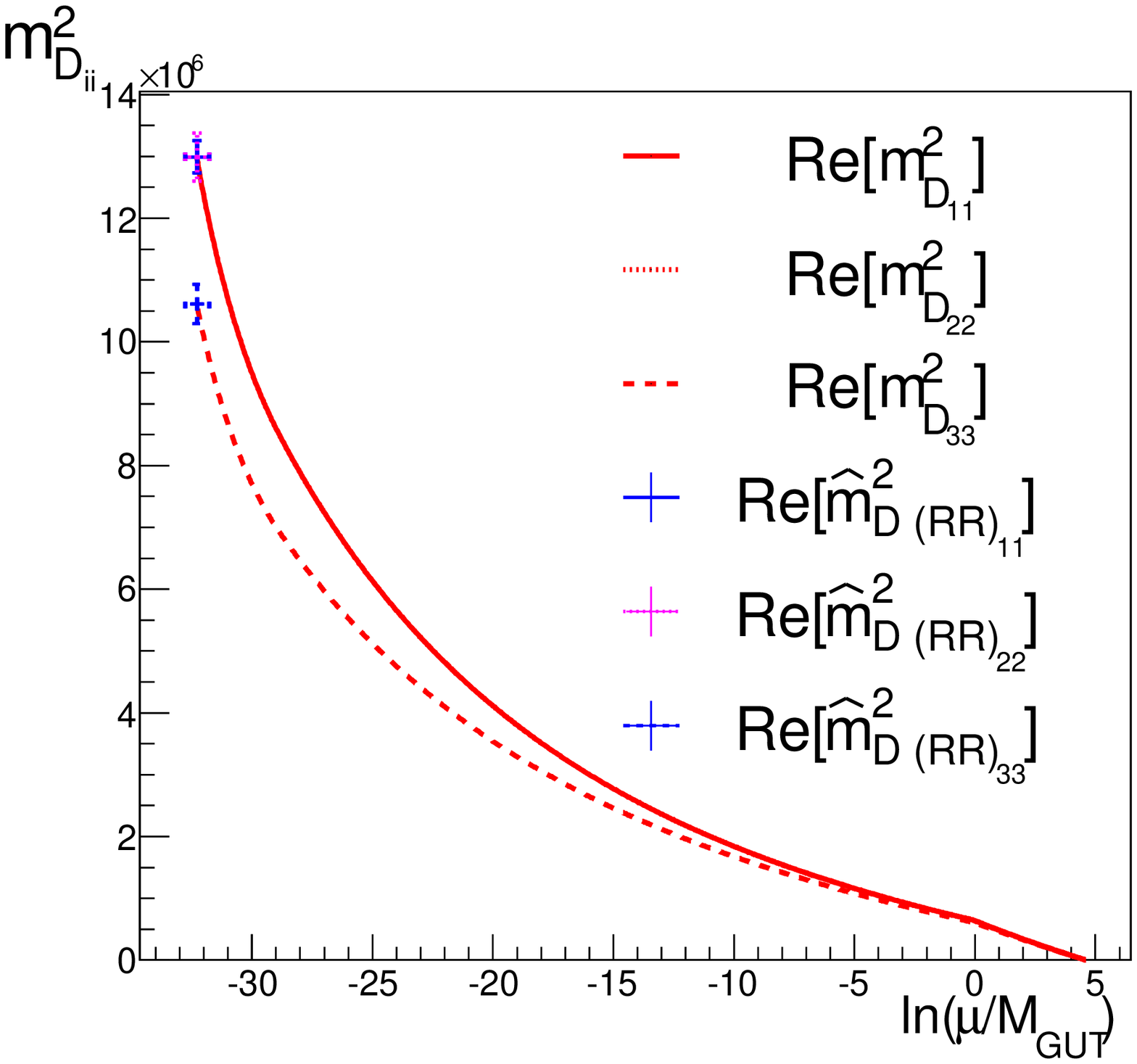}
\includegraphics[width=8.1cm, height=5.5cm]{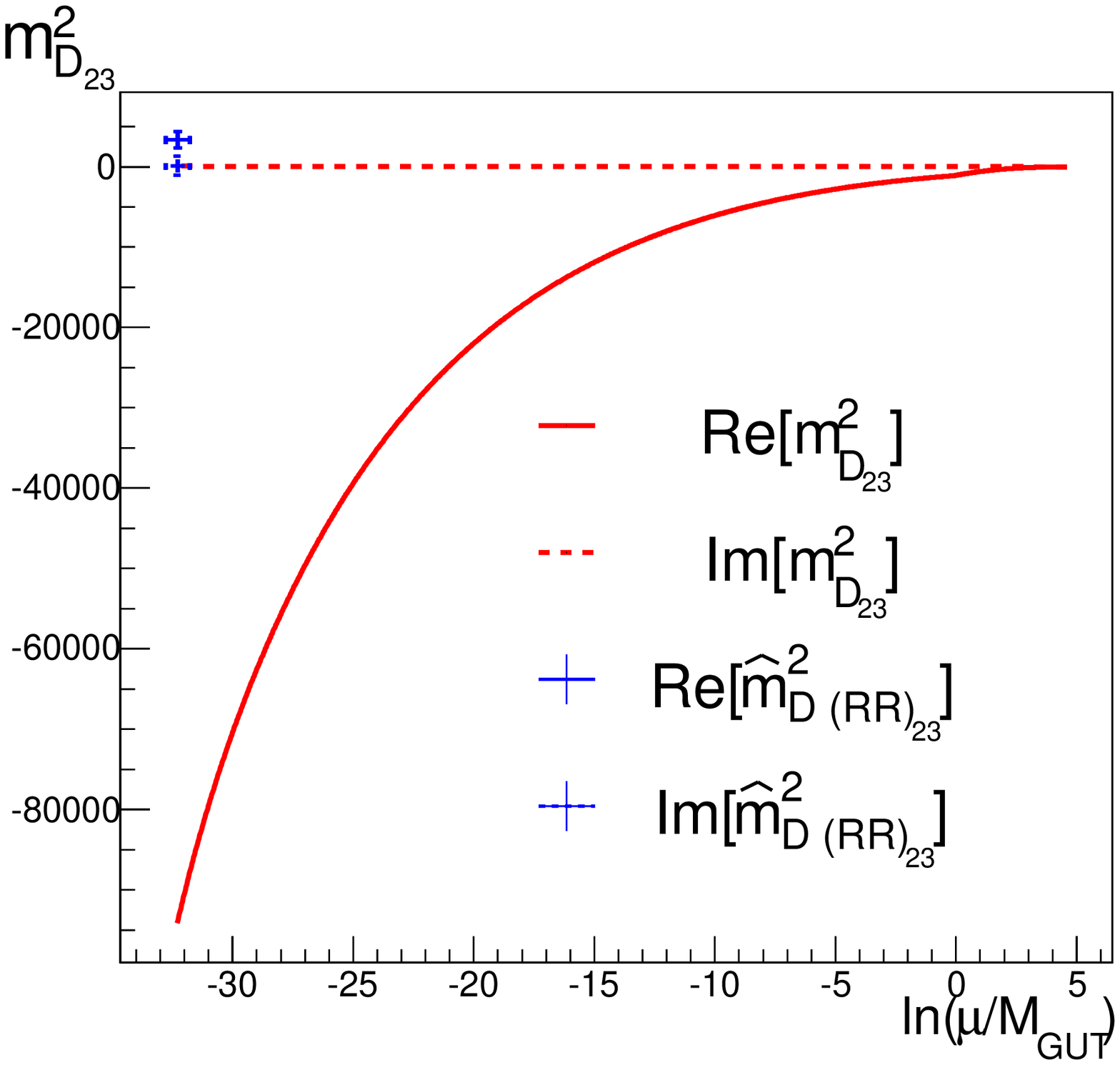}
\includegraphics[width=8.1cm, height=5.5cm]{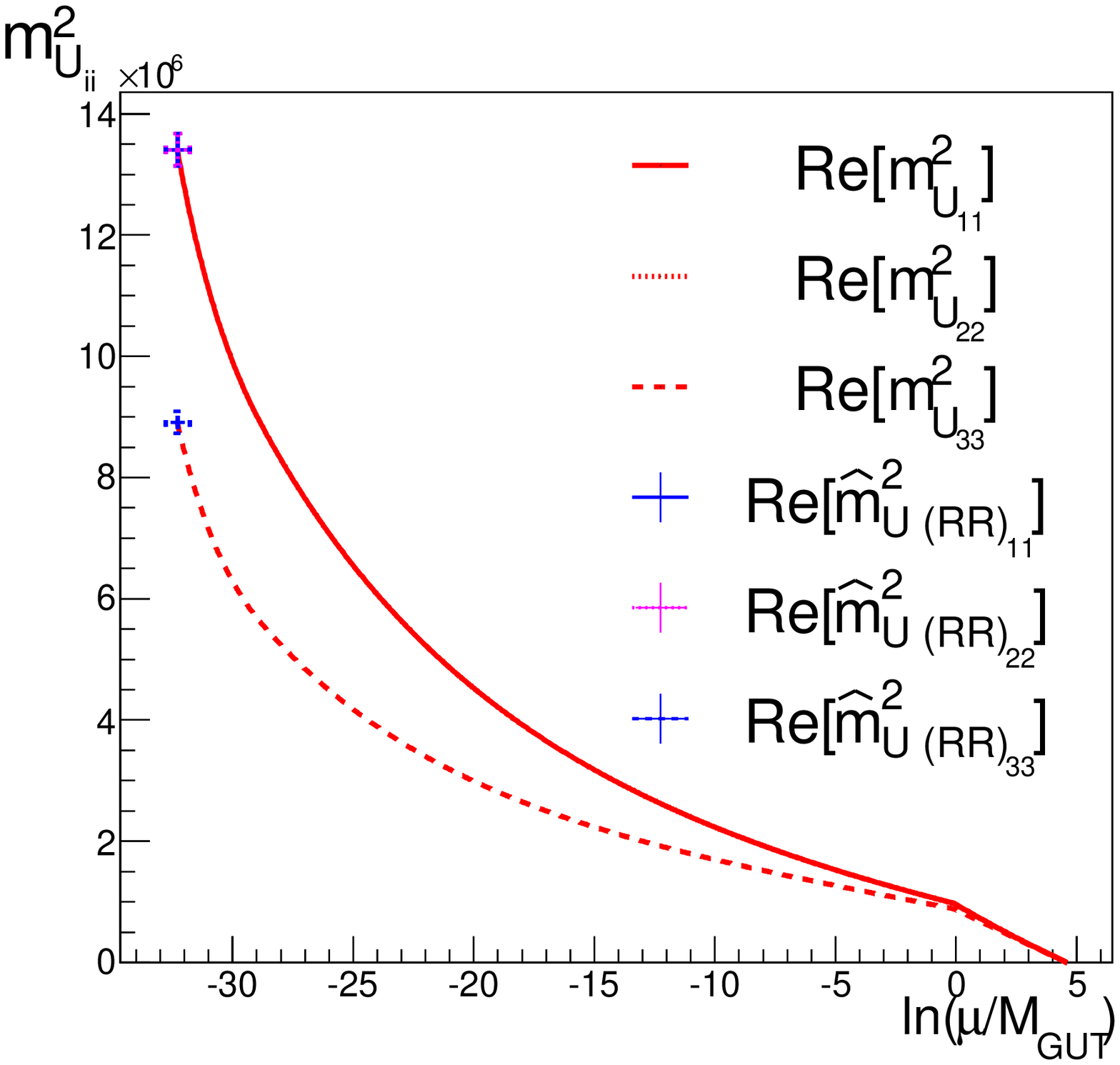}
\includegraphics[width=8.1cm, height=5.5cm]{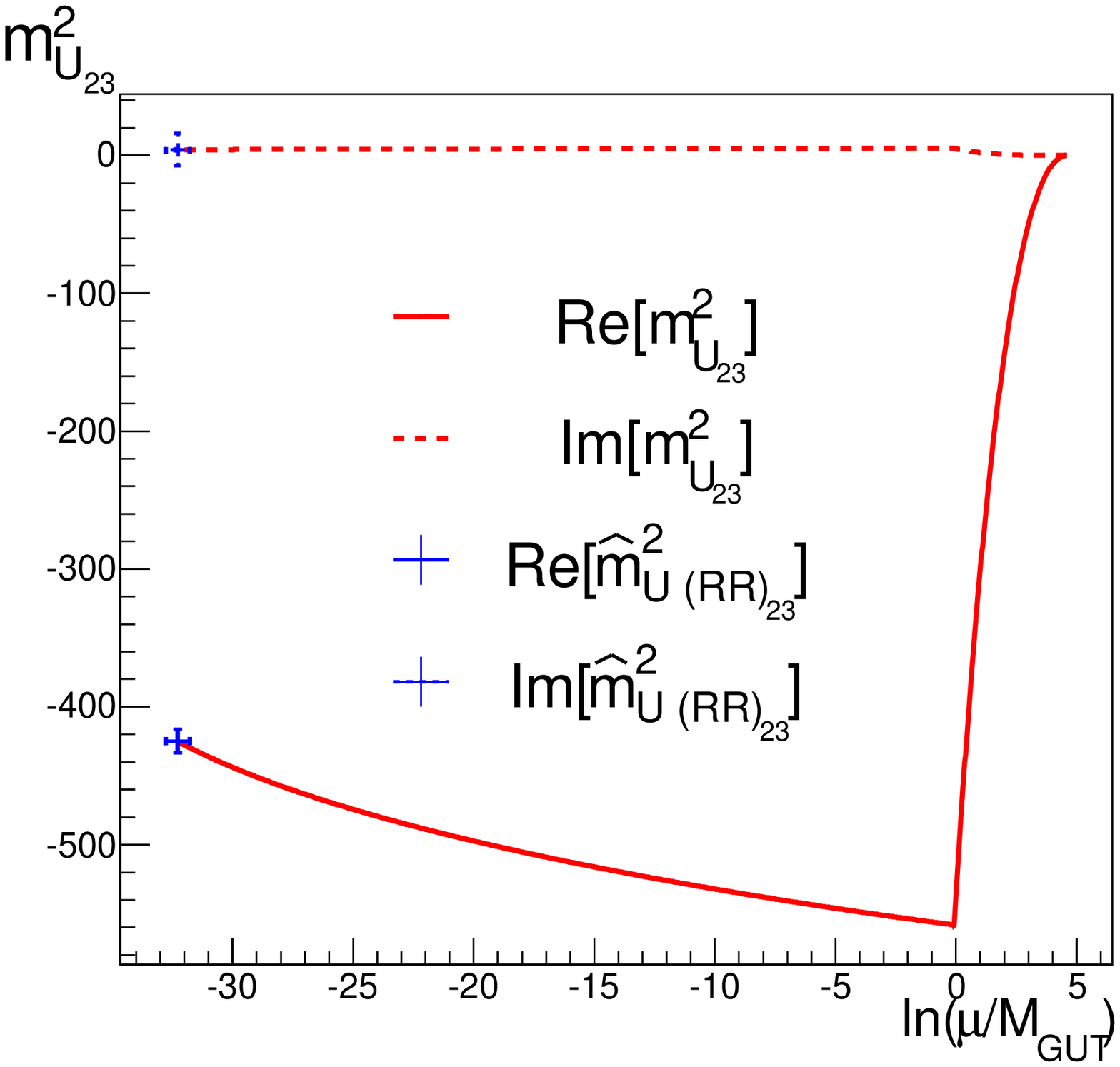}
\caption{\it{The runnings of soft-squared masses for the MSSM parameters for Ansatz A1. 
The horizontal axis is $\ln(\mu/\mgut)$, and the vertical axis shows the mass-squared in units of [GeV]$^2$. 
For the diagonal soft-squared masses, the split between first and second generation is not appreciable on the scale of the plot.
In each panel we show the runnings of  ${\mQs}_{ii}$, $i=1, 2, 3$, ${\mQs}_{23}$, ${\mds}_{ii}$, $i=1, 2, 3$, ${\mds}_{23}$, ${\mus}_{ii}$, $i=1, 2, 3$ and ${\mus}_{23}$.
%$(\mfs)_{23}$ for $f=Q, D, U, L$, and $E$ respectively. 
The red lines in the plots run from the input scale, 
$\Mi$, down to $\Mw$ (with the appropriate matching at $\Mg$) and are
given in the basis where Yukawa couplings {\it{are not}} diagonal.
The blue crosses show the result (at the electroweak scale) of running in the basis where the Yukawa matrices are diagonal. 
The states labeled without a hat are the states given in the basis
where Yukawa couplings {\it{are not}} diagonal. The SU(5) parameters were
specified in the basis where the Yukawa couplings are not diagonal.
\label{fig:MSSMevol23}}}
\end{figure}

\begin{figure}[!h]
\centering
\includegraphics[width=8.1cm, height=5.5cm]{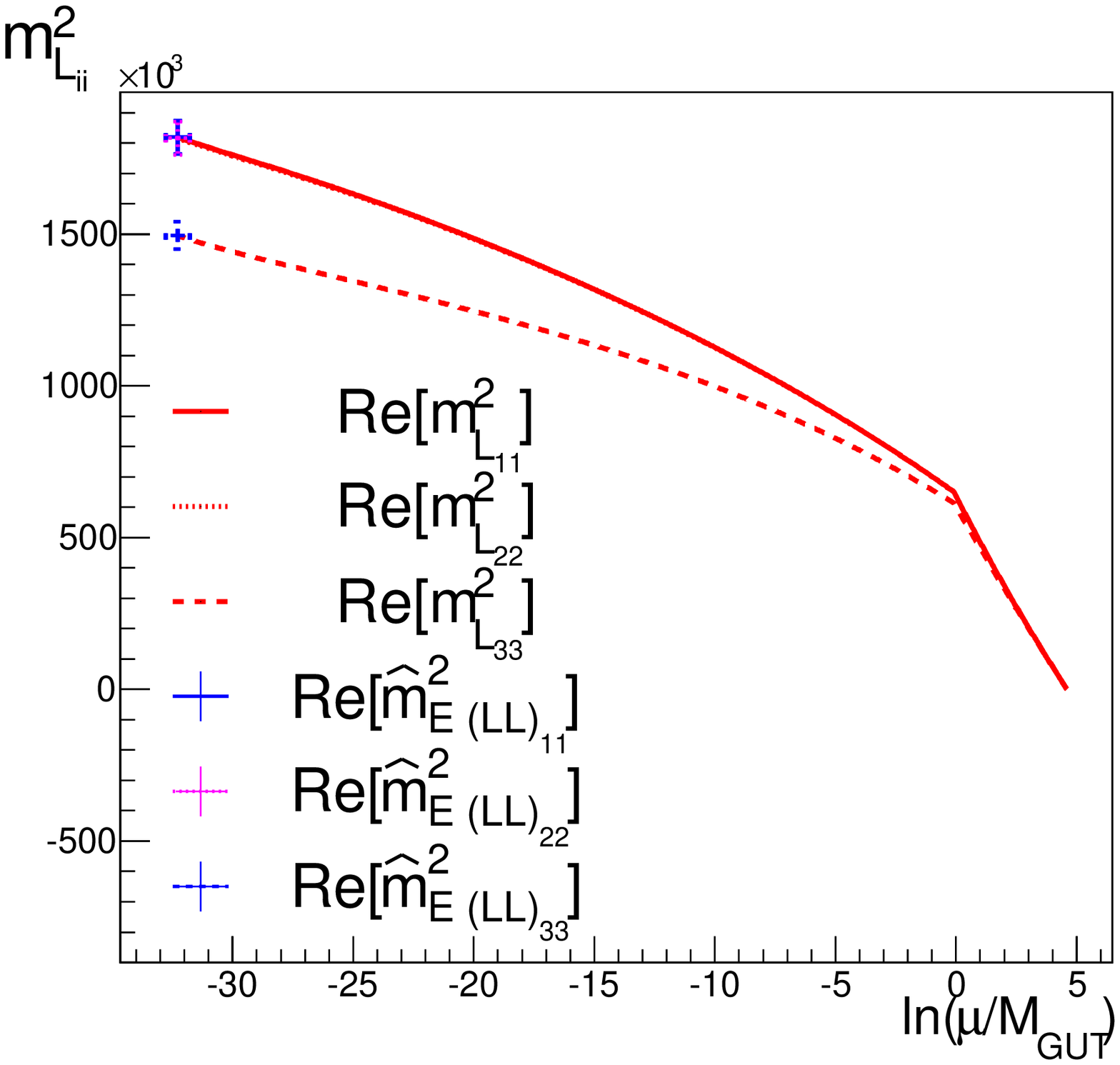}
\includegraphics[width=8.1cm, height=5.6cm]{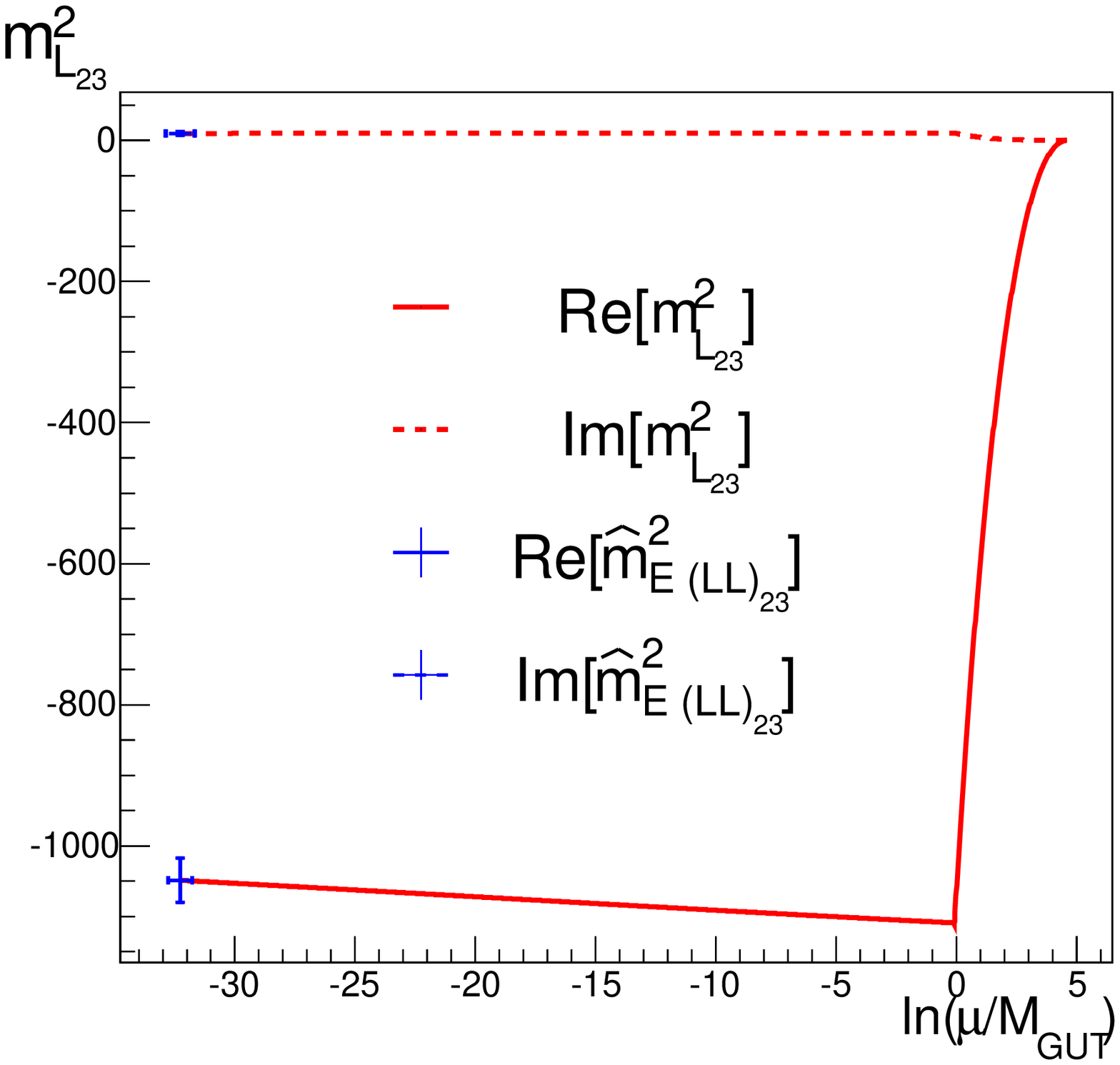}
\includegraphics[width=8.1cm, height=5.6cm]{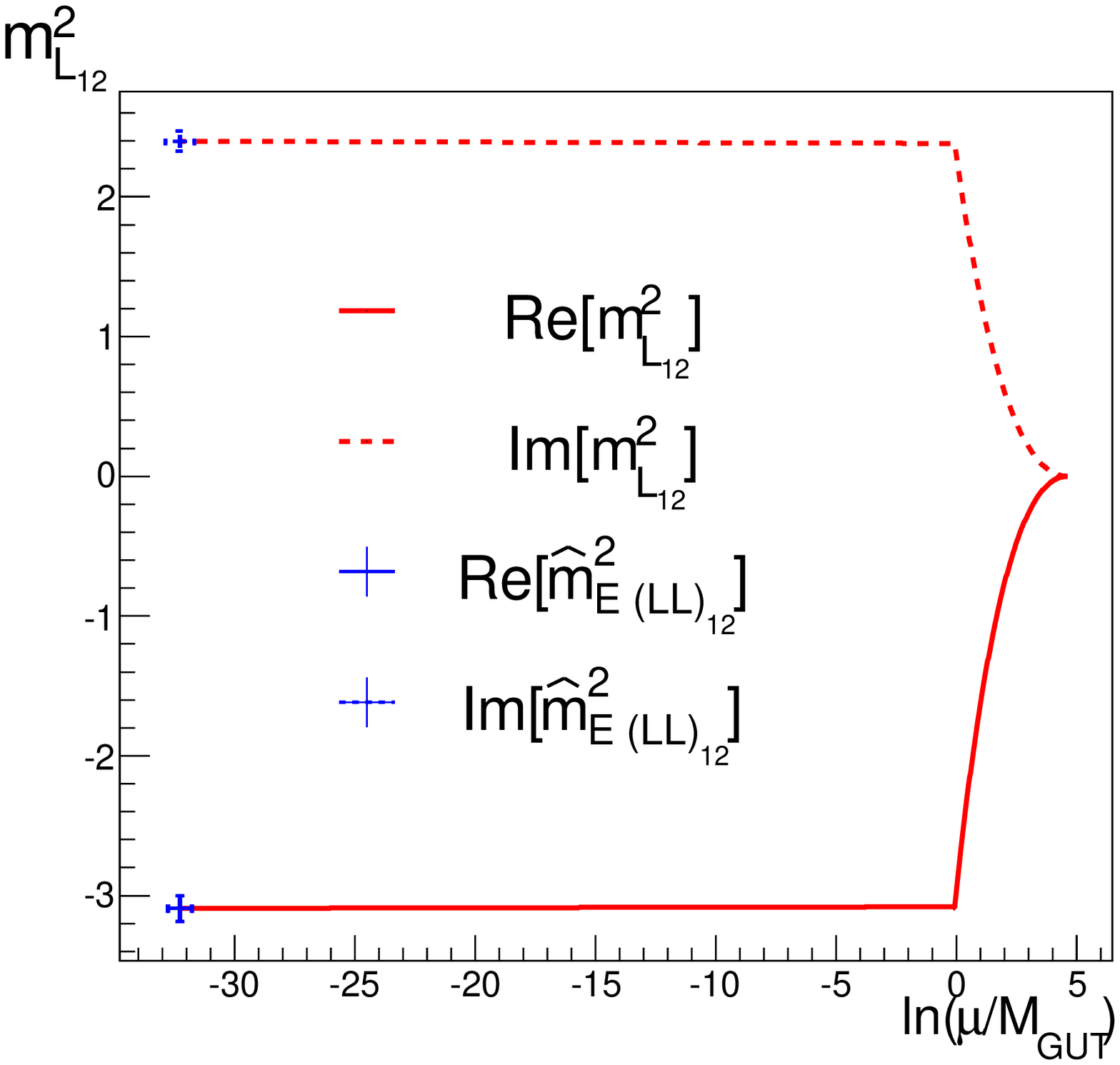}\\
\includegraphics[width=8.1cm, height=5.5cm]{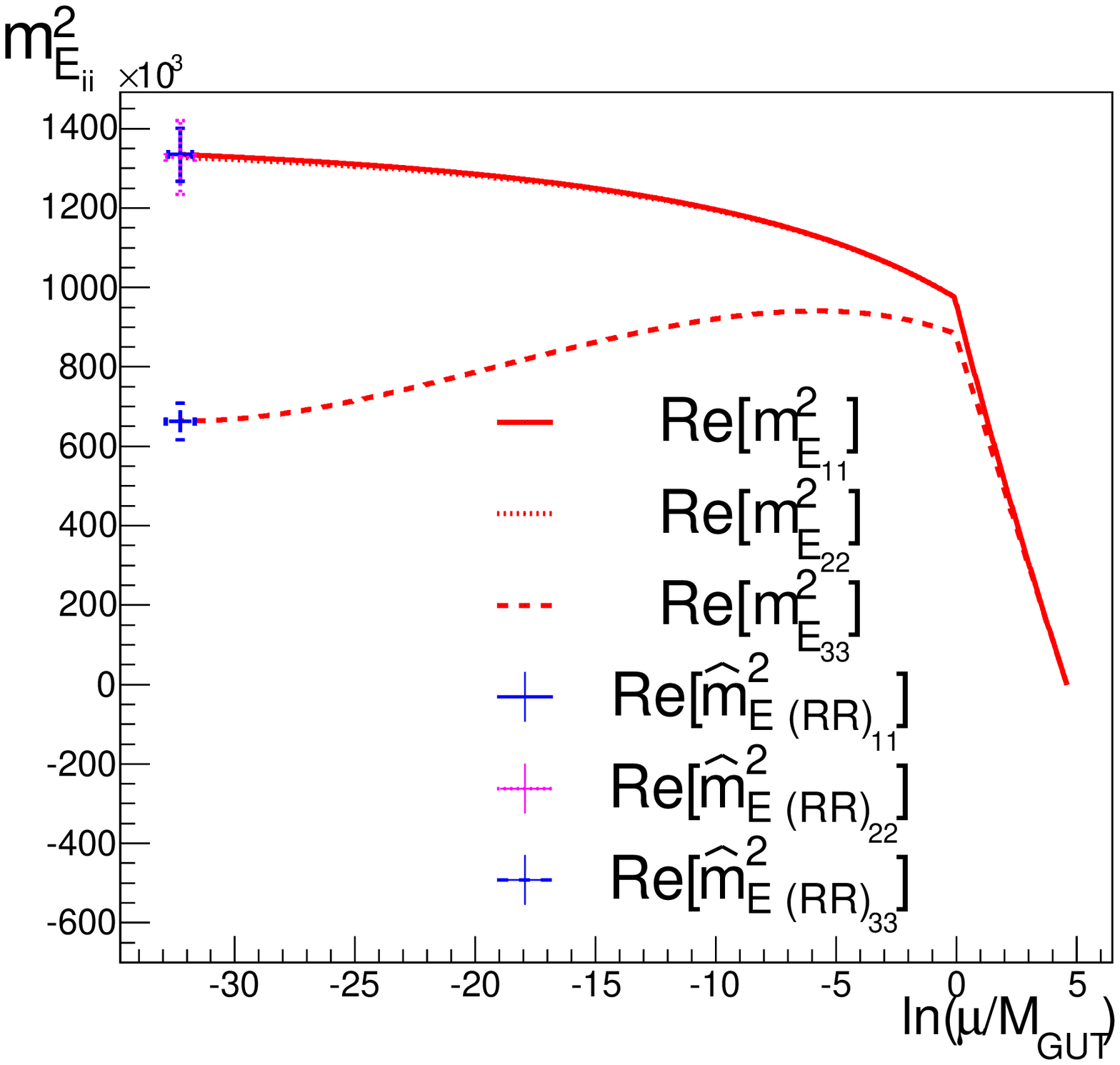}
\includegraphics[width=8.1cm, height=5.6cm]{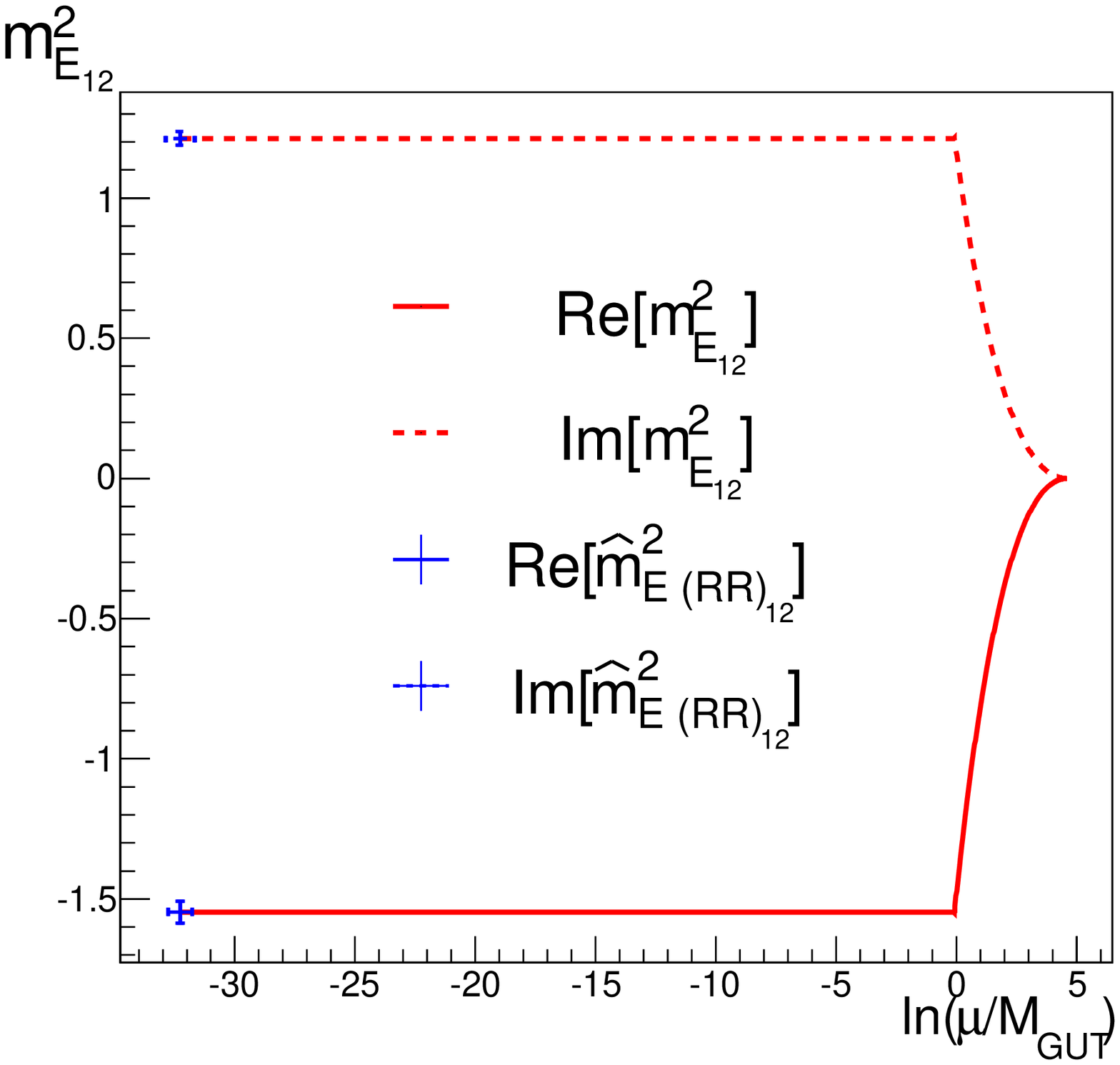}
\caption{\it{
As in \Figref{fig:MSSMevol23}, but for  ${\mls}_{ii}$, $i=1, 2, 3$, ${\mls}_{23}$, ${\mls}_{12}$, ${\mes}_{ii}$, $i=1, 2, 3$, and ${\mes}_{12}$.  %${\mfs}_{1,2}$,  for $f=Q, D, U, L$, and $E$.  
The horizontal axis is again $\ln(\mu/\mgut)$. 
}
\label{fig:MSSMevol12}}
\end{figure}

\begin{figure}[!h]
\centering
\includegraphics[width=8.1cm, height=5.6cm]{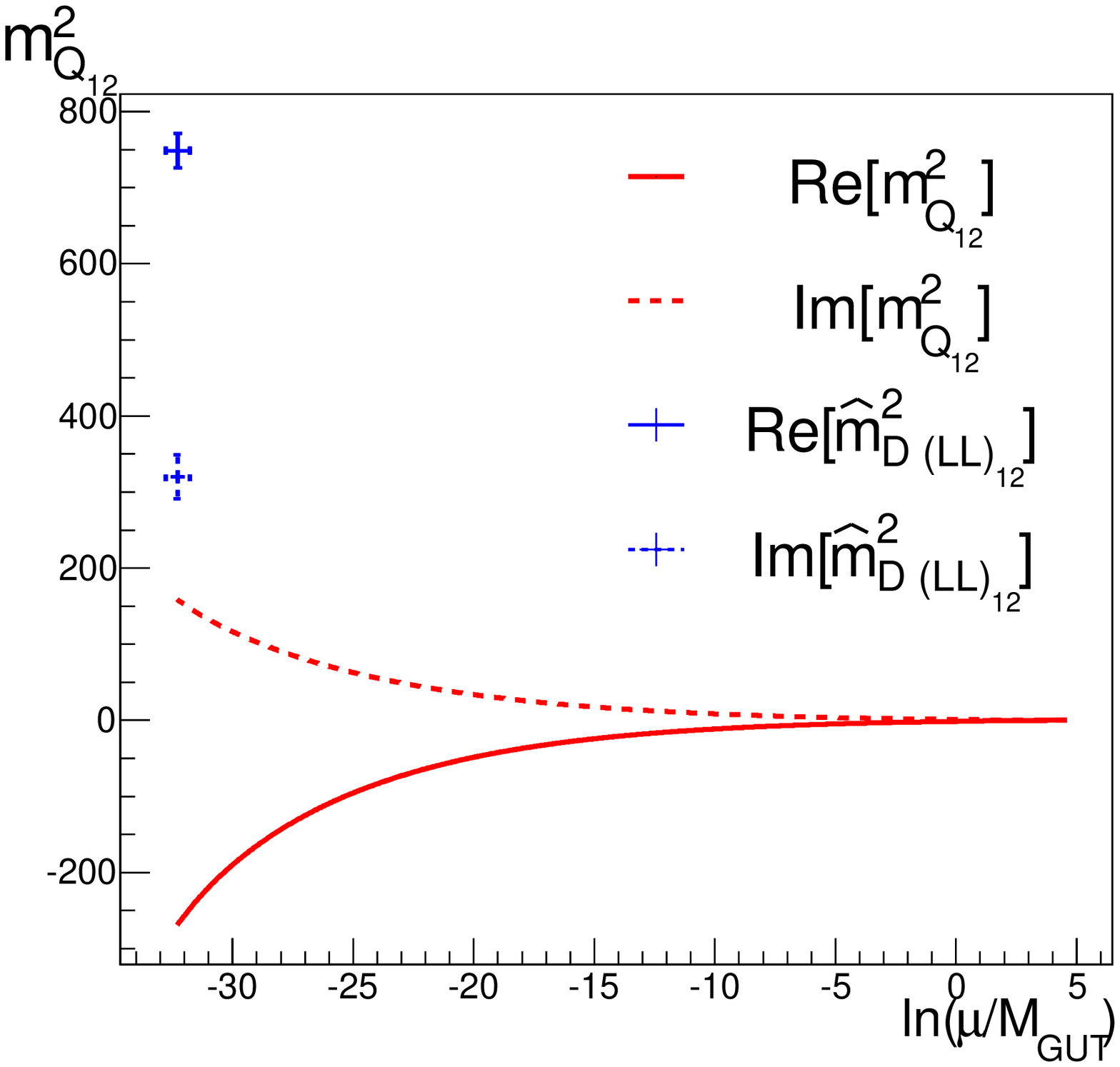}	
\includegraphics[width=8.1cm, height=5.6cm]{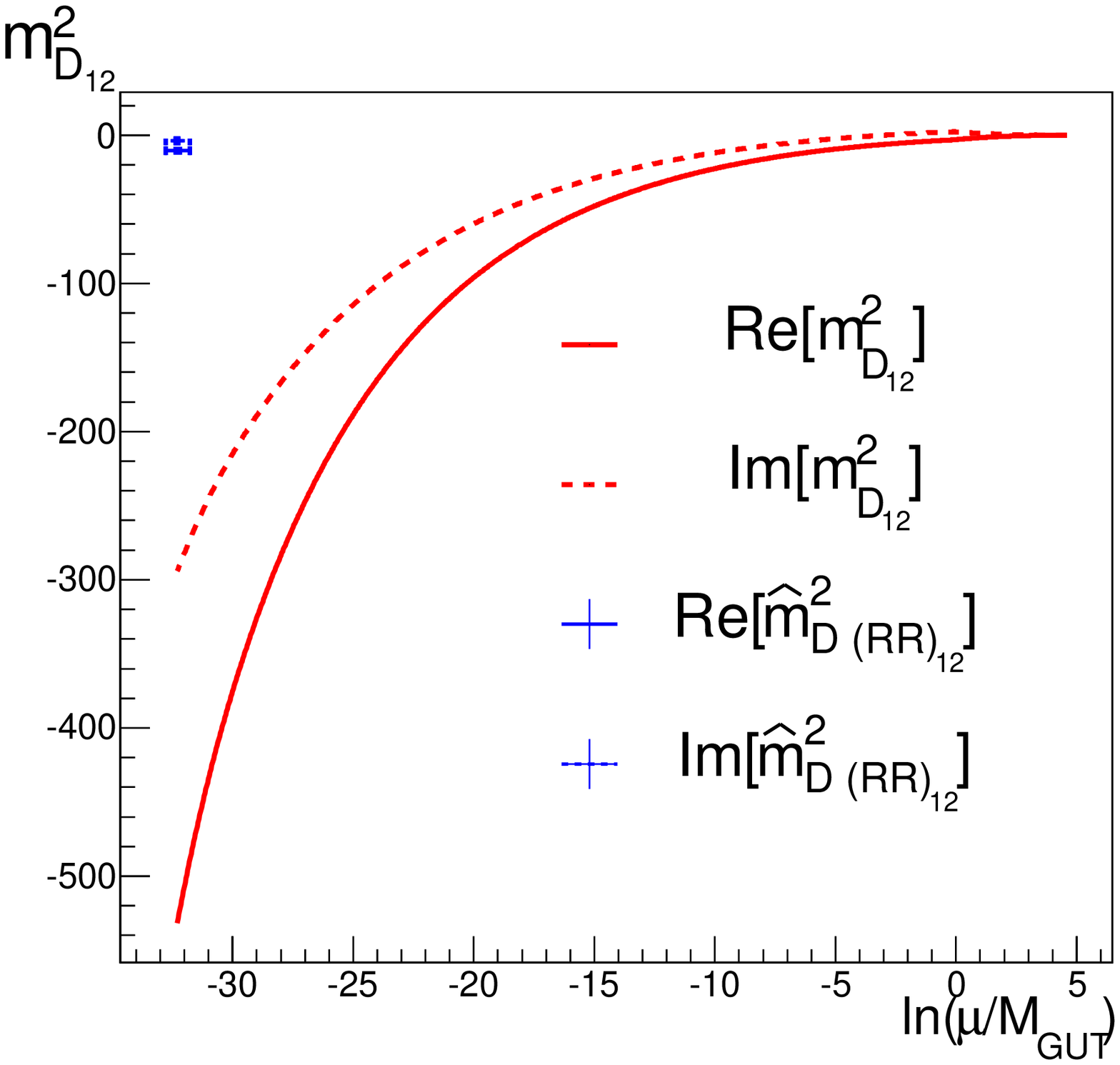}
\includegraphics[width=8.1cm, height=5.6cm]{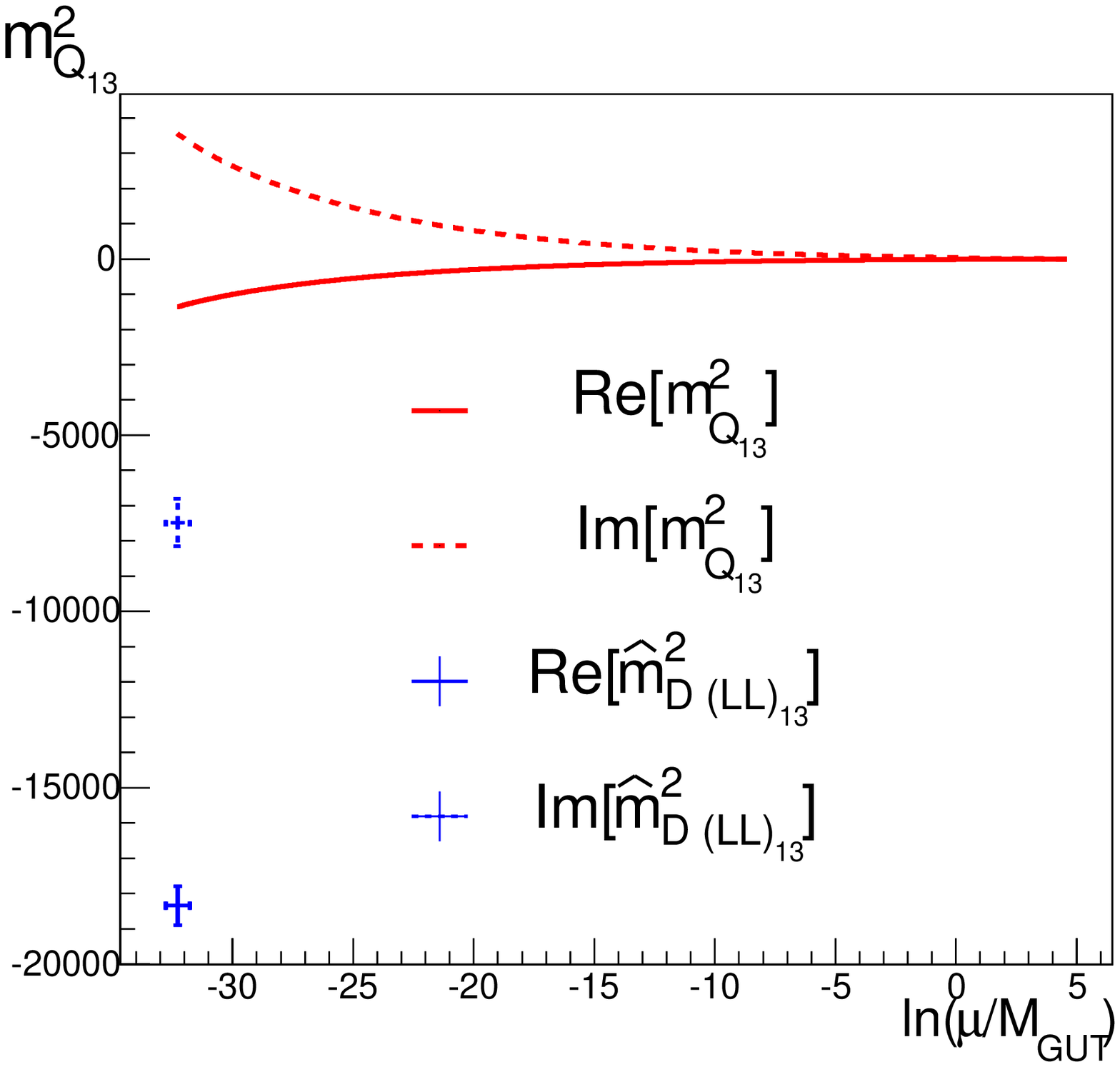}
\includegraphics[width=8.1cm, height=5.6cm]{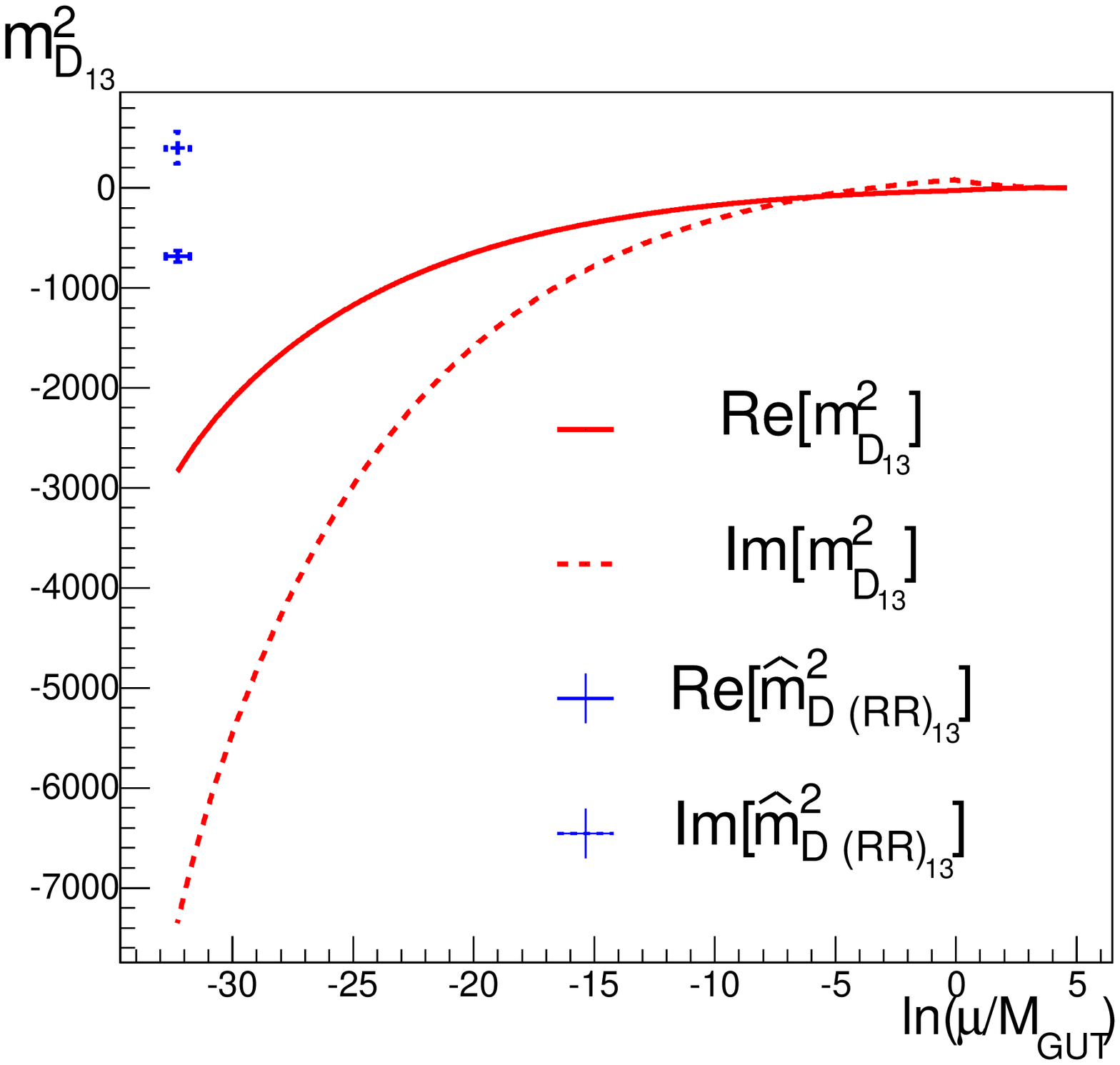}
\includegraphics[width=8.1cm, height=5.6cm]{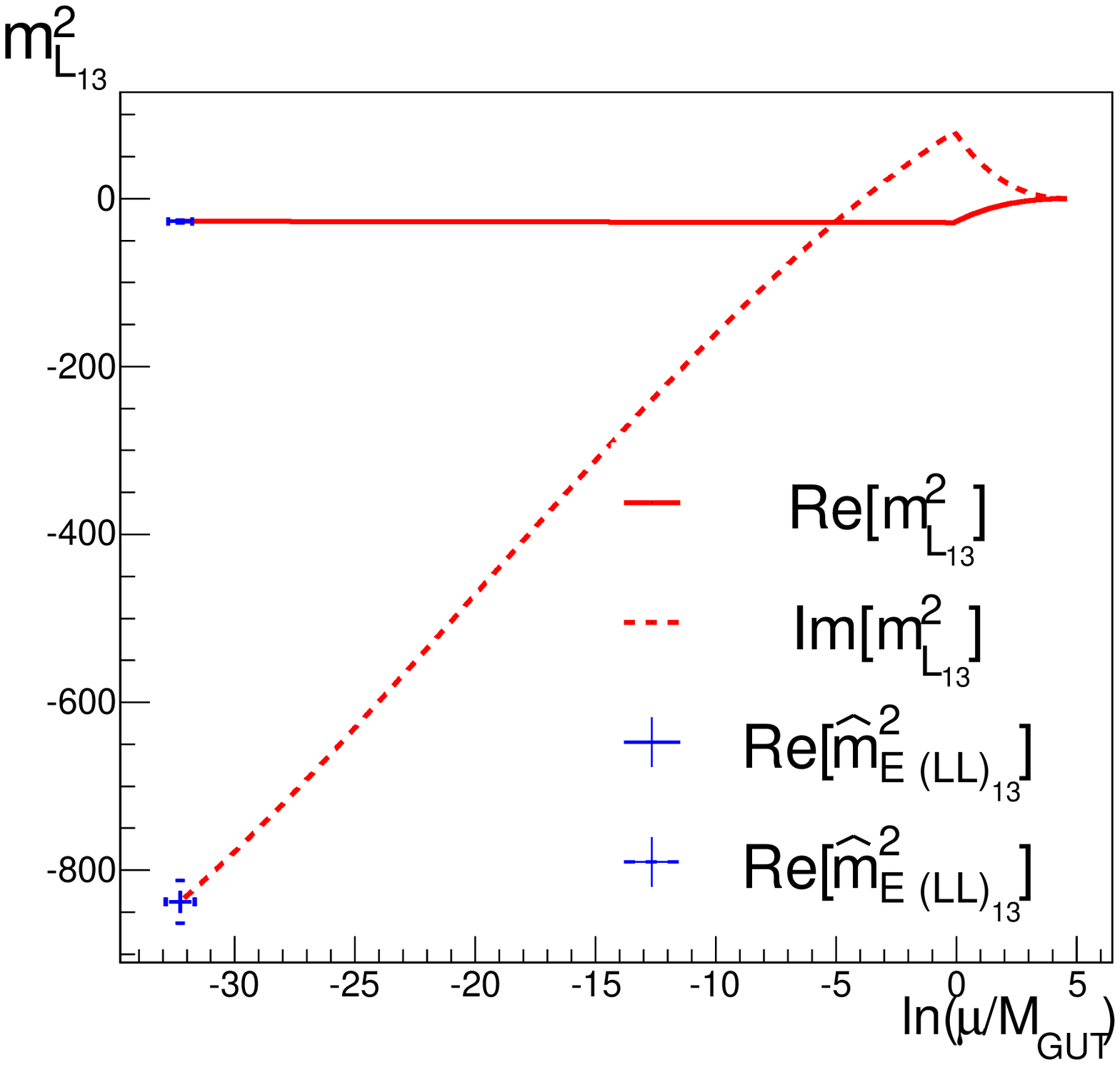}
\includegraphics[width=8.1cm, height=5.6cm]{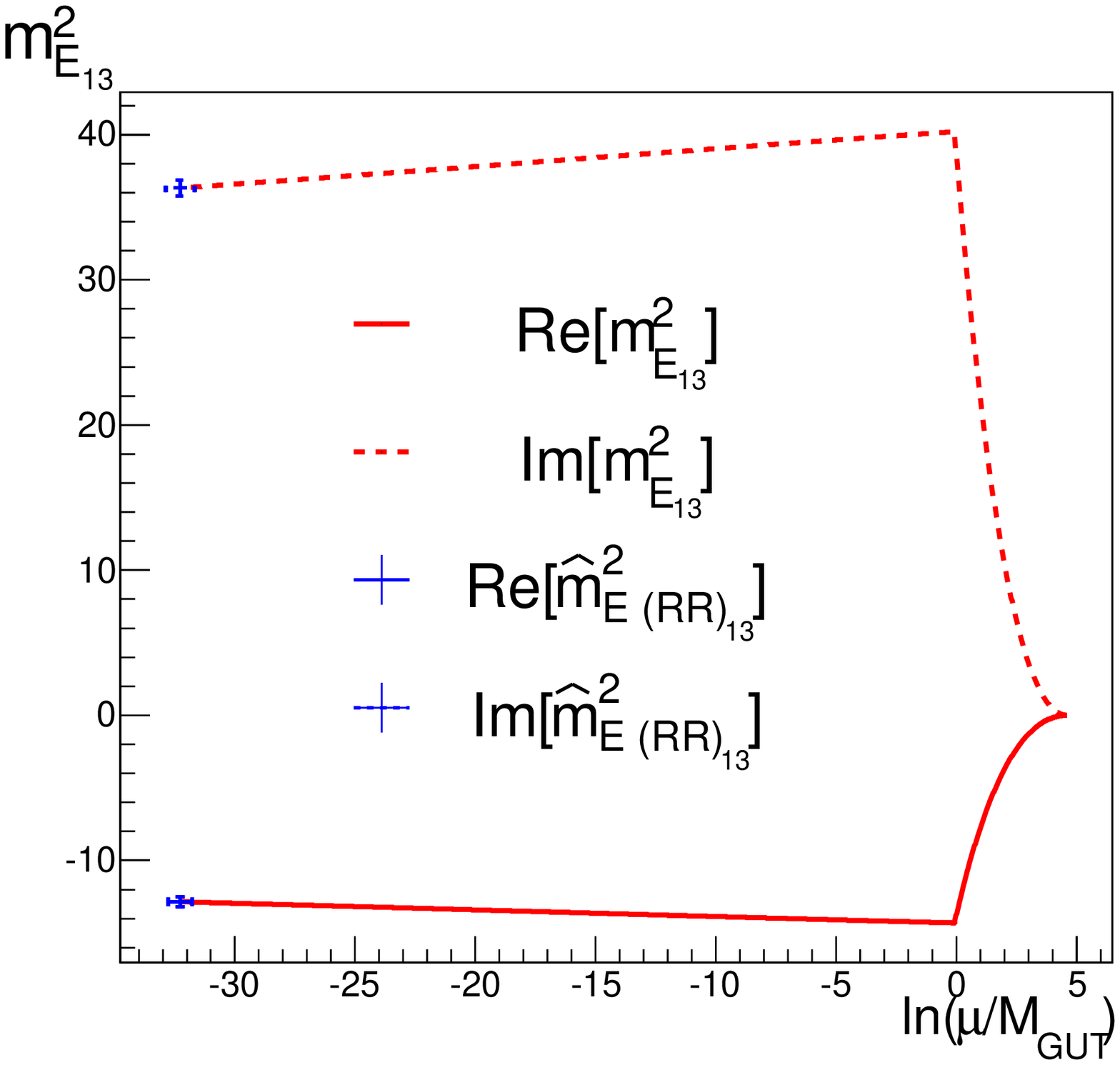}
\caption{\it{
As in \Figref{fig:MSSMevol12} but now for ${\mQs}_{12}$, ${\mds}_{12}$ and ${\mfs}_{13}$, $f=Q, D, L, E$. The horizontal axis is again $\ln(\mu/\mgut)$. 
}
\label{fig:MSSMevol13}}
\end{figure}

%\paragraph{Comments about Ansatz A1}
The matching of couplings and masses at $\mgut$, using \eq{eq:bcatMGUT}, is made in the basis where the Yukawa couplings are not diagonal, i.e., we assume ansatz A1. The transformations to the super-CKM (SCKM)
basis,  where the Yukawa couplings are diagonal, are given by
\bea
\label{eq:SCKMtransf}
\hat{m}^2_{ D (LL)} &=&V^{D\dagger}_L m^2_{Q}  V^D_L,\nn\\
 %=V^\dagger_{\rm{CKM}} m^2_{\tilde Q} V_{\rm{CKM}}\nn\\
\hat{m}^2_{ E (LL)} &=&V^{E\dagger}_L m^2_{ L} V^E_L,\nn\\
\hat{m}^2_{ f (RR)} &=&V^{f\dagger}_R m^2_f V^f_R,\quad  f=D,E,\nn\\%\mfs V^f_R,\quad  f=D,E,\nn\\
\hat{m}^2_{ f (LR)} &=&-\hat{a}_f  v_f  + \mu \tan\beta \ \hat m_{f_i}\delta_{ij},\quad
\nn\\
%&&%\afT=V^{f\dagger}_L \af V^{f}_R,  \quad f=D,E,
\hat{a}_f & = & V^{f\dagger}_L a_f V^{f}_R,  \quad f=D,E,
\eea
where the matrices $V^{f\dagger}_X$ ($X=L,R$) are defined in (\ref{eq:VdandVe}) and the
$\hat m_{f_i}$ are the fermion masses, $i=1,2,3$. 
 We could also define $\hat{m}^2_{ U (LL)}=V^{U\dagger}_L m^2_{Q}  V^U_L$, but we are taking $V^U_L$ to be diagonal.
We note that $SU(2)_L$ invariance in the SCKM basis is preserved trivially, because then  $m^2_{ Q}=V^D_L \hat{m}^2_{D (LL)} V^{D\dagger}_L$=
$V^U_L \hat{m}^2_{U (LL)} V^{U\dagger}_L$ and so $\hat{m}^2_{ U (LL)}=V_{\rm{CKM}} \hat{m}^2_{D (LL)}V^{\dagger}_{\rm{CKM}} $.
A more complete set of 
transformation rules are given in Appendix \ref{sec:tr}.
 The choice of Ansatz in Eqs. (\ref{eq:Az1}) to (\ref{eq:Az4}) induces off-diagonal entries at $\mgut$ in the SCKM basis, which are 
 constrained by flavor-violating processes. In particular, for Ans\"atze 3 and 4, we find that 
 $\Bmueg$ is too large and, in addition, for Ansatz 4 some electric dipole moments (EDMs) are too large.
 On the other hand, both the Ans\"atze A1 and A2 induce acceptable amounts of flavor violation. 
 Ansatz 1 is interesting because both the right- and left-diagonalization matrices $V_{L, R}$
 are CKM-like. We plot the runnings of the MSSM soft-squared parameters for this Ansatz
 in Figs. \ref{fig:MSSMevol23} to \ref{fig:MSSMevol13}.

We plot in the first panel of Fig.~\ref{fig:MSSMevol23} the running of some squark soft masses-squared.  
The horizontal axis is $x=\ln(\mu/\mgut)$, and the red curves represent the evolution of states in the basis where 
Yukawa couplings {\it{are not}} diagonal.  That is, starting with our boundary conditions:
$m_0 = 0$ at $x \simeq 4.6$, 
the sfermion masses are run down to $\mgut$ where they are matched to the MSSM
sfermion masses. As the running to $\mgut$ has induced some off-diagonal entries,
these are also run down to the weak scale ($x \simeq -32$). 
This running is shown by the set of red curves.
For example, at the electroweak scale, the running of the diagonal left-handed squark masses-squared
${\mQs}_{33}$ reaches $11\times 10^6$ GeV$^2$, while  ${\mQs}_{ii\ne3}$ reaches 
$14\times 10^6$ GeV$^2$ . The split between the first and second generation is not appreciable on the scale 
displayed on the plot. We can diagonalize the mass matrices at the weak scale using
Eq. (\ref{eq:SCKMtransf}) with the $V$'s defined by A1. That result is shown by the blue crosses. 
 In the upper left panel it makes no difference whether we run in the diagonal (SCKM) basis
\eq{eq:SCKMtransf} or in the non-diagonal Yukawa basis,
as the blue crosses sit at the endpoint
of the red curves.

In the second to sixth panels, we show the running of ${\mQs}_{23}$, ${\mds}_{ii}$, $i=1,2,3$, ${\mds}_{23}$,  ${\mus}_{ii}$, $i=1, 2, 3$, and ${\mus}_{23}$.
 %$(\mfs)_{23}$ for $f=Q, D, U, L$ and $E$, respectively. 
We again show as blue crosses the endpoints of the running at the weak scale when the states are
in the basis where Yukawa couplings are diagonal. The positions of these show the impact of the changes
in size of off-diagonal parameters. We see that the blue crosses differ only slightly from the running shown by the 
red curves for the imaginary parts of $Q_{23}$ and $D_{23}$. Since the Yukawa couplings
are chosen to be diagonal for $U$, %$L$, and $E$,
 the blue crosses are found at the endpoints of the red curves
in these cases.  

In Figs.  \ref{fig:MSSMevol12} and \ref{fig:MSSMevol13} we show the same for the runnings of ${\mls}_{ii}$, $i=1,2,3$, ${\mls}_{23}$, ${\mls}_{12}$,  ${\mes}_{ii}$, $i=1, 2, 3$, ${\mes}_{12}$,
${\mQs}_{12}$, ${\mds}_{12}$ and ${\mfs}_{13}$, $f=Q, D, L, E$. We do not show the runnings of ${\mes}_{23}$, ${\mus}_{12}$ and ${\mus}_{13}$ because, given the matching
conditions at $\mgut$, \eq{eq:bcatMGUT}, and the fact that in the $E$ and $U$ sectors the Yukawa couplings are chosen to be diagonal, their runnings are  similar to ${\mus}_{23}$, ${\mes}_{12}$
and ${\mes}_{13}$, respectively.

%$(\mfs)_{12}$ for $f=Q, D, U, L$ and $E$ and $(\mfs)_{13}$ for $f=Q, D, U, L$ and $E$, respectively. 
As one can see in Figs.~\ref{fig:MSSMevol12} and \ref{fig:MSSMevol13},
there is considerably less running for the (12) components of the squark mass matrices. For the (13) sectors, looking at  Fig. \ref{fig:MSSMevol13}, we find that, depending on the sector, the running can be
more or less important. As expected for ${\mes}_{13}$, the running is negligible because in this sector the Yukawa couplings are chosen to be diagonal. Note that, due to the CP violating phase of the
CKM, the running of the imaginary parts of ${\mds}_{13}$ and ${\mls}_{13}$ become appreciable. 
 In the $D$ sector, however, once the SCKM transformations, \eq{eq:SCKMtransf}, are taken into account, some of the CP violation is rotated away,
 as a result of the first condition of Ansatz~1, \eq{eq:Az1}.

\subsection{Flavor-violating parameters}
We define the flavor-violating parameters
%New definition
\bea
\deltaXY \equiv \frac{ \hat{m}^2_{ f  (XY)_{ij}}   } {\sqrt{
\hat{m}^2_{ f (XX)_ {ii}} \hat{m}^2_{ f (YY)_{jj} }}},\quad f=D,E, 
\label{eq:deltfv}
\eea
where the mass matrices appearing on the right-hand side of the equation are defined in \eq{eq:SCKMtransf} and  $X = L, R$.
Flavor-violating parameters are often defined in the absence of a particular model in which the running can be performed
explicitly. However, general limits on flavor-violating parameters cannot be obtained, 
because the forms in which they enter into observables  are in general quite model-dependent, 
see for example \cite{Jager:2008fc} and \cite{Altmannshofer:2009ne}. There are dependences both on 
the mass scales of the supersymmetric particles involved in a particular process -
in a particular model not all the supersymmetric particles may be relevant - and on the specific
underlying flavor framework. However, a few observables can severely constrain the parameters of \eq{eq:deltfv}
and give clean bounds on them, particularly for the sleptonic parameters. They still depend on the mass
scale and assumptions of the underlying flavor model, but can be used as an indication, 
provided the model satisfies the conditions under which the bounds are derived. In particular,
in \cite{Arana-Catania:2013nha} we find a set of conditions compatible with our assumptions,
and we use them to compare to the lepton-flavour-violating parameters of \eq{eq:deltfv}.

The main purpose here in using the parameters of \eq{eq:deltfv} is to compare
our different models and to examine the different runnings and the contributions from the different sectors 
to a particular observable. Using the bounds of  \cite{Arana-Catania:2013nha},
we make comparisons and comment on cancellations in the models.

\begin{figure}[!h]
\centering
\includegraphics[width=8.1cm, height=6cm]{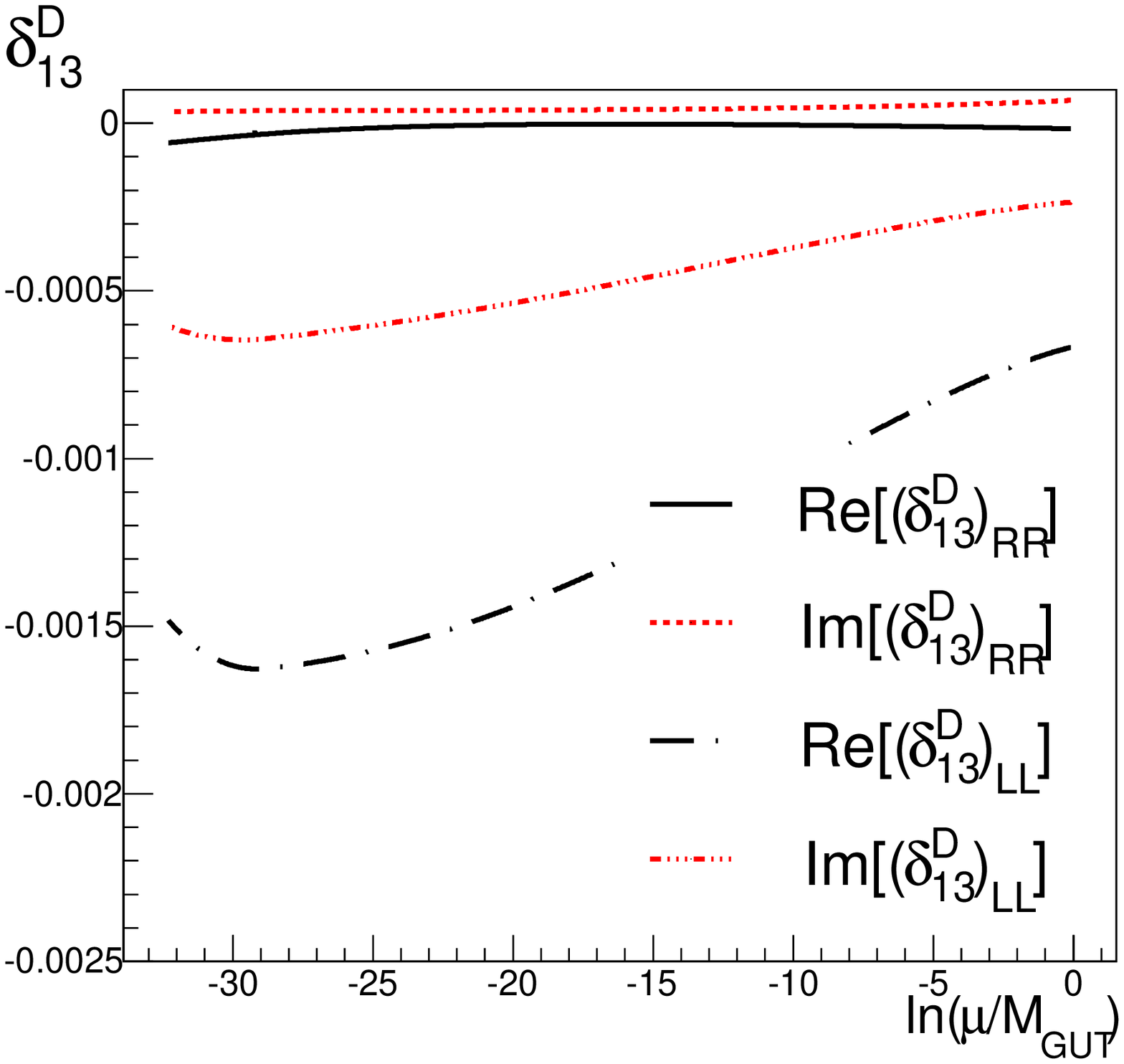}
\includegraphics[width=8.1cm, height=6cm]{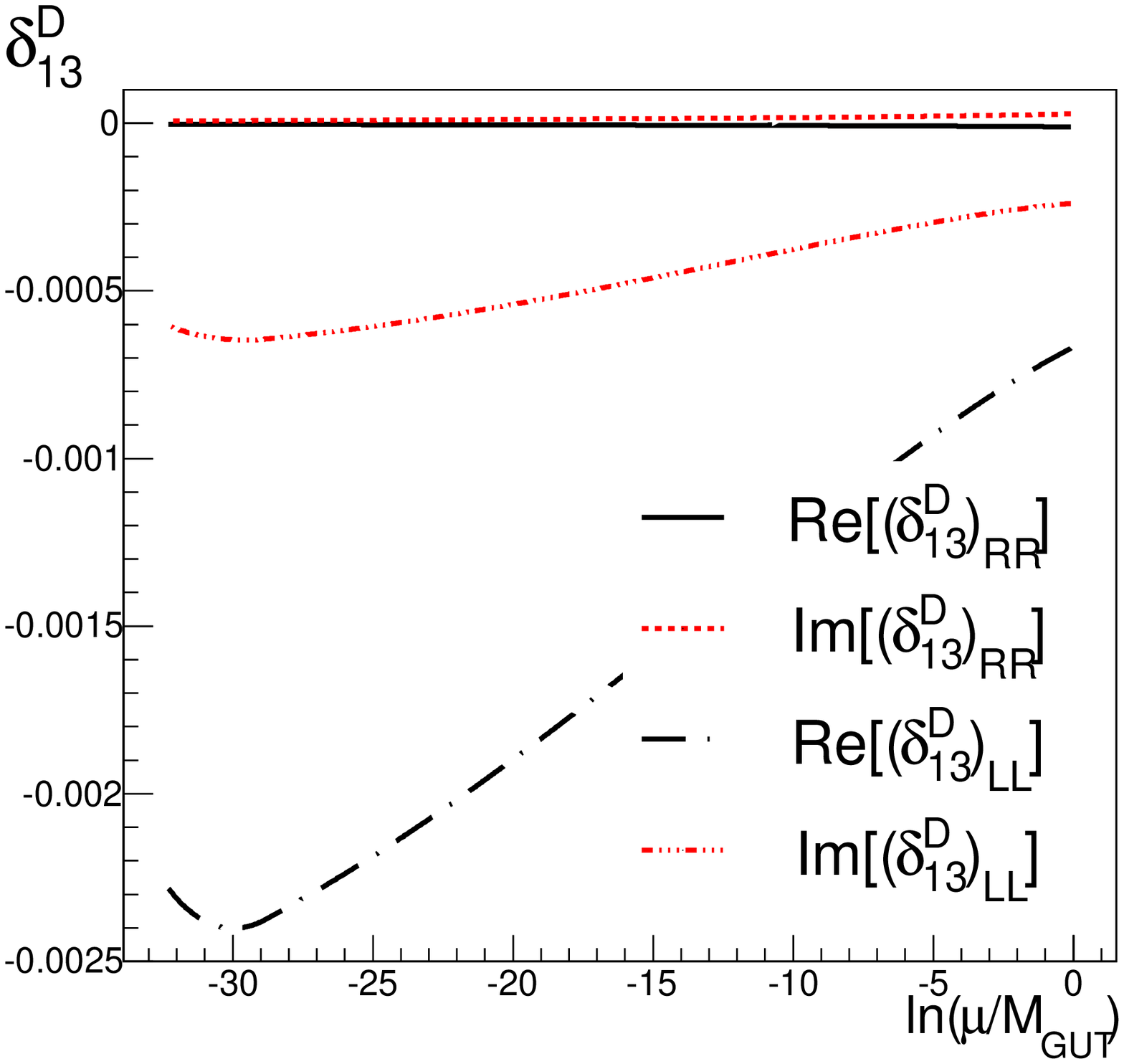}
\includegraphics[width=8.1cm, height=6cm]{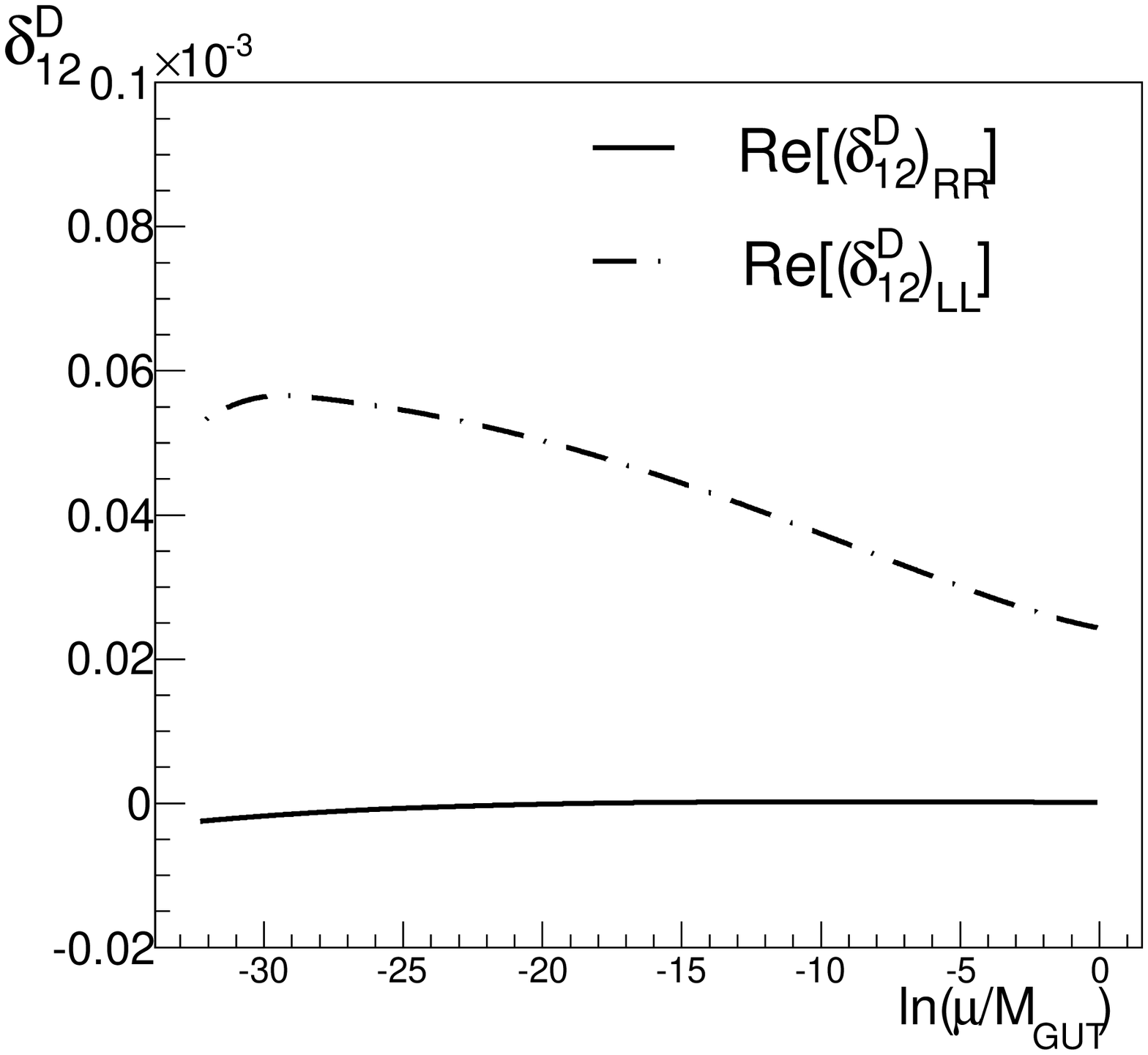}
\includegraphics[width=8.1cm, height=6cm]{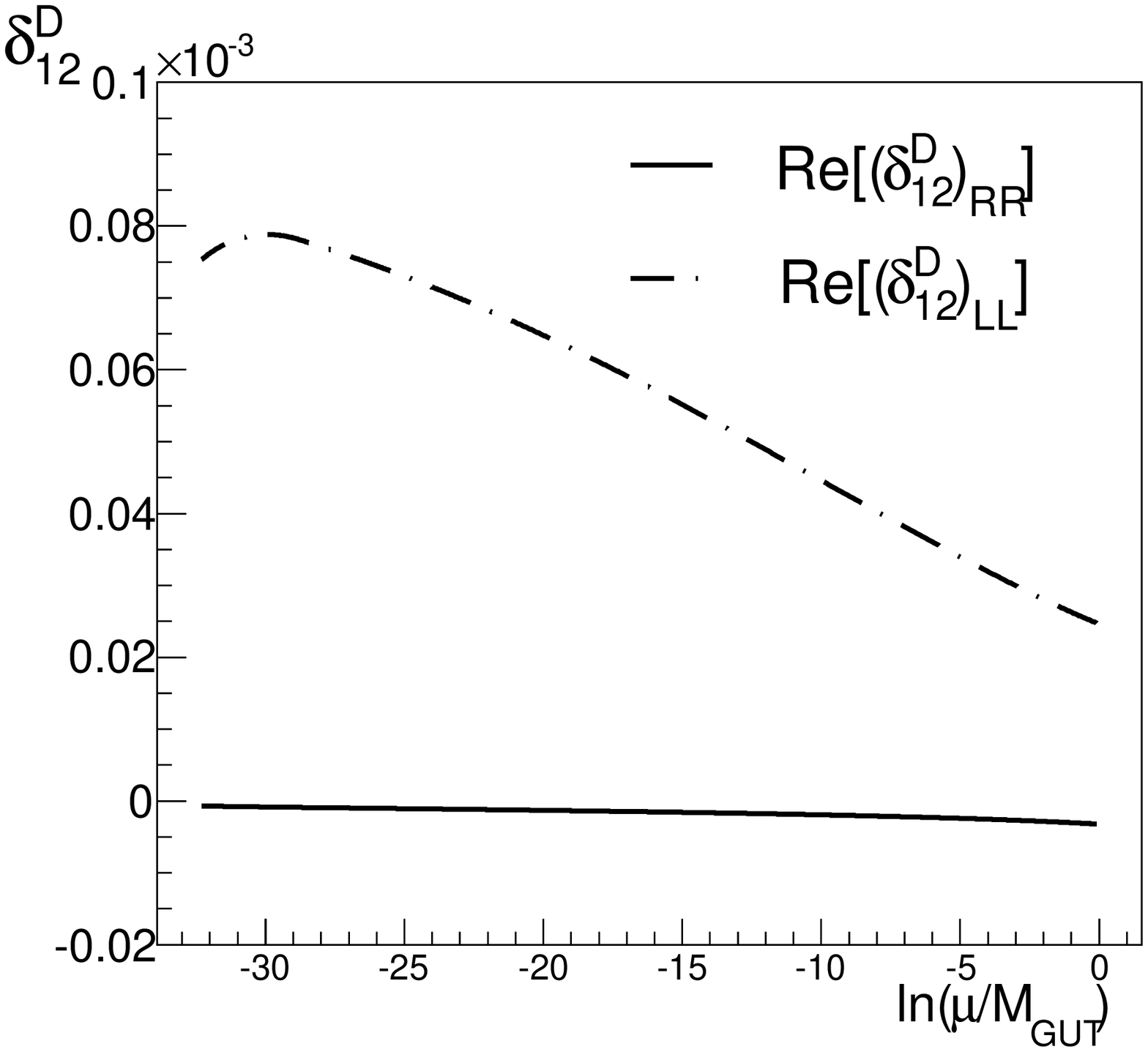}
\caption{\it{Comparison of the runnings of the $D$-quark flavor-violating parameters
$\delta$ for A1 (left panels) and A2 (right panels), for $(\delta_{1i}^{D})_{\{ XX \}}$,  $i=2,3$, $X=L,R$. } \label{fig:delflavA1A2_1} }
\end{figure}

\begin{figure}[!h]
\centering
\includegraphics[width=8.1cm, height=6cm]{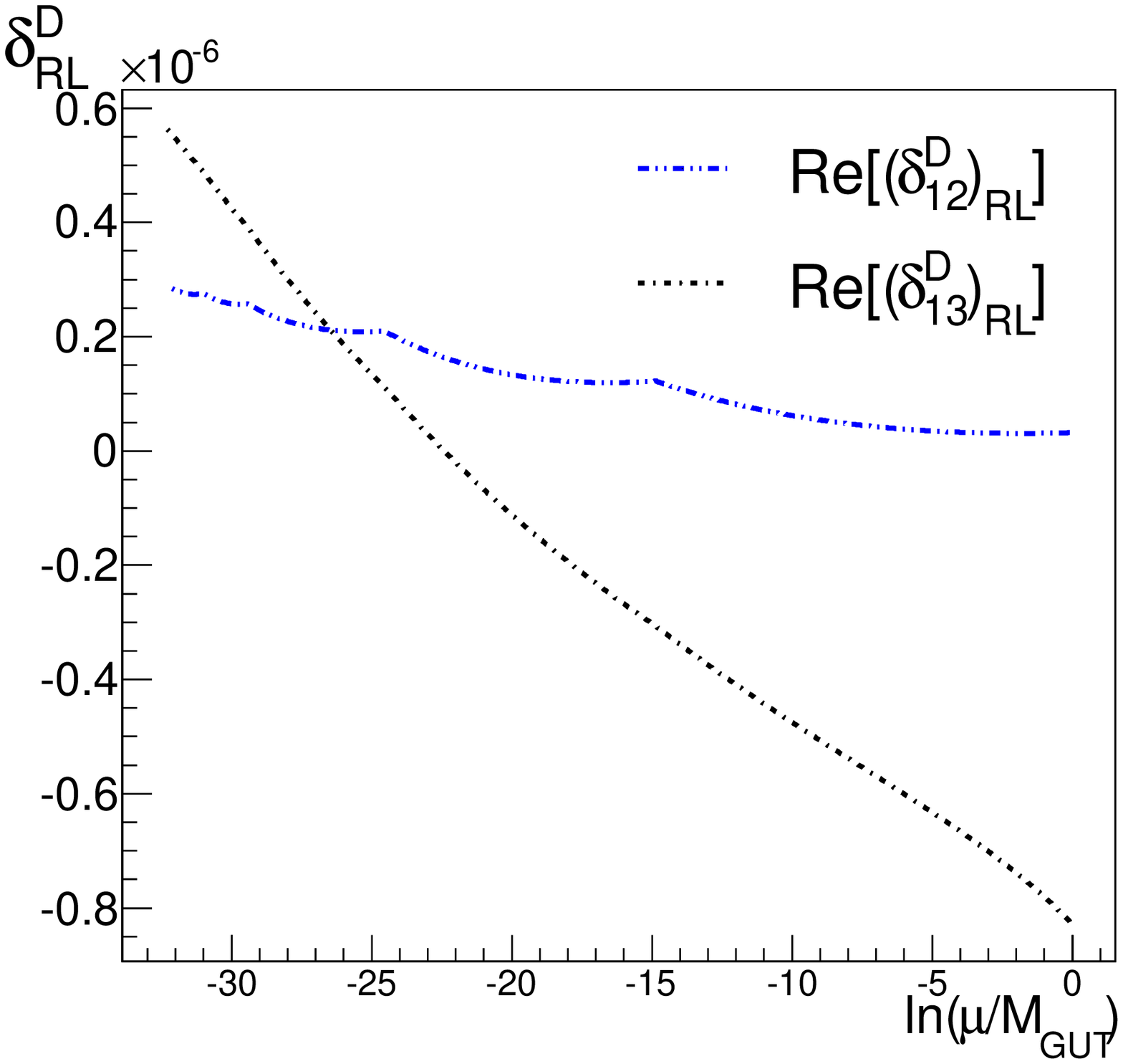}
\includegraphics[width=8.1cm, height=6cm]{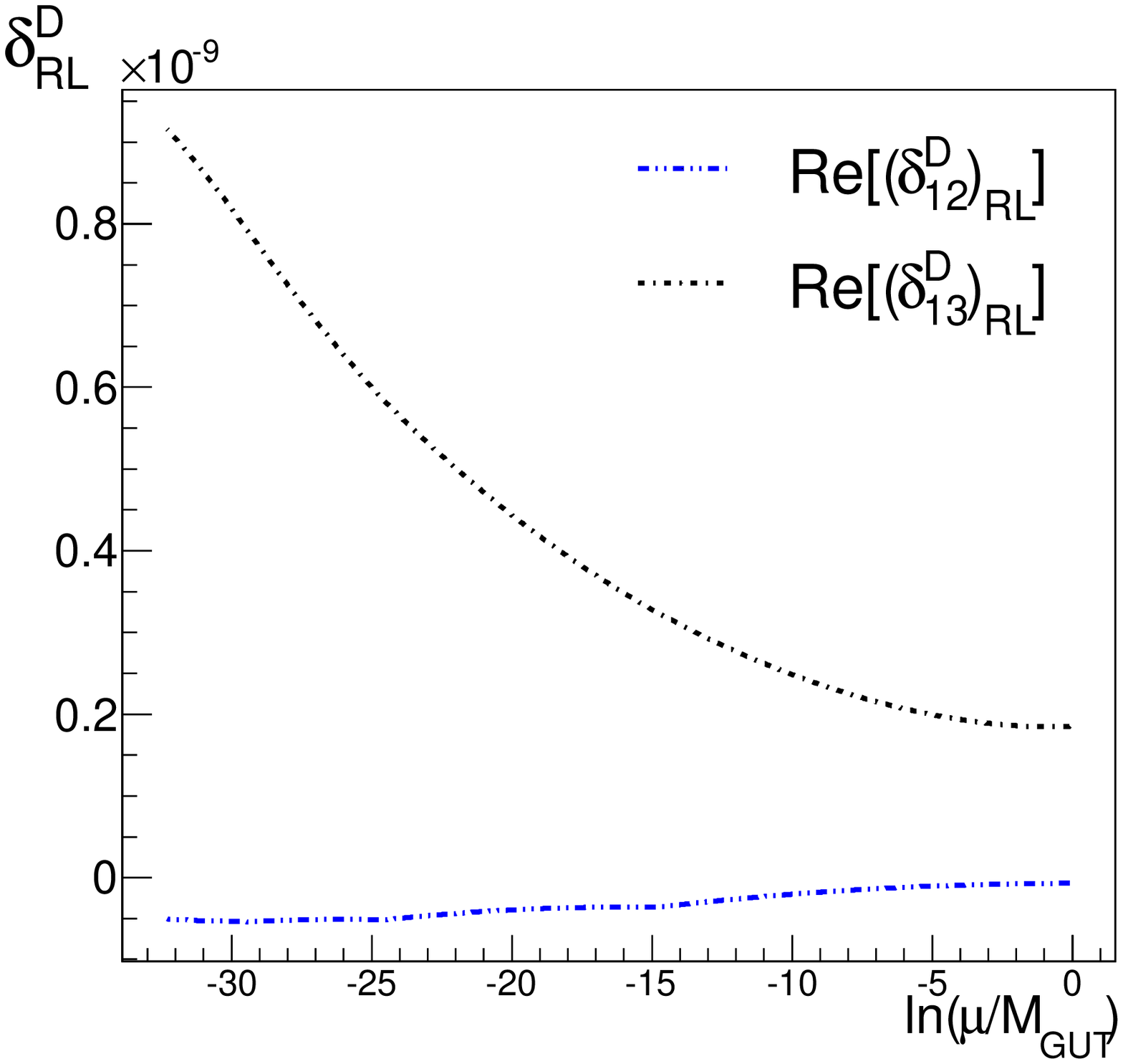}
\includegraphics[width=8.1cm, height=6cm]{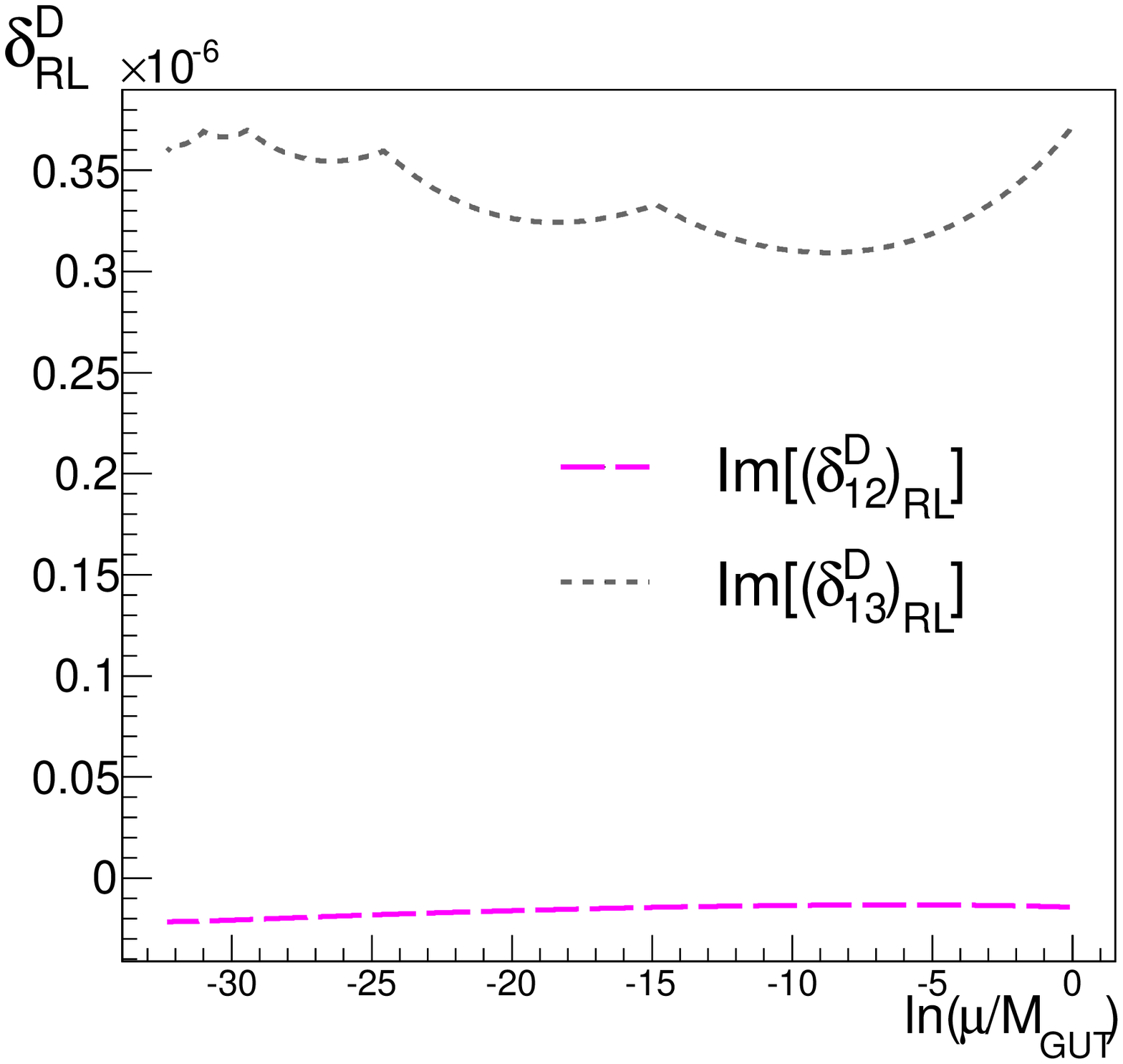}
\includegraphics[width=8.1cm, height=6cm]{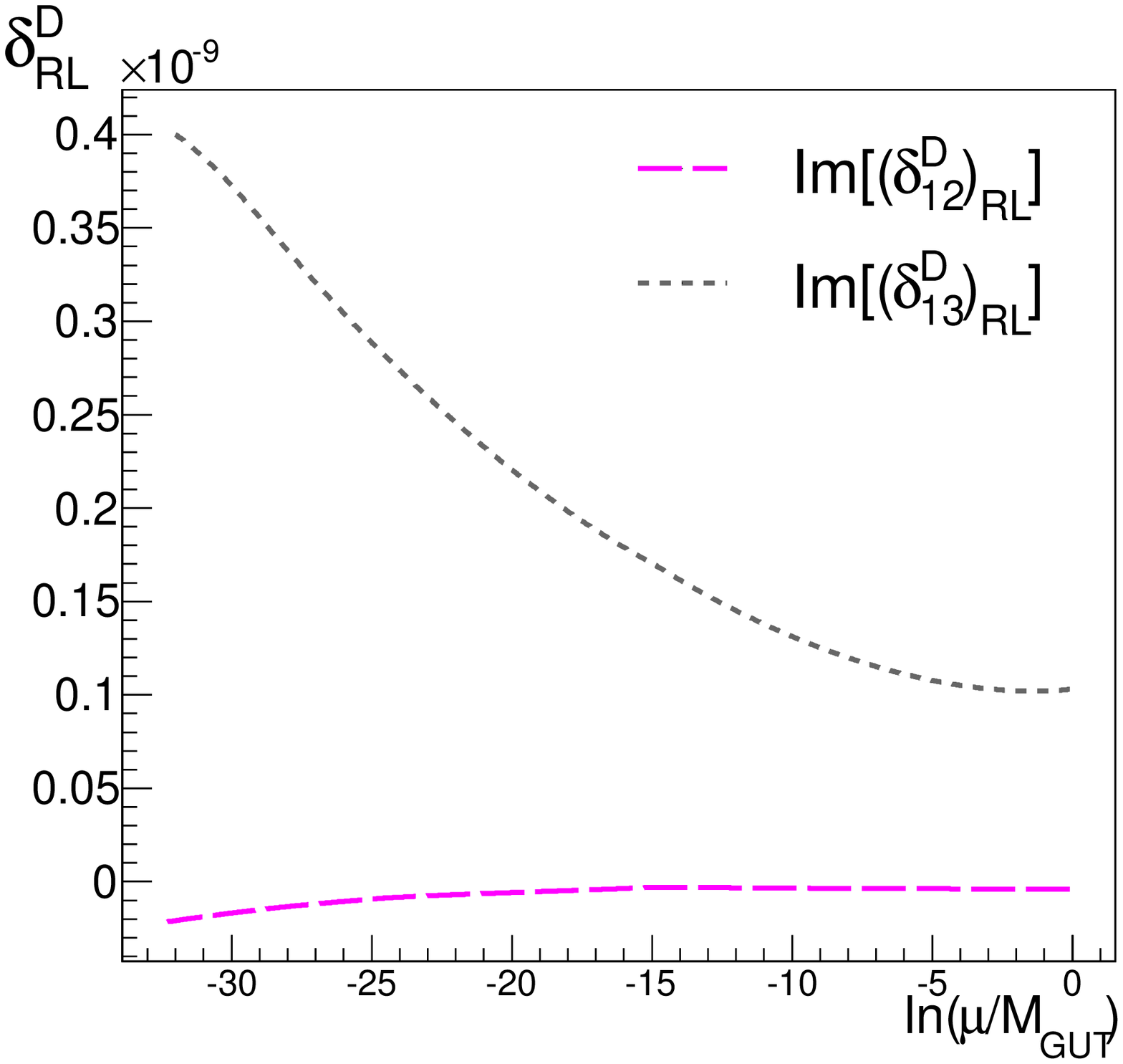}
\caption{\small{\it Comparison of the runnings of the $D$-quark flavor-violating parameters
$\delta$ for A1 (left panels) and A2 (right panels),  for $(\delta_{1i}^{D})_{LR}$,  $i=2,3$.}
%and $(\delta_{23}^{D})_{XY}$, $X,Y=2,3$.} 
 \label{fig:delflavA1A2_2}
}
\end{figure}

\paragraph{Comparison between A1 and A2}  ~\\
The only difference between A1 and A2 is in the $D$-quark sector, Eqs. (\ref{eq:Az1}-\ref{eq:Az4}), 
so we concentrate our comparison on the %$Q$ and $d$-squark sectors.
 $D$-squark sector (left and right). Our interest in comparing these Ans\"atze is to assess the relevance of switching off (A2) the effect of the CKM matrix in the right-handed $D$ sector, and
    to check potential differences in the observables. 
The runnings of the flavor-violating parameters from the GUT scale to the weak scale is shown in 
 Figs.  \ref{fig:delflavA1A2_1} and  \ref{fig:delflavA1A2_2} for models A1 and A2. 
At $\Mg$,  the initial values for most of the $\delta$ parameters are quite similar for both these Ans\"atze. 
In some cases the sign differs, but they have a comparable absolute value. 

The parameters that differ the most are the LR flavor-violating parameters in the (13) flavour sector,
as can be seen from Tables \ref{tbl:A1A2REdlepE} and \ref{tbl:A1A2IMdDsq}. The behaviors of the real and imaginary parts are plotted in
Fig. \ref{fig:delflavA1A2_2}, notice the different scales of the plots. The differences between Ans\"atze A1 and A2
are to be expected, as they arise from the different choices for $V^D_R$. Looking at the terms that enter into the beta function
of $a_D$, shown in \eq{eq:transfsckmaD} in Appendix B, we see that all terms involving $a_D$ are 
sensitive to the change of $V^D_R$, which affect directly the LR parameters.
In contrast, we see from \eq{eq:transfsckm} of Appendix B that only some of the soft mass-squared
 transformations that contribute to the LL and RR parameters are sensitive to the choice of $V^D_R$. 
As a result, we do not expect the flavor-violating parameters coming entirely from the soft masses-squared to be very
sensitive to the change of  $V^D_R$. Indeed, looking at Tables \ref{tbl:A1A2REdlepE} and \ref{tbl:A1A2IMdDsq}, 
we see that the difference is at most one order of magnitude in the (13) RR
and LL sectors for A1 and A2. In the (12) sector, we again see the strong dependence on $V^D_R$
in the LR parameters and relatively small changes in the RR and LL parameters.  In the (23) sector, 
none of the terms are greatly affected by the choice of $V^D_R$. 
Because of the difference between the Yukawa couplings  $h_{D_{23}}$ and  $h_{D_{32}}$,  
the parameters $a_{D_{23}}$ and $a_{D_{32}}$ will have different values at the EW scale, but their
absolute values are similar because their runnings are dominated by the largest Yukawa coupling,
i.e., terms $\propto \hat{h}^2_{D_{3}}$ in \eq{eq:transfsckmaD}, and
contain $V^{D\dagger}_{L,R ii}$ elements. This is not the case for the lighter sectors, because their
Yukawa couplings are smaller.
By way of comparison, we plot the runnings of some of the (12)
flavor parameters in Fig. \ref{fig:delflavA1A2_1}.

%In Ansatz A1, $h_D$ is symmetric, but not in  Ansatz A2, so in the latter case we have
Due to the running of the $a_D$ beta function due in particular to the third and fourth terms in \eq{eq:transfsckmaD}, $a_D$ will not evolve symmetrically 
even if $h_D$ is symmetric, and therefore
\bea
{\rm{Re}}[(\delta^D_{ij})_{RL}] &\neq& {\rm{Re}}[(\delta^D_{ji})_{RL}] \, , \nonumber\\
{\rm{Im}}[(\delta^D_{ij})_{RL}] &\neq& {\rm{Im}}[(\delta^D_{ji})_{RL}] \, .
\eea
If, in addition, $h_D$ is not symmetric (as in Ansatz A2), this effect can be quite noticeable (in particular for $(\delta^D_{12})_{RL}$,
$(\delta^D_{21})_{RL}$, $(\delta^D_{13})_{RL}$ and $(\delta^D_{31})_{RL}$).
%and we do not expect the same evolution of these parameters in these Ans\"atze. 
However, the behaviors of the other flavor-violating parameters tells us that their evolutions
in the SCKM basis do not differ too much, which is explained by the forms of the 
transformations in \eq{eq:transfsckm}, so we have not plotted them in Figs.  \ref{fig:delflavA1A2_1} and  \ref{fig:delflavA1A2_2}. 
The order-of-magnitude differences in the real and imaginary 
parts of $(\delta_{ij}^f)_{\{ XY \}}$ are given in Tables \ref{tbl:A1A2REdlepE} and \ref{tbl:A1A2IMdDsq}, respectively.

\begin{table}[!h]
\begin{center}
\begin{tabular}{| l | l | l  | l |}
%\hline
%\multicolumn{4}{l}{Comparison of $|\rm{Re}[\deltaXY]|$ for $f=Q,D$ in A1 and A2. }\\
\hline
\hline
 & A1 & A2 & Bound (Process) \cite{AbdusSalam_VS:2016} \\
 \hline
%$(\delta_{ij}^f)_{\{ XY \}} $                           &      & \\
$\rm{Re}[(\delta_{12}^D)_{ RR }]$                & $-8 \times 10^{-7}$ &$-2\times 10^{-7}$ &  $10^{-2}$ ($\Delta M_K$) \\
$\rm{Re}[(\delta_{12}^D)_{ LL }]$                  &  $5 \times 10^{-5}$ & $8\times 10^{-5}$ & $10^{-2}$ ($\Delta M_K$)\\
$\rm{Re}[(\delta_{12}^D)_{ \{  LR, RL  \} }]$  & $3 \times 10^{-7}$ &  $-5 \times 10^{-11}$  & -\\
%Added line
$\rm{Re}[(\delta_{21}^D)_{ \{  LR, RL  \} }]$ & $3 \times 10^{-7}$ &  $3 \times 10^{-7}$  & -\\
$\rm{Re}[(\delta_{13}^D)_{ RR }]$                &  $-6\times 10^{-5}$ &  $-2\times 10^{-6}$ & -\\
$\rm{Re}[(\delta_{13}^D)_{ LL }]$                  & $-1\times 10^{-3}$ & $-2\times 10^{-3}$ &-\\
$\rm{Re}[(\delta_{13}^D)_{ \{  LR, RL  \} }]$  & $5 \times 10^{-7}$  &  $9\times 10^{-10}$  & - \\
%Added line
$\rm{Re}[(\delta_{31}^D)_{ \{  LR, RL  \} }]$ & $-5 \times 10^{-7}$ & $-5 \times 10^{-7}$ & -\\
$\rm{Re}[(\delta_{23}^D)_{ RR }]$                &  $3 \times 10^{-4}$ &  $-8 \times 10^{-3}$ & $10^{-2}$ ($\BBsm$) \\
$\rm{Re}[(\delta_{23}^D)_{ LL }]$                  &  $7\times 10^{-3}$ & $1\times 10^{-2}$   & $10^{-2}$ ($\BBsm$) \\
$\rm{Re}[(\delta_{23}^D)_{ \{  LR, RL  \} }]$ &   $-2\times 10^{-6}$ & $-3 \times 10^{-6}$ & -\\
%Added line
$\rm{Re}[(\delta_{32}^D)_{ \{  LR, RL  \} }]$ &   $2\times 10^{-6}$ & $2 \times 10^{-6}$ & -\\
\hline
\end{tabular}
\end{center}
\caption{\it{Comparison  of the real parts of the $D$-squark parameters  $(\delta_{ij}^f)_{\{ XY \}}$ in Ans\"atze A1 and A2.  \label{tbl:A1A2REdlepE}}}
\end{table}
\begin{table}[!h]
\begin{center}
\begin{tabular}{| l | l |  l  | l }
%\hline
%\multicolumn{4}{c}{Comparison of $|\rm{Im}[\deltaXY]|$, $f=Q,D$ for Ans\"atze A1 and A2. }\\
\hline
\hline
 & A1 & A2  \\
 \hline
%$(\delta_{ij}^f)_{\{ XY \}} $                           &      & \\
$\rm{Im}[(\delta_{12}^D)_{ RR }]$                & $-3\times 10^{-7}$ &$2\times 10^{-7}$  \\
$\rm{Im}[(\delta_{12}^D)_{ LL }]$                  &  $2\times 10^{-5}$ & $3\times 10^{-5}$  \\
$\rm{Im}[(\delta_{12}^D)_{ \{  LR, RL  \} }]$  & $-2\times 10^{-8}$ &  $-2\times 10^{-11}$  \\
%New line
$\rm{Im}[(\delta_{21}^D)_{ \{  LR, RL  \} }]$  & $5\times 10^{-9}$ &  $1 \times 10^{-8}$  \\
$\rm{Im}[(\delta_{13}^D)_{ RR }]$                &  $3 \times 10^{-5}$ & $6\times 10^{-6}$  \\
$\rm{Im}[(\delta_{13}^D)_{ LL }]$                  & $-6\times 10^{-4}$ & $-6\times 10^{-4}$  \\
$\rm{Im}[(\delta_{13}^D)_{ \{  LR, RL  \} }]$  & $4\times 10^{-7}$  &  $4\times 10^{-10}$  \\
%New line
$\rm{Im}[(\delta_{31}^D)_{ \{  LR, RL  \} }]$  & $-1\times 10^{-8}$  &  $-1\times 10^{-8}$  \\
$\rm{Im}[(\delta_{23}^D)_{ RR }]$                &  $1\times 10^{-5}$ &  $2\times 10^{-6}$  \\
$\rm{Im}[(\delta_{23}^D)_{ LL }]$                  &  $-1 \times 10^{-4}$ & $-1\times 10^{-4}$  \\
$\rm{Im}[(\delta_{23}^D)_{ \{  LR, RL  \} }]$ &   $8 \times 10^{-8}$ & $3\times 10^{-9}$   \\
%New line
$\rm{Im}[(\delta_{32}^D)_{ \{  LR, RL  \} }]$ &   $-5 \times 10^{-8}$ & $-5\times 10^{-8}$   \\
\hline
\end{tabular}
\end{center}
\caption{\it{Comparison  of the imaginary parts of the $D$-squark parameters  $(\delta_{ij}^f)_{\{ XY \}}$ in Ans\"atze A1 and A2. 
The bounds on the imaginary parts in these scenarios cannot be taken directly from the literature, but must instead
be constructed from different observables (see Section \ref{Subsc:compobs}).
\label{tbl:A1A2IMdDsq}}}
\end{table}

\paragraph{Comparison between A1 and A4} ~\\
The difference between A1 and A4 is due to the replacement of $ V^{E*}_R=V^E_L=1$
 by $V^{E*}_R=V^E_L=V_{\rm{CKM}}$, 
so we expect a significant increase in the  leptonic $L,R$ flavor-violating parameters,
as they are directly linked to the trilinear couplings, which are enhanced by the running of the Yukawa couplings.

\begin{figure}[!ht]
\centering
\includegraphics[width=8.1cm, height=6cm]{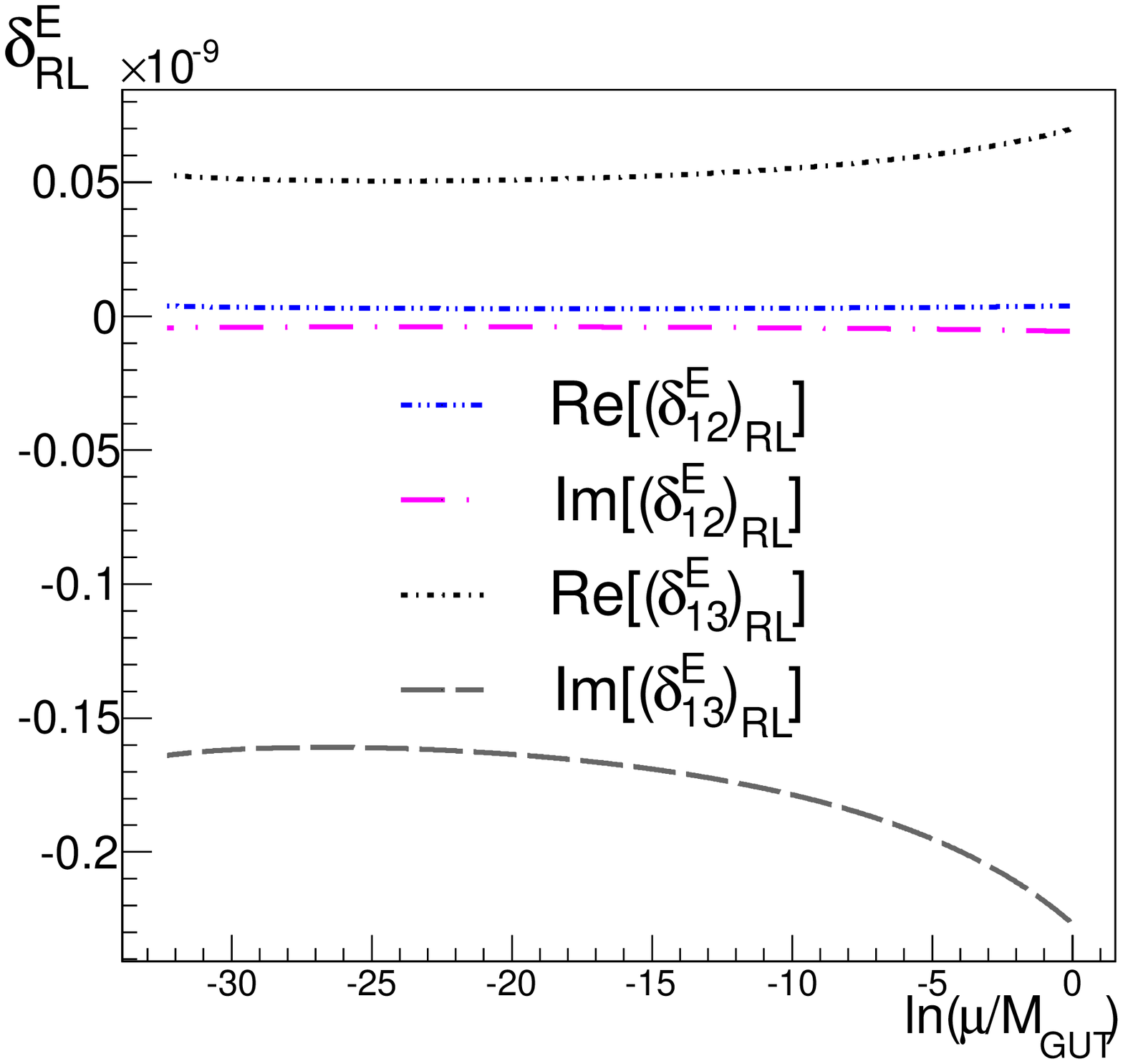}
\includegraphics[width=8.1cm, height=6cm]{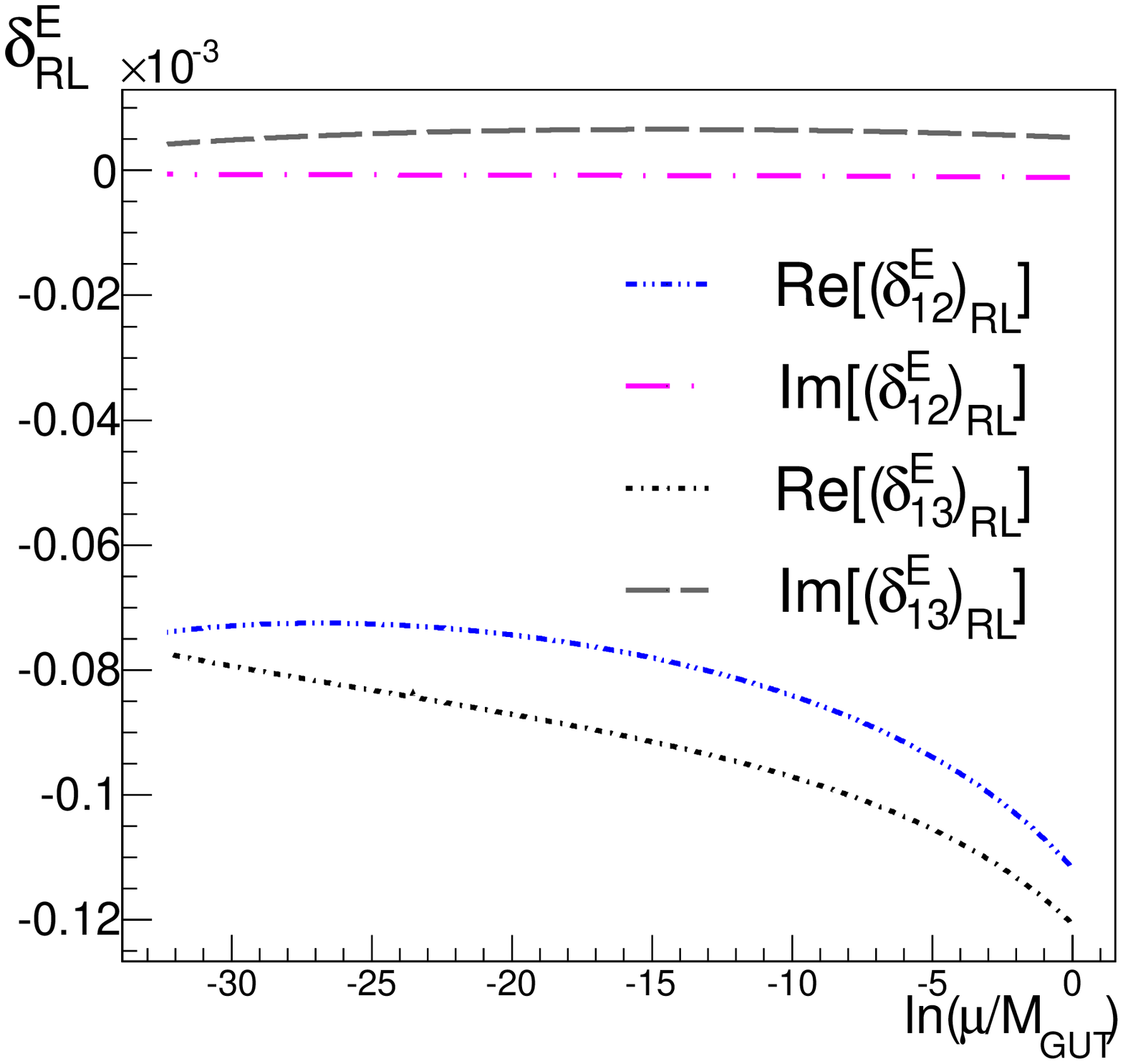}\\
\includegraphics[width=8.1cm, height=6cm]{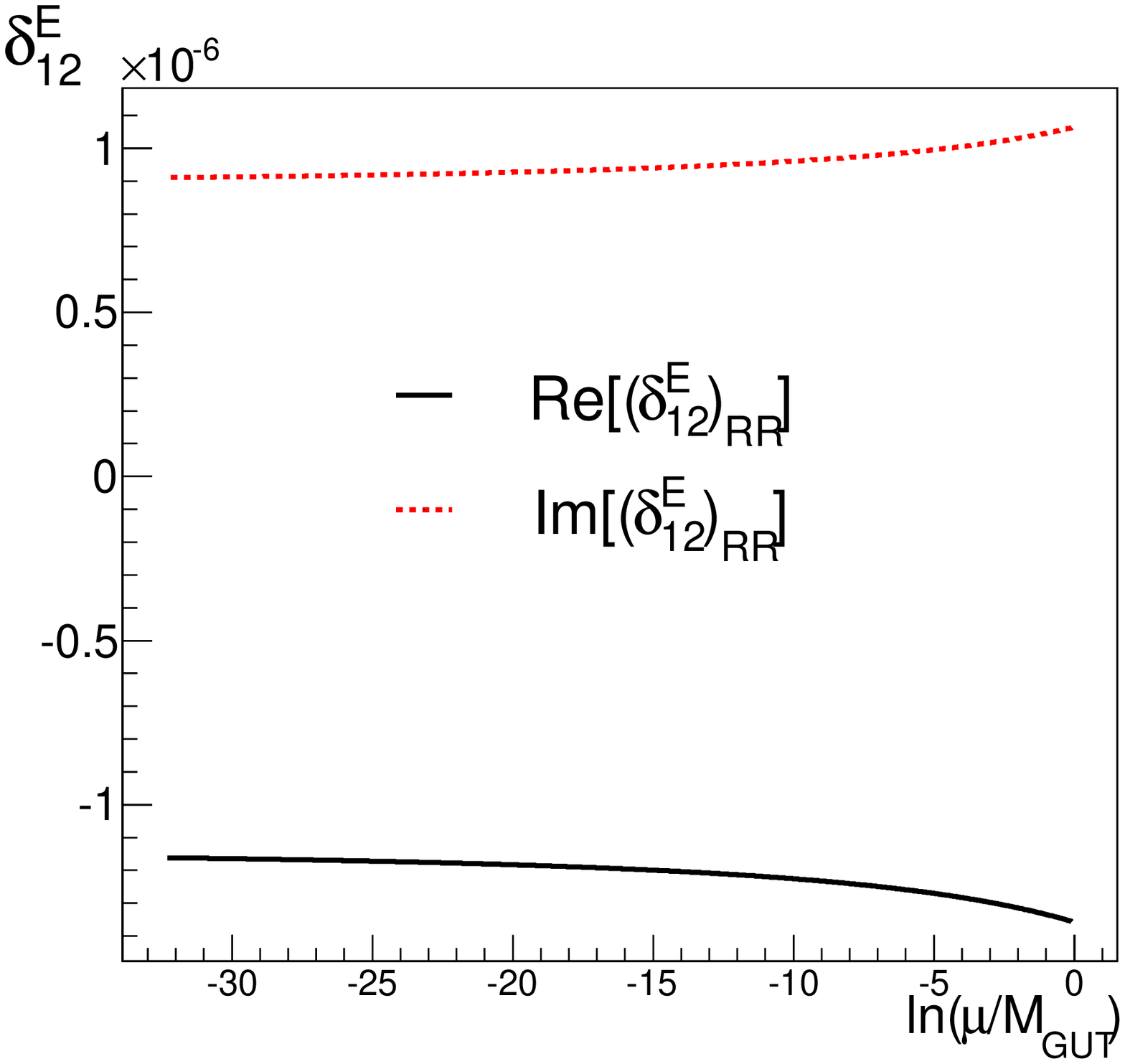}
\includegraphics[width=8.1cm, height=6cm]{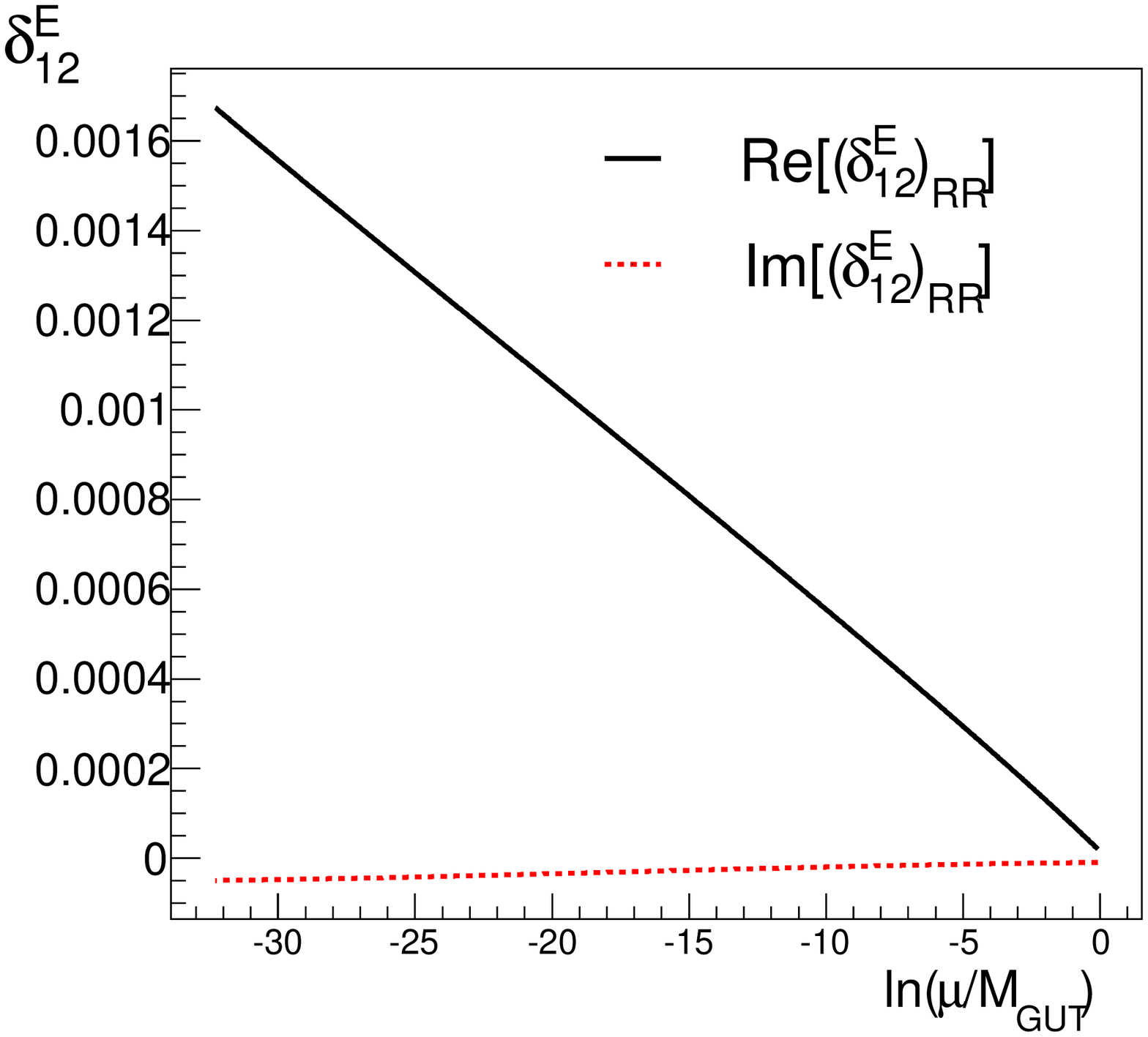}
\includegraphics[width=8.1cm, height=5cm]{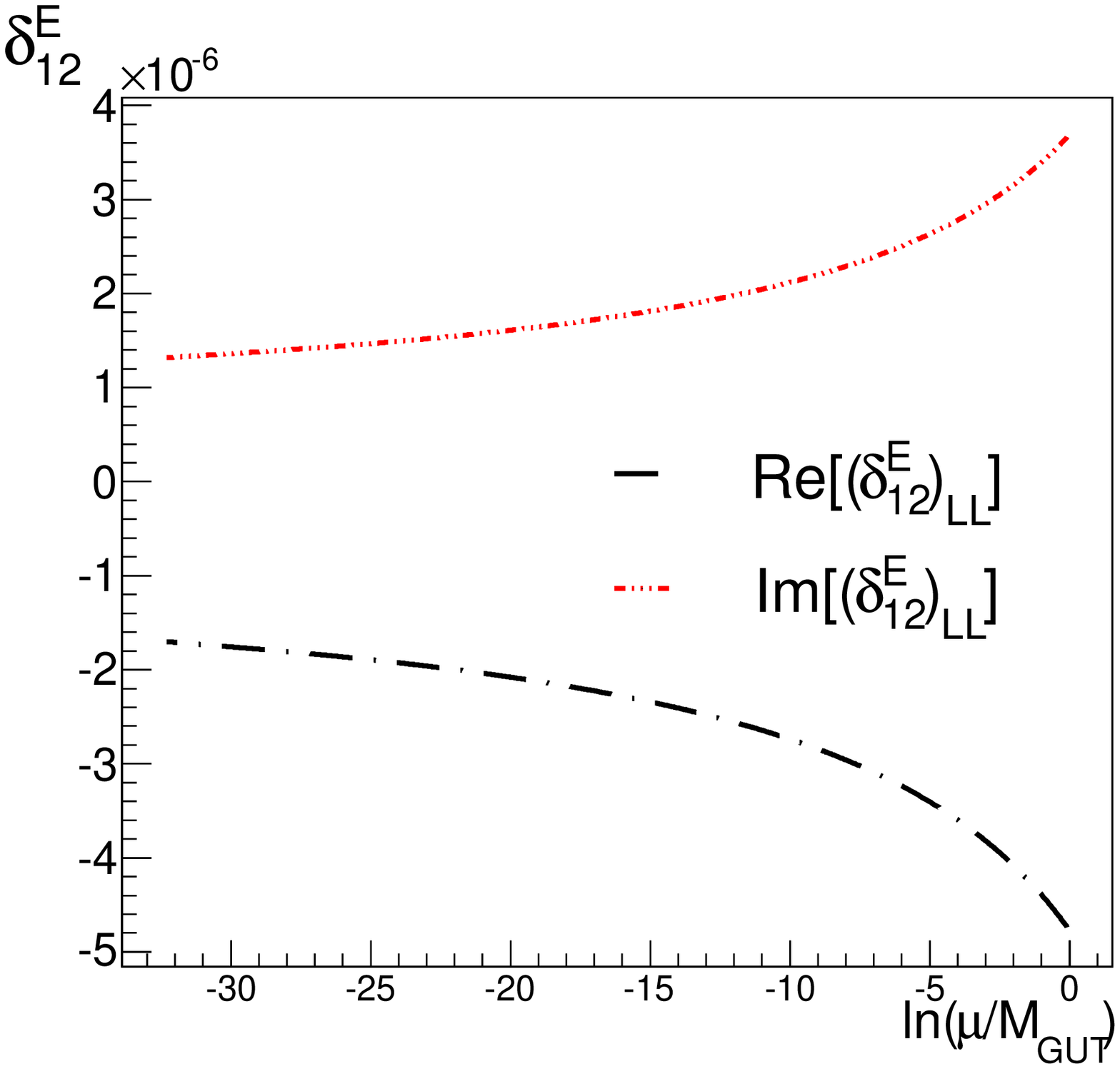}
\includegraphics[width=8.1cm, height=5cm]{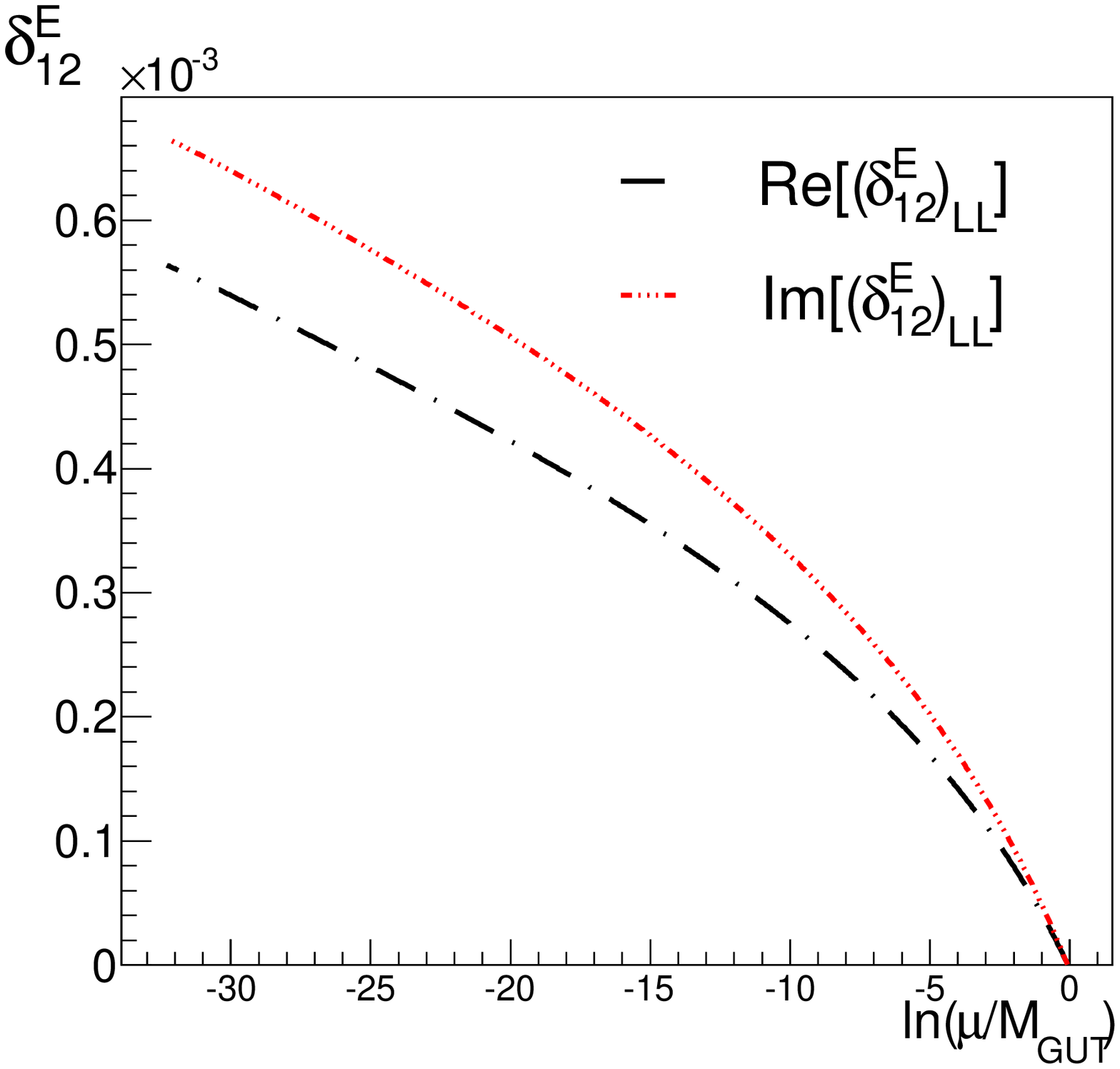}
\caption{\it{Comparison of the runnings of the $E$-lepton flavor-violating parameters
$\delta$ for A1 (left panels) and A4 (right panels),  for $(\delta_{1i}^{E})_{LR}$,  $i=2,3$, and $(\delta_{12}^{E})_{XX}$, $X=L,R$.}
\label{fig:MSSM_EA2}}
\end{figure}

From Table \ref{tbl:A4A1REdlepE} we can see that this is the case for all the real parts of the flavor-violating parameters, specially for $(\delta_{1j}^E)_{\{LR,RL\}} $ for $j=2,3$. This can again
be understood in terms of the evolution of the different terms entering into the beta function of the $E$ trilinear terms.  These have the same form of \eq{eq:transfsckmaD} with the replacement
$D\rightarrow E$, and no analogous  $U$ terms since we are not considering neutrinos. We can see that in general all flavor-violating parameters coming entirely from the soft-squared masses (i.e., $RR$ and $LL$)
have a milder
change than from the $LR$ counterparts, as expected from the analogous form of the terms entering into the $m^2_E$ beta function (analogous to \eq{eq:transfsckm} with the proper replacements). 

For A4, the parameter $|\rm{Re}[(\delta_{12}^E)_{ LL }]|$ exceeds the limit of the general analysis of
\cite{Arana-Catania:2013nha}, and most of the other parameters are at their limits. 
 The different orders of magnitude of the real and imaginary parts of $(\delta_{ij}^{L,E})_{\{ XY \}}$ are given in Tables \ref{tbl:A4A1REdlepE} and \ref{tbl:A4A1IMdlepE}, respectively.
\begin{table}[!h]
\begin{center}
\begin{tabular}{| l | l |  l  | l |}
%\hline
%\multicolumn{4}{c}{Comparison with of $|\rm{Re}[\deltaXY]|$ $f=L,E$ in Ans\"atze A1 and A4. }\\
\hline
\hline
 & A1 & A4 & Bound (Process) \cite{Arana-Catania:2013nha}\\
 \hline
%$(\delta_{ij}^f)_{\{ XY \}} $                           &      & \\
$\rm{Re}[(\delta_{12}^E)_{ RR }]$               & $-1 \times 10^{-6}$  & $2 \times 10^{-3}$ & $10^{-3}$ ($\mu\rightarrow e \gamma$) \\
$\rm{Re}[(\delta_{12}^E)_{ LL }]$                & $-2 \times 10^{-6}$  &  $6\times 10^{-4}$ &$10^{-5}$ ($\mu\rightarrow e \gamma$) \\
$\rm{Re}[(\delta_{12}^E)_{ \{  LR, RL  \} }]$   & $4\times 10^{-12}$   & $-7\times 10^{-5}$ & $10^{-5}$ ($\mu\rightarrow e \gamma$) \\
%Added Line
$\rm{Re}[(\delta_{21}^E)_{ \{  LR, RL  \} }]$   & $2\times 10^{-9}$   & $-7\times 10^{-5}$  & $10^{-5}$ ($\mu\rightarrow e \gamma$) \\
$\rm{Re}[(\delta_{13}^E)_{ RR }]$              & $-1\times 10^{-5}$  &  $-5\times 10^{-3}$ & $10^{-2}$ ($\tau\rightarrow e \gamma$) \\
$\rm{Re}[(\delta_{13}^E)_{ LL }]$                & $-2\times 10^{-5}$  & $-2\times 10^{-3}$ &$10^{-3}$ ($\tau\rightarrow e \gamma$) \\
$\rm{Re}[(\delta_{13}^E)_{ \{  LR, RL  \} }]$ &  $5\times 10^{-11}$ & $-8 \times 10^{-5}$   & $10^{-2}$ ($\tau\rightarrow e \gamma$) \\
%Added Line
$\rm{Re}[(\delta_{31}^E)_{ \{  LR, RL  \} }]$ &  $3\times 10^{-9}$ & $-9 \times 10^{-5}$  & $10^{-2}$ ($\tau\rightarrow e \gamma$) \\
$\rm{Re}[(\delta_{23}^E)_{ RR }]$              &  $-5\times 10^{-4}$   &  $2\times 10^{-2}$ & $10^{-2}$ ($\tau\rightarrow \mu \gamma$) \\
$\rm{Re}[(\delta_{23}^E)_{ LL }]$                & $-6\times 10^{-4}$  &  $7\times 10^{-3}$   &$10^{-2}$ ($\tau\rightarrow \mu \gamma$) \\
$\rm{Re}[(\delta_{23}^E)_{ \{  LR, RL  \} }]$ & 0 &   $5\times 10^{-4}$ &$10^{-2}$ ($\tau\rightarrow \mu \gamma$)  \\
%Added Line
$\rm{Re}[(\delta_{32}^E)_{ \{  LR, RL  \} }]$ & 0 & $5\times 10^{-4}$   &$10^{-2}$ ($\tau\rightarrow \mu \gamma$)  \\
\hline
\end{tabular}
\end{center}
\caption{\it{Comparison  of the real parts of the leptonic parameters  $(\delta_{ij}^E)_{\{ XY \}}$ in Ans\"atze A1 and A4. \label{tbl:A4A1REdlepE}}}
\end{table}

\begin{table}[!t]
\begin{center}
\begin{tabular}{| l | l |  l  | }
\hline
%\multicolumn{3}{c}{Comparison of $|\rm{Im}[\deltaXY]|$ for $f=L,E$ in Ans\"atze A1 and A4. }\\ %Commented, Feb 5, 16 (L)
\hline
\hline
 & A1 & A4 \\
 \hline
%$(\delta_{ij}^f)_{\{ XY \}} $                           &      & \\
$\rm{Im}[(\delta_{12}^E)_{ RR }]$               &$9\times 10^{-7}$ & $-5\times 10^{-5}$ \\
$\rm{Im}[(\delta_{12}^E)_{ LL }]$                  & $1\times 10^{-6}$ &  $7\times 10^{-4}$ \\
$\rm{Im}[(\delta_{12}^E)_{ \{  LR, RL  \} }]$  &  $-4\times 10^{-12}$ & $-6\times 10^{-7}$   \\
%Added Line
$\rm{Im}[(\delta_{21}^E)_{ \{  LR, RL  \} }]$  &  $2\times 10^{-9}$ & $-6\times 10^{-6}$   \\
$\rm{Im}[(\delta_{13}^E)_{ RR }]$              & $4\times 10^{-5}$  &  $4\times 10^{-4}$ \\
$\rm{Im}[(\delta_{13}^E)_{ LL }]$               & $-5\times 10^{-4}$   & $-3\times 10^{-3}$ \\
$\rm{Im}[(\delta_{13}^E)_{ \{  LR, RL  \} }]$  &  $-2\times 10^{-10}$ & $4\times 10^{-6}$ \\
%Added Line
$\rm{Im}[(\delta_{31}^E)_{ \{  LR, RL  \} }]$  &  $3\times 10^{-5}$ & $2\times 10^{-4}$ \\
$\rm{Im}[(\delta_{23}^E)_{ RR }]$              &  $5\times 10^{-6}$  &  $6\times 10^{-5}$ \\
$\rm{Im}[(\delta_{23}^E)_{ LL }]$                & $6\times 10^{-6}$   &  $-8\times 10^{-4}$  \\ 
$\rm{Im}[(\delta_{23}^E)_{ \{  LR, RL  \} }]$   & 0&   $9\times 10^{-7}$ \\
%Added Line
$\rm{Im}[(\delta_{23}^E)_{ \{  LR, RL  \} }]$   & 0&   $4\times 10^{-5}$ \\
\hline
\end{tabular}
\end{center}
\caption{\it{Comparison  of the imaginary parts of the leptonic parameters  $(\delta_{ij}^f)_{\{ XY \}}$ in Ans\"atze A1 and A4.  \label{tbl:A4A1IMdlepE}}}
\end{table}

In \Figref{fig:MSSM_EA2} we compare the runnings of the $E$-lepton flavor-violating parameters
$\delta$ for A1 (left panels) and A4 (right panels),  for $(\delta_{1i}^{E})_{LR}$,  $i=2,3$, and $(\delta_{12}^{E})_{XX}$, $X=L,R$. Although the transformation to the SCKM basis, where
flavor-violating parameters are computed, has the effect of canceling partially
the effect of the running of soft-squared masses, it is not enough to suppress sufficiently  $\Bmueg$, and in fact there is a significant
increase in $\Btaueg$ and $\Btaumug$ with respect to Ansatz A1, see Tables \ref{Tbl:relobsII}-\ref{Tbl:relobsI}.

\begin{figure}
\centering
\begin{picture}(600,110)(210,100)%115
%top picture
\Line(360,120)(400,120)
\Line(400,120)(480,120){3}
\Line(470,120)(530,120)
\DashCArc(440,120)(40,360,180){3}
%\PhotonArc(440,120)(40,0,180){3}{8}
%LR cross vertex
\Vertex(440,160){2}
%dLbL cross
\Line(405,152)(415,142)
\Line(415,152)(405,142)
%
%dRbR cross
%
\Line(465,152)(475,142)
\Line(475,152)(465,142)
\Text(442,110)[1]{$\tilde g$}
\Text(420,165)[1]{$\tilde{b}_L$}
\Text(460,165)[1]{$\tilde{b}_R$}
\Text(390,130)[1]{$\tilde{d}_L$}
\Text(490,130)[1]{$\tilde{d}_R$}
\Text(380,110)[1]{$d$}
\Text(505,110)[1]{$d$}
\Photon(480,140)(520,160){3}{3}
\Text(505,163)[1]{$\gamma$}
\end{picture}
\caption{\it{Contribution of flavor-violating process to the d-quark EDM.}\label{fig:edm_flavor_cont}}
\end{figure}
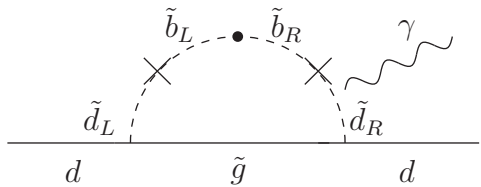

In addition, from the imaginary parts of $|\rm{Im}[(\delta_{ij}^{D,E})_{ XY }]|$ we can get a significant contribution to the EDMs \cite{Pospelov:2005pr}. We have used the {\tt SUSY\_FLAVOR} code~\cite{susyflavor} to compute the EDMs,
but we can understand easily how the imaginary parts of the flavor-violating parameters are constrained.    The constrained combinations are 
\bea
\label{eq:EDMflavconstdel}
\delta^f_{131}\equiv \rm{Arg}\left[(\delta^f_{13})_{LL} (\delta^f_{33})_{LR}  (\delta^f_{31})_{RR} \right], \quad f=D,E.
\eea
Since A1 and A2 differ in their D sectors, a direct difference in the values of the neutron EDM  is expected (as reflected in Table~\ref{Tbl:relobsII}).  Note that, in Fig. \ref{fig:edm_flavor_cont}, both the
real and imaginary parts of $(\delta^f_{13})_{LL}$  and $(\delta^f_{13})_{RR}$ become important.  Specifically, the leading term of the imaginary part of $(\delta^f_{13})_{LL} (\delta^f_{33})_{LR}
(\delta^f_{31})_{RR}$ is
\bea
\label{eq:leadingcont_d131}
\rm{Im}[(\delta^f_{13})_{LL}] \rm{Re}[(\delta^f_{33})_{LR}]  \rm{Re}[(\delta^f_{31})_{RR}] +
\rm{Re}[(\delta^f_{13})_{LL}] \rm{Re}[(\delta^f_{33})_{LR}]  \rm{Im}[(\delta^f_{31})_{RR}].
\nn
\eea
In   \cite{Pospelov:2005pr} sensitivities for the quantities $\delta^f_{131}$ defined in \eq{eq:EDMflavconstdel} 
were computed using the mass insertion approximation with a common scale for soft masses of 1~TeV. It was found, in particular,  
that $\delta^f_{131}\sim 10^{-4}-10^{-3}$. Since our model has specific and
correlated values for the soft parameters, we can compare the impacts of $(\delta^f_{13})_{LL}$ and $(\delta^f_{13})_{RR}$ 
directly to the neutron EDM.  We note that the
imaginary parts of  $(\delta^f_{13})_{LL}$ and $(\delta^f_{13})_{RR}$ in A2 found in Table \ref{tbl:A1A2IMdDsq} are approximately 
one order of magnitude smaller than the corresponding parameters in A1. Hence the neutron EDM is slightly
decreased (by less than an order of magnitude), as seen in Table \ref{Tbl:relobsII}. Using these results,
in Section~\ref{BNSS} we place bounds on $(\delta^f_{13})_{LL}$ and $(\delta^f_{13})_{RR}$ by saturating the EDM bound.

\subsection{Comments on the results and comparison to observables  \label{Subsc:compobs}}

Ans\"atze A1 and A2 predict acceptable flavor violation, while Ans\"atze A3 and A4 do not.  We recall that the properties of the
$D$ and $L$ sectors are controlled by the $\mathbf{\bar{5}}$ sector of $SU(5)$, whereas
the  $Q$, $U$ and $E$ sectors are controlled by the $\mathbf{10}$ sector. The premise of the Ansatz A4 for the 
Yukawa couplings,  \eq{eq:Az4}, was that if soft mass-squared sectors were transformed to the SCKM basis by the same transformations
as the corresponding Yukawa sectors, then the off-diagonal parameters of the corresponding sectors would be  
suppressed because the off-diagonal elements of 
$\mfs$ would be mainly rotated away with the same matrices that make the Yukawa couplings diagonal. 
This is largely the case for the $D$ sector:  in the cases of
$(\mfs)_{13}$ and $(\mfs)_{23}$, for $f=D$, the rotation to the SCKM matrix produces a smaller matrix
element than in the non-SCKM basis. However, this is not the case for other sectors,
where too much flavor violation is produced. On the other hand, in the cases of  Ans\"atze  A2 and A1, 
the rotation to the SCKM basis effectively rotates away any large flavor violation. %Better?

For A1, with the $E_{LL}$ sector we expected a similar behavior  (because the sector is directly linked to the diagonalizing matrices of $h_E$),  but the rotation away of parameters is not as successful as in the $D$ sector (see  Fig.~\ref{fig:MSSM_EA2}). For the sectors $Q$, $U$ and $E$, associated with the $10$ sector of $SU(5)$, which is treated as having diagonal matrices, the off-diagonal elements of $\mfs$ for $f=\ Q,\ D,\ U$  appear as a consequence of the off-diagonal elements of $\mfs$ for $f=E, L$. So they are not related, in principle, but since the Yukawa couplings $h_d$ and $h_e$ are related, we expected that the SCKM transformations of the soft sectors $Q$ and $E$ would tend also to suppress the off diagonal elements. This is however not the case for them, specially for the $E$ sector, where the off-diagonal elements may be considerably enlarged, as seen in Fig.~\ref{fig:MSSM_EA2}.

In Tables \ref{Tbl:relobsII} and \ref{Tbl:relobsI}  we compare the values of the relevant observables
predicted by the Ans\"atze \eq{eq:Az1}-\eq{eq:Az4},  and the corresponding current experimental values.

The CP-violating parameter $\epsilon'$ could give important constraints on models where LR flavor-violating
contributions are much bigger than their RR and LL counterparts: $|(\delta^D_{ij})_{RL}|\gg$
$|(\delta^D_{ij})_{XX}|$, $X=R, L$. This is not the case in our framework, where both the
real and imaginary parts of $[(\delta^D_{ij})_{XX}]$ are much bigger than $[(\delta^D_{ij})_{RL}]$. Furthermore,  chirality-conserving mass
insertion parameters turn out to be more stringently constrained from $\Delta m_K$ and 
$\epsilon_K$~\cite{Gabrielli:1995bd}, which enter as combinations of the
real and imaginary parts of $(\delta^D_{ij})_{XX}$ for the combinations  $ij=\{12,21\}$ \cite{Kersten:2012ed}. 
However, regarding $\Delta m_K$, even in the Standard Model precise computations are not 
possible due to unknown long-distance contributions. Hence, we compare the best estimate obtained from the 
short-distance (SD) contributions~\cite{Brod:2011ty}, denoted by $\Delta m^{SD}_K=(3.1\pm 1.2 )\times
10^{-15}$, to the experimental value, see Table~\ref{Tbl:relobsI}. We see that the central value of $\Delta m^{SD}_K$ 
accounts for 86\% of the experimental central value, and its uncertainty can
easily account for the reported experimental value within $1\sigma$. The value that we obtain in our model also lies
comfortably within $1\sigma$ of the experimental value, taking
into account the SD uncertainty. For $\Delta m_{B_s}$, the SM value is $(17.70\pm 15\%)$, 
and the value that we obtain is in better  agreement with the experimental value.

{\scriptsize{
\begin{table}[p]
\begin{center}
\begin{tabular}{|c|c|c|c|}
\hline
\multicolumn{4}{c}{Relevant observables for A1 and A2}\\
\hline
\hline
 & Experimental Values & & \\
 \hline
EDMs [e cm]& &  A1 &A2\\
\hline
Electron EDM     &  $<8.7  \times10^{-29}$\cite{rpp}               &   {$1.11\times 10^{-30}$}  & $1.19\times 10^{-30}$ \\
Muon EDM         &  $-(0.1 \pm 0.9) \times 10^{-19}$\cite{rpp}    &  $3.49\times 10^{-30}$ & $3.63\times 10^{-30}$\\
Tau EDM            &  $-1.27\times 10^{-25}$                 &   $6.92\times 10^{-30} $ &  $7.13\times 10^{-30}$ \\
Neutron EDM     &  $\begin{array}{c}  <2.9 \times 10^{-26} \cite{rpp}  \\O(10^{-28}) \cite{Hewett:2012ns} 
%<1.94\times 10^{-28} \cite{Altarev:2009zz}
\end{array}$      
          &  $-1.17\times 10^{-28}$  & $-7.18\times 10^{-29}$\\
\hline
M. Anomalies & & &\\
$\Delta a_e$ &  $8.2\times 10^{-13} \cite{Aoyama:2012wj}$ & $1.43\times 10^{-14}$ & $ 1.43 \times 10^{-14}$\\
$\Delta a_\mu$ &  $(2.87 \pm 0.80)\times 10^{-8}$  & $6.15\times 10^{-10}$   &  $ 6.16 \times 10^{-10}$ \\
$\Delta a_\tau$ & $(-5.3 \times 10^{-2}, 1.2 \times 10^{-3})$ & $1.80\times 10^{-7}$  & $ 1.80\times 10^{-7} $\\
\hline
$l^j \rightarrow l^i \gamma$ decays & &  &\\
\hline
$\Bmueg$ & $< 5.7  \times 10^{-13} $ \cite{Adam:2013mnn}  & $5.3 \times 10^{-16}$& $5.4  \times 10^{-16}$ \\
$\Btaueg$ & $< 3.3\times 10^{-8}$ \cite{Aubert:2009ag}&   $1.2 \times 10^{-13}$   &  $1.2 \times 10^{-13}$\\
$\Btaumug$ &  $< 4.4\times 10^{-8}$ \cite{Aubert:2009ag}&$6.0 \times 10^{-12}$  &  $6.2 \times 10^{-12}$\\
\hline
 B decays & &  &\\
\hline
%$\BBsm$& $(3.2^{+1.5}_{-1.2} \times 10^{-9})$ \cite{Aaij:2012nna} &  $3.4\times 10^{-9}$  & $3.4\times 10^{-9}$\\
$\BBsm$& $(2.9\pm 0.7 \times 10^{-9})$ \cite{bmm} &  $3.4\times 10^{-9}$  & $3.4\times 10^{-9}$\\
Untagged & & $3.75\times 10^{-9}$ & $3.76\times 10^{-9}$ \\ 
$\BBdm$ & $(3.6^{+1.6}_{-1.4}) 10^{-10}$ \cite{bmm} & $1.1\times 10^{-10}$ &  $1.0\times 10^{-10}$ \\
$\BBsme$&  $<2.0 \times 10^{-7}$  & $2.09\times 10^{-27}$  & $2.13\times 10^{-27}$\\
$\Btaunu$&  $(1.20 \pm 0.25)\times 10^{-4}$  & $7.43 \times 10^{-5}$ & $7.43 \times 10^{-5}$\\
$\Bsg$ &  $\begin{array}{c}(3.55 \pm 0.24\pm 0.09)\times 10^{-4}\\ \cite{Amhis:2012bh}\end{array} $
 & $3.92 \times 10^{-4}$ &$3.92 \times 10^{-4}$\\
\hline
$\nu$ Kaon decays & &  &\\
\hline
$\BRKLpinunu$ &   $< 2.6\times 10^{-8}$ &$2.32\times 10^{-11}$   & $2.32\times 10^{-11}$ \\
$\BRKPpinunu$ &  $(1.7\pm 1.1) \times 10^{-10}$  & $7.64\times 10^{-11}$  & $7.64\times 10^{-11}$\\
\hline
$\begin{array}{c}
\rm{KK\ mixing} % \\
%(\Delta m\ \rm{in}\ \rm{[GeV]})
\end{array}$
& &  &\\
\hline 
$|\epsilon_K|$ & $(2.223\pm 0.010)10^{-3}$ & $1.81\times 10^{-3}$  & $1.81\times 10^{-3}$ \\
$\Delta m_K$ [GeV] &  {$(3.63\pm 0.0059) \times 10^{-15}$ } &  $ 2.63 \times 10^{-15}$  & $ 2.63 \times 10^{-15}$\\
\hline
BB mixings & &  &\\
\hline
$\Delta m_{B_d}$ & $(3.36\pm 1.97 \times 10^{-2})\times 10^{-13}$  & $3.05\times 10^{-13}$  & \\
$\Delta m_{B_s}$ & $(1.164\pm 1.4\times 10^{-3})\times 10^{-11}$  &  $0.98\times 10^{-11}$ & \\
\hline
\end{tabular}
\end{center}
\caption{\small{\it Comparison of the predictions for Ans\"atze A1 and A2 with
    the experimental values. The values have been obtained with
    the {\tt SUSY\_FLAVOR} code. Without flavor violation, we obtain
    $\Bsg=4.01\times 10^{-4}$. The value in the Table, which includes
    flavor violation, is consistent with experiment within $2\sigma$, taking into
    account the Standard Model uncertainty of $0.23\times 10^{-4}$.  The NNLO Standard Model
    value of $|\epsilon_K|$ is $(1.81\pm 0.28)\times 10^{-3}$.  See Section~\ref{Subsc:compobs} for notes regarding the values of $\Delta m_K$ and $\Delta m_{B_s}$ . }
\label{Tbl:relobsII}}
\end{table}
}}
{\scriptsize{
\begin{table}[p]
\begin{center}
\begin{tabular}{|c|c|c|c|}
\hline
\multicolumn{4}{c}{Relevant observables for A3 and A4}\\
\hline
\hline
 & Experimental Values & & \\
 \hline
EDMs [e cm]& &  A3 &A4\\
\hline
Electron EDM     &   $<8.7  \times10^{-29}$\cite{rpp}               &   {$6.42\times 10^{-28}$}  & {$2.7\times 10^{-27}$} \\
Muon EDM         &  $-(0.1 \pm 0.9) \times 10^{-19}$\cite{rpp}    &  $-1.40\times 10^{-28}$ & $-3.0\times 10^{-27}$\\
Tau EDM            &  $-1.27\times 10^{-25}$                 &   $4.57\times 10^{-28} $ &  $3.7\times 10^{-28}$ \\
Neutron EDM     &  $\begin{array}{c}  <2.9 \times 10^{-26} \cite{rpp}  \\O(10^{-28}) \cite{Hewett:2012ns} 
%<1.94\times 10^{-28} \cite{Altarev:2009zz}
\end{array}$           &  $-3.59\times 10^{-29}$   & $3.7\times 10^{-30}$\\
\hline
M. Anomalies & & &\\
$\Delta a_e$ &  $8.2\times 10^{-13} \cite{Aoyama:2012wj}$ & $3.2\times 10^{-14}$ & $ 1.2 \times 10^{-14}$\\
$\Delta a_\mu$ &  $(2.87 \pm 0.80)\times 10^{-8}$  & $1.4\times 10^{-9}$   &  $ 6.1 \times 10^{-10}$ \\
$\Delta a_\tau$ & $(-5.3 \times 10^{-2}, 1.2 \times 10^{-3})$ & $4.1\times 10^{-7}$  & $ 1.8\times 10^{-7} $\\
\hline
$l^j \rightarrow l^i \gamma$ decays & &  &\\
\hline
$\Bmueg$ & $< 5.7  \times 10^{-13} $ \cite{Adam:2013mnn}  & {$1.7 \times 10^{-9}$} & ${7.49  \times 10^{-10}}$ \\
$\Btaueg$ & $< 3.3\times 10^{-8}$ \cite{Aubert:2009ag}&   $6.6 \times 10^{-13}$   &  $1.02 \times 10^{-12}$\\
$\Btaumug$ &  $< 4.4\times 10^{-8}$ \cite{Aubert:2009ag}&$1.05 \times 10^{-11}$  &  $1.52 \times 10^{-12}$\\
\hline
 B decays & &  &\\
\hline
$\BBsm$& $(2.9\pm 0.7 \times 10^{-9})$ \cite{bmm}&  $3.3\times 10^{-9}$  & $3.4\times 10^{-9}$\\
Untagged & & $3.66\times 10^{-9}$ & $3.75\times 10^{-9}$ \\
$\BBdm$ &  $(3.6^{+1.6}_{-1.4}) 10^{-10}$ \cite{bmm} & $1.4\times 10^{-10}$ &  $1.1\times 10^{-10}$ \\
$\BBsme$&  $<2.0 \times 10^{-7}$  & $9.1\times 10^{-23}$  & $2.3\times 10^{-21}$\\
$\Btaunu$&  $(1.20 \pm 0.25)\times 10^{-4}$  & $7.37 \times 10^{-5}$ & $7.43 \times 10^{-5}$\\
$\Bsg$ &  $\begin{array}{c}(3.55 \pm 0.24\pm 0.09)\times 10^{-4}\\ \cite{Amhis:2012bh}\end{array} $
 & $4.22 \times 10^{-4}$ &$3.9 \times 10^{-4}$\\
\hline
$\nu$ Kaon decays & &  &\\
\hline
$\BRKLpinunu$ &   $< 2.6\times 10^{-8}$ &$2.32\times 10^{-11}$   & $2.3\times 10^{-11}$ \\
$\BRKPpinunu$ &  $(1.7\pm 1.1) \times 10^{-10}$  & $7.6\times 10^{-11}$  & $7.6\times 10^{-11}$\\
\hline
$\begin{array}{c}
\rm{KK\ mixing} % \\
%(\Delta m\ \rm{in}\ \rm{[GeV]})
\end{array}$
& &  &\\
\hline 
$|\epsilon_K|$ & $(2.223\pm 0.010)10^{-3}$ & $1.82\times 10^{-3}$  & $1.81\times 10^{-3}$ \\
$\Delta m_K$ [GeV] &  {$(3.483\pm 0.0059)\times 10^{-15}$ } &  $2.63\times 10^{-15}$  & $2.63\times 10^{-15}$ \\
\hline
\end{tabular}
\end{center}
\caption{\small{\it Comparison of the predictions for our Ans\"atze A3 and A4
    with the experimental values. 
\label{Tbl:relobsI}}}
\end{table}
}}

\section{Beyond No-Scale GUTs: Maximal Flavour Violation \label{BNSS}}

In this Section we explore the possibilities for deformations of the pure no-scale boundary
conditions with non-vanishing scalar masses. We study the conditions under which flavour violation
could be maximal, in the sense that off-diagonal entries in the scalar mass matrices could be
as large as the diagonal entries at some super-GUT input scale $\Mi >
\Mg$. {We emphasize that, as in the previous Section, we are not trying to explain any
observed effects, but rather to set limits on the sizes of the possible off-diagonal terms using current 
measurements.
When these are as large as the diagonal terms, we are in the regime we have called
MaxSFV.}

This analysis builds upon that in the previous Sections, in which we analysed the most
powerful flavour constraints in representative no-scale SU(5) GUT scenarios.
To this end, we consider scalar mass-squared matrices with universal flavour-diagonal entries
and consider the effects of switching real off-diagonal real entries
on, one at a time. {{There are in fact far too many possible parameter combinations in general.
Thus, in order to see the effect of the additional (off-diagonal) parameters and 
to be concrete, we make some simplifying assumptions and test these one by one (in terms of generations).}}

Thus we consider matrices of the form
\bea
m^2_{\bar{5}}=\left(
\begin{array}{ccc}
a & b & 0\\
b & a & 0\\
0 & 0 & a
\end{array}
\right),
\quad
m^2_{\bar{5}}=\left(
\begin{array}{ccc}
a & 0 & b\\
0 & a & 0\\
b & 0 & a
\end{array}
\right),
\quad
m^2_{\bar{5}}=\left(
\begin{array}{ccc}
a & 0 & 0\\
0 & a & b\\
0 & b & a
\end{array}
\right),\label{m2b5BNS}
\eea
and 
\bea
m^2_{10}=\left(
\begin{array}{ccc}
c & d & 0\\
d & c & 0\\
0 & 0 & c
\end{array}
\right),
\quad
m^2_{10}=\left(
\begin{array}{ccc}
c & 0 & d\\
0 & c & 0\\
d & 0 & c
\end{array}
\right),
\quad
m^2_{10}=\left(
\begin{array}{ccc}
c & 0 & 0\\
0 & c & d\\
0 & d & c
\end{array}
\right),\label{m210BNS}
\eea
for $a\neq b$ and $c\neq d$, with all four parameters real. The rest of the boundary conditions at $\Mi$ are taken to be the
same as in \Secref{subsec:BCMin}.
We start by considering separately the boundary conditions (\ref{m2b5BNS}) and  (\ref{m210BNS}),
treating each sector separately. 

\subsection{Off-diagonal entries in $\mathbf{m}^2_{\bar{\mathbf{5}}}$ and $\mathbf{m}^2_{{\mathbf{10}}}$}

\subsubsection{The $(1,2)$ sector}

\begin{figure}
\centering
\begin{picture}(680,170)(215,0)%115
%top picture
\Line(360,120)(400,120)
\DashLine(400,120)(480,120){3}
\Line(480,120)(520,120)
\CArc(440,120)(40,360,180)
\PhotonArc(440,120)(40,0,180){3}{8}
%mi cross
%
\Line(435,125)(445,115)
\Line(435,115)(445,125)
\Text(432,110)[1]{$\tilde \ell_m$}
\Text(455,110)[1]{$\tilde \ell_n$}
\Text(420,165)[1]{$\tilde \chi^0$}
\Text(380,110)[1]{$\ell_2=\mu$}
\Text(505,110)[1]{$\ell_1=e$}
\Photon(470,150)(510,170){3}{3}
\Text(500,150)[1]{$\gamma$}
%bottom pictures
  %
\Vertex(340,70){2}
%\Line(335,175)(345,165)
%\Line(335,165)(345,175)
%
\Line(260,30)(300,30)
\Text(320,75)[1]{$\tilde B$}
\Text(360,75)[1]{$\tilde B$}
\DashLine(300,30)(380,30){3}
\CArc(340,30)(40,360,180)
\PhotonArc(340,30)(40,0,180){3}{8}
\Line(380,30)(420,30)
\Vertex(325,30){2}
\Line(350,35)(360,25)
\Line(350,25)(360,35)
\Text(280,20)[1]{$\mu_L$}
\Text(310,20)[1]{$\tilde\mu_L$}
\Text(345,20)[1]{$\tilde\mu_R$}
\Text(370,20)[1]{$\tilde e_R$}
\Text(405,20)[1]{$e_R$}
\Text(343,01)[1]{$\deltaRR \neq 0 \Rightarrow \amuegR \neq 0$}
%\Line(40,30)(80,30)
%\CArc(120,30)(40,360,180)
%\PhotonArc(120,30)(40,0,180){3}{8}
%\Vertex(120,70){2}
\Text(320,75)[1]{$\tilde B$}
\Text(360,75)[1]{$\tilde B$}
%other
%\Text(540,05)[1]{$ a_{\mu e \gamma}$}
%
\Vertex(540,70){2}
%\Line(335,175)(345,165)
%\Line(335,165)(345,175)
%
\Line(460,30)(500,30)
\Text(520,75)[1]{$\tilde B$}
\Text(560,75)[1]{$\tilde B$}
%\DashArrowLine(300,130)(380,130){3}
\DashLine(500,30)(580,30){3}
\CArc(540,30)(40,360,180)
\PhotonArc(540,30)(40,0,180){3}{8}
\Line(580,30)(620,30)
\Vertex(525,30){2}
\Line(550,35)(560,25)
\Line(550,25)(560,35)
\Text(480,20)[1]{$\mu_R$}
\Text(510,20)[1]{$\tilde\mu_R$}
\Text(545,20)[1]{$\tilde\mu_L$}
\Text(570,20)[1]{$\tilde e_L$}
\Text(605,20)[1]{$e_L$}
\Text(543,01)[1]{$\deltaLL \neq 0 \Rightarrow \amuegL \neq 0$}
\end{picture}
\caption{\it Top: Loop diagram representing the transition $\ell_2 \rightarrow \ell_1 \gamma$ in mass eigenstates, with $\ell_i$ being the lepton mass eigenstates, in our Ansatz 1 $\ell_2=\mu$ and
  $\ell_1=e$, and $\tilde \ell_m$ the sleptons eigenstates, $m=1,2,\hdots, 6$. Bottom: Leading contributions to $\amuegG$ in the case of large $\mu$, represented in the flavor basis.}
\label{eq:spNeut_diagrams}
\end{figure}
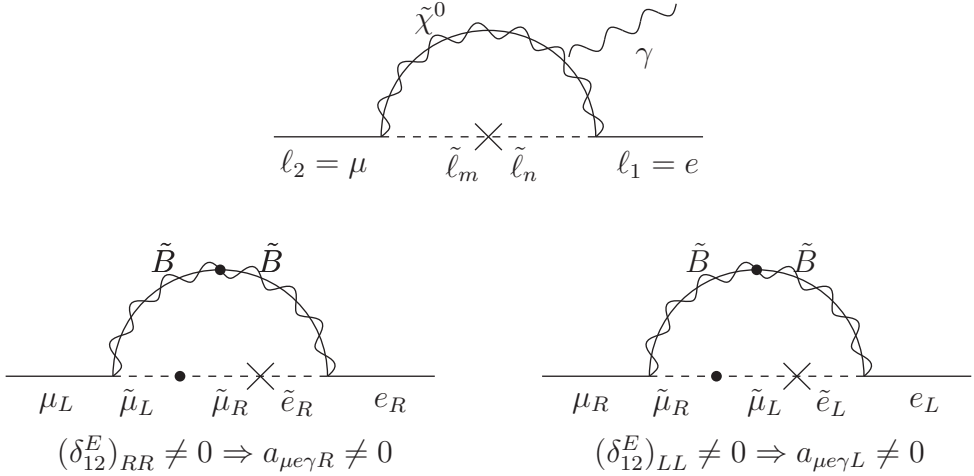

The inputs (\ref{m2b5BNS}) imply the following universal matching conditions at $\Mg$:
\bea
m^2_{D _{12}}=m^2_{L _{12}}=m^2_{\bar{5}_{12}} \, ,
\eea
which will have the strongest impact on observables where $m^2_{D _{12}}$, $m^2_{L _{12}}$ enter at one loop level. In the lepton sector, this is the case for the amplitude of the process
$\mu\rightarrow e \gamma$, shown in Fig.~\ref{eq:spNeut_diagrams}. This amplitude can be written as  $\mathcal{M}_{\mu e \gamma}$$= \frac{e}{2m_\mu}\, \epsilon^{*\alpha}
\bar{u}_e(k+q)
\,[ i \sigma_{\beta\alpha} q^\beta
( a_{\mu e \gamma R} P_L + a_{\mu e \gamma L} P_R) ] \, u_\mu (k)$
where 
$\sigma_{\alpha\beta} = i/2 \, [\gamma_\alpha,\gamma_\beta]$,
$\epsilon^\alpha$ is the
photon polarization vector, $k$ and $k+q$ are on-shell momenta, 
$u_\mu$, $\bar u_\mu$ are spinors that satisfy the Dirac equation and  $P_{R,L}=(1\pm \gamma_5)/2$. The $L/R$ index in
$a_{\mu e \gamma L/R}$ refers to the electron chirality. 
In terms of $a_{\mu e \gamma L}$ and $a_{\mu e \gamma R}$, we have
\begin{equation}
\label{def:BRemugamma}
\text{BR} (\mu\rightarrow e \gamma)= \frac{3\pi^2 e^2}{G_F^2 m^4_\mu} \, ( |a_{\mu e \gamma L}|^2 + |a_{\mu e \gamma R}|^2 ).
\end{equation}
In fact, if $\tan\beta$ is large and $\mu>M_2>M_1$, the dominant contribution to $\text{BR} (\mu\rightarrow e \gamma)$  is mediated by bino exchange,
which is represented in the flavor basis by the mass-insertion diagrams of the second row in
Fig.~\ref{eq:spNeut_diagrams}.
The contributions to $a_{\mu e \gamma L,R}$  in the mass eigenstate basis are schematically as follows  \cite{Kersten:2014xaa} 
\begin{eqnarray}
\amuegL &\approx&  g_1^2 \frac{m_\mu}{48\pi^2} \sum_m K^*_{m1} K_{m5} \,
% g_1^2 \frac{m_\mu}{48\pi^2} \sum_m  f( (\delta^{E}_{1m})^*_{LL} (\delta^E_{m2})_{LL}) \,
\frac{M_1}{m^2_{L_m}} \, F_2^N\left(\frac{m^2_{\tilde\chi^0_1}}{m_{L_m}^2} \right) ,
\nonumber\\
\amuegR &\approx& g_1^2 \frac{m_\mu}{48\pi^2} \sum_m K^*_{m4} K_{m2} \,
%g_1^2 \frac{m_\mu}{48\pi^2} \sum_m f( (\delta^E_{1m})^*_{RR} (\delta^E_{m2})_{RR} \,
\frac{M_1}{m^2_{L_m}} \,F_2^N\left(\frac{m^2_{\tilde\chi^0_1}}{m_{L_m}^2} \right), \label{eq:amuegLRLargeMu}
\end{eqnarray}
where the matrices $K$ diagonalize the squared-mass matrices, \eq{eq:diagmatKl}. 
The mass eigenstate index $1$ in $K^*_{m1}K_{m5}$, corresponds roughly to $\tilde e_L$ while the index $5$ to
$\tilde \mu_L$.  Analogously in  $K^*_{m4}K_{m2}$, the index $4$ roughly corresponds to $\tilde \mu_R$,
while the index $2$ refers to $\tilde e_R$, and $F_2^N$ denotes the loop function involved in each diagram.  All
other contributions involve higgsinos and are therefore suppressed for
large $\mu$. We note that in Eqs. (\ref{eq:amuegLRLargeMu}), all indices $m=1,2,\hdots, 6$
give contributions to $\Bmueg$.  
When $\hat{m}^2_{E (RR)_{12}}$ is significantly suppressed with respect to $\hat{m}^2_{E
  (LL)_{12}}$, which is our case since the seed for $\hat{m}^2_{E (RR)}$ is $m^2_{10}$, whereas the seed  for  $\hat{m}^2_{E (LL)}$ is $m^2_{5}$, then 
  $\amuegL$ dominates. Its main 
    contributions come from mass
    eigenstates mainly containing $\tilde e_L$, $\tilde \mu_L$ and $\tilde \mu_R$, provided that flavor-violating $LR$ parameters are smaller than $LL$ and $RR$, which is also our case.  Then the most
    important contributions in  $\amuegL$  come when $\tilde \ell_m=$$\tilde e_L$, $\tilde \mu_L$ and $\tilde \mu_R$, and for each of them
\bea
&&K^*_{m1}K_{m5}\approx \left.t_{\tilde l_m} (\delta^E_{22})_{LR} (\delta^E_{12})_{LL} \right|_{\tilde\ell_m\approx \tilde e_L, \tilde \mu_L, \tilde \mu_R},
\eea
where $t_{\tilde l_m} $ is a number of $O(1)$ and is different for each of the terms above \cite{Kersten:2014xaa}.
Hence, once $m^2_{\bar{5}_{12}}$ is non-zero at the input scale, the parameter  $m^2_{L _{12}}$ is significantly bigger at the GUT scale than in the no-scale supergravity case, and this directly impacts the
increase of $\amuegL$ and hence $\Bmueg$. In the quark sector, the analogous increase of 
$ m^2_{D _{12}}$ at the GUT scale will have an impact in the analogous amplitude to $\amuegR$, that is $a_{sd\gamma R}$, but this time the Standard Model contribution to the decay 
$s\rightarrow d \gamma$  will be
the dominant one, with supersymmetry making only a tiny correction \cite{Mertens:2011ts}. 
We observe that $m^2_{D _{12}}$ affects the observable  $\BBdm$ because it enters through
a penguin diagram with Higgsinos and sleptons in the loop, but also in this case the contribution is tiny in comparison to the SM contribution \cite{Dedes:2008iw}. Finally, $\Delta m_K$ and hence
$\epsilon_K$ are affected by $m^2_{D _{12}}$ but in this case the contributions coming from the small flavor-violating parameters in Ansatz 1 do not have an effect, as these parameters should typically
be of $O(10^{-2})$ to have an impact in changing the values of $\epsilon_K$ and $\Delta m_K$ \cite{Kersten:2012ed}.

\begin{figure}[t]
\centering
\includegraphics[width=8.1cm, height=7cm]{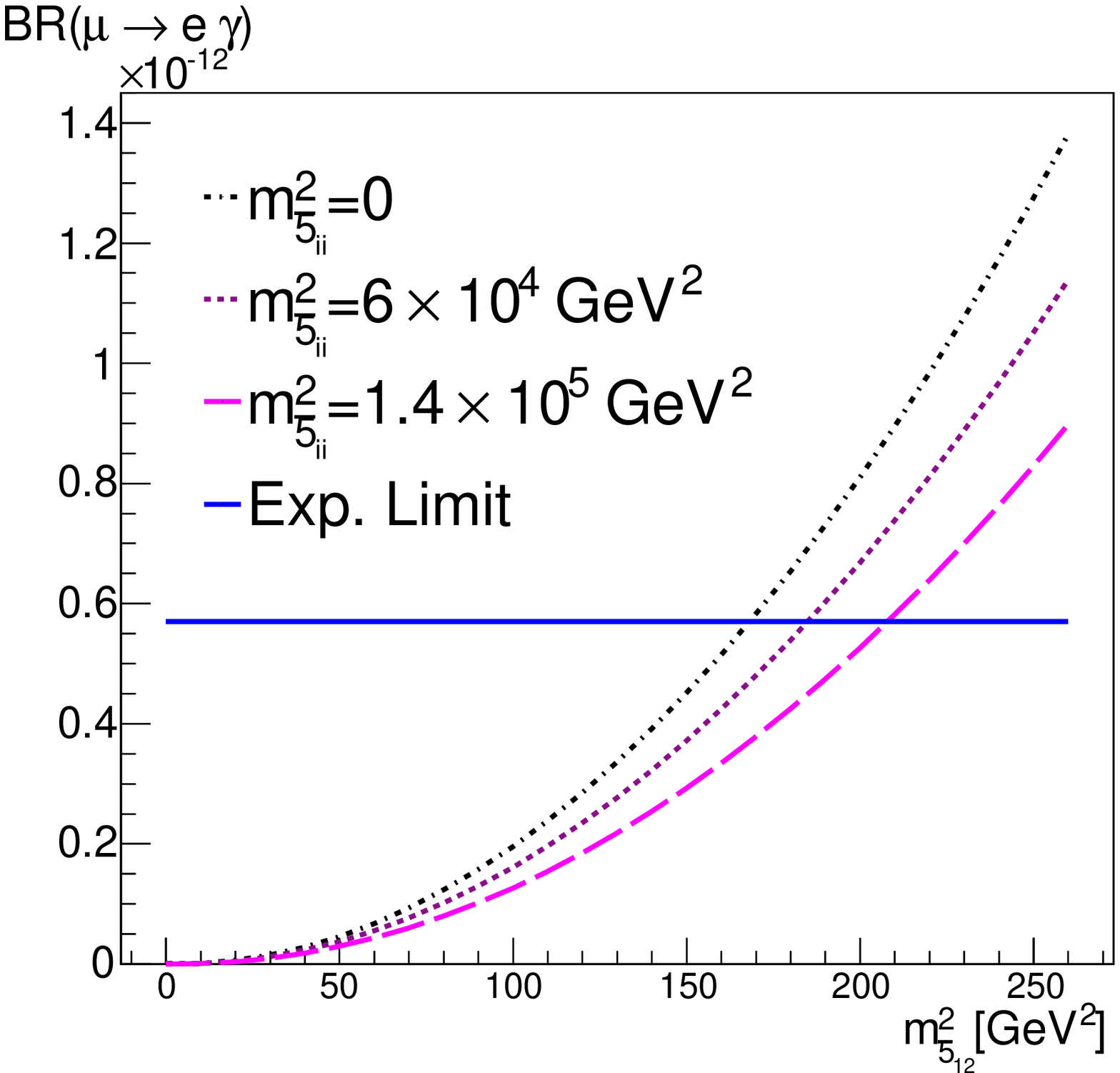}
\includegraphics[width=8.1cm, height=7cm]{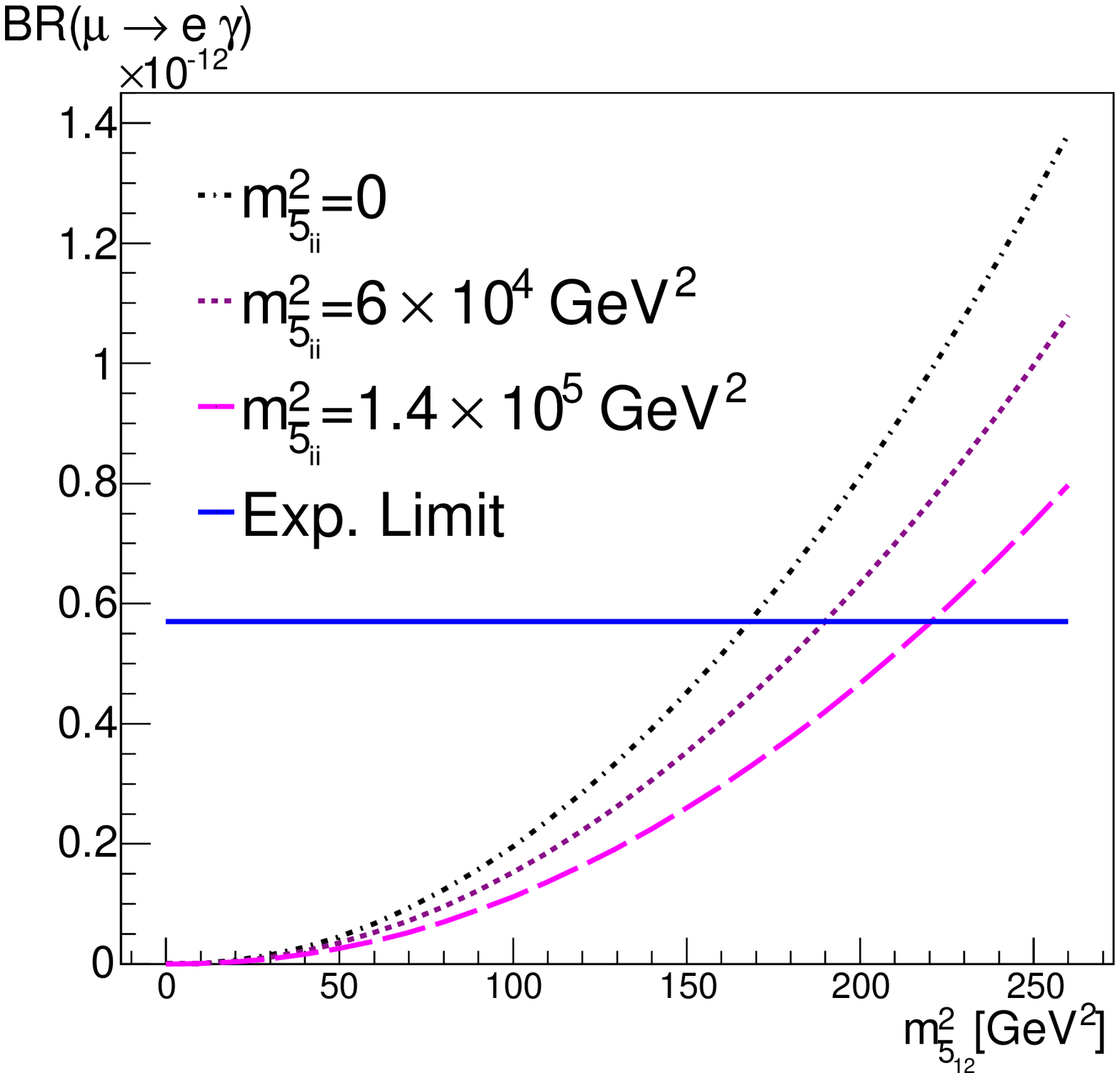}
\includegraphics[width=8.1cm, height=7cm]{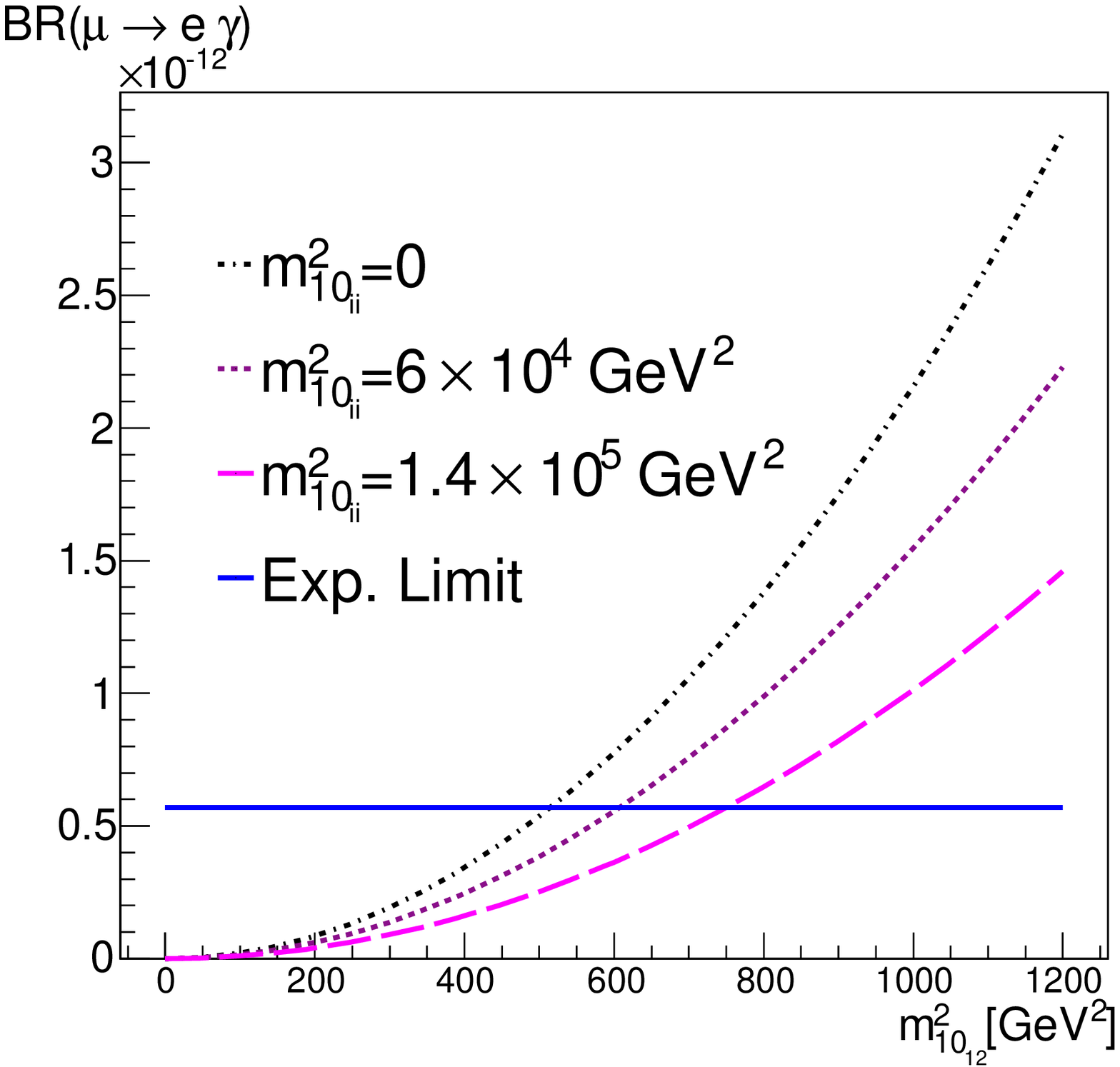}
\caption{\small{\it Top: $b = m^2_{\bar{5}_{12}}$ vs. $\Bmueg$ for $a = m^2_{\bar{5}_{ii}}=0$, $6\times 10^4$ GeV$^2$  and $1.4\times 10^5$ GeV$^2$.  In the first panel, $m^2_{10}=0$  ($c=d=0$), while in the second 
 $m^2_{10_{ii}}=m^2_{\bar{5}_{ii}}$  ($c=a$, $d=0$). 
Bottom:  $d = m^2_{10_{12}}$ vs. $\Bmueg$, where only off-diagonal elements of $m^2_{10}$ were set using $m^2_{10_{ii}}=m^2_{\bar{5}_{ii}}$  (that is $c=a$, $b=0$).
%LAST PLOT: CHOICE 4. 
}} \label{fig:m2512vsBRmuegamma} 
\end{figure}

We plot in \Figref{fig:m2512vsBRmuegamma}
the value of $\Bmueg$ as a function of $b = m^2_{\bar{5}_{12}}$, for  the three choices
$a = m^2_{\bar{5}_{ii}}=0$, $6\times 10^4$ GeV$^2$ and $1.4\times 10^5$ GeV$^2$.
In the first panel we take $m^2_{10_{ii}}=0$ ($c=d=0$), whereas in the second panel
$m^2_{10_{ii}}=m^2_{\bar{5}_{ii}}$ ($c=a$, $d=0$). The results are quite similar, though there are differences in the
supersymmetric spectra. In both cases we find that $\Bmueg$ requires $m^2_{\bar{5}_{12}} = b \lesssim 170$~GeV$^2$
for small $m^2_{\bar{5}_{ii}} = a \to 0$. This means that flavour violation among the scalar
masses-squared in the $(1, 2)$ sector could be maximal, i.e., $b/a \sim 1$, if the universal diagonal entry
$m^2_{\bar{5}_{ii}} = a \lesssim 170$~GeV$^2$. As one would expect, larger values of $m^2_{\bar{5}_{12}}$
would be allowed for larger values of $m^2_{\bar{5}_{ii}}$, but the ratio $b/a$ could not reach unity.
We consider the choice $m^2_{\bar{5}_{ii}}=1.4\times 10^5$ GeV$^2$ as an upper limit for $a$
since, for higher values, the spectrum is  sufficiently different from our original benchmark {\bf B} that 
it no longer satisfies the constraint on the relic density.  In this case, the upper limit 
on $b$ is 210 (220) GeV$^2$ for  $m^2_{10_{ii}}=0$ ($= a$), so 
we  must require $b \ll a$. In the third panel of \Figref{fig:m2512vsBRmuegamma}
we consider the effects of the off-diagonal components in $m^2_{10}$.
Here, we have set $m^2_{10_{ii}}=m^2_{\bar{5}_{ii}}$ with $m^2_{\bar{5}_{12}} =0$ ($c=a$, $b=0$), and have
plotted the branching ratio as a function of $d = m^2_{10_{12}}$. As one can see, the constraint on 
$d$ is much weaker than the analogous constraint on $b$.  

We conclude that the MaxSFV scenario is possible in the (1, 2) sector if $m^2_{\bar{5}_{12}} = b \lesssim 170$~GeV$^2$, and if $m^2_{10_{12}} = d \lesssim 520$~GeV$^2$ as summarized in Table \ref{tbl:MaxSFV}.

\subsubsection{The $(1,3)$ sector \label{sbsc:13BNS}}

We see from Eq.~\ref{eq:su5betafuncts} that, once the Yukawa couplings are complex, then the soft-squared masses and trilinear terms 
also become complex. We remind the reader that the only source of CP
violation stems from the CKM phase, which through \eq{eq:Az1} translates into the complex Yukawa couplings, \eq{eq:VdandVe}.
 Then, once an imaginary seed for $m^2_{\bar{5}_{13}}$ is set, a real non-zero entry at $\Mi$ for $m^2_{\bar{5}_{13}}$ will increase faster for both the imaginary and real parts of
 $m^2_{\bar{5}_{13}}$, than in the case of the pure no-scale set up.

\begin{figure}[!h]
\centering
\includegraphics[width=8.1cm, height=7cm]{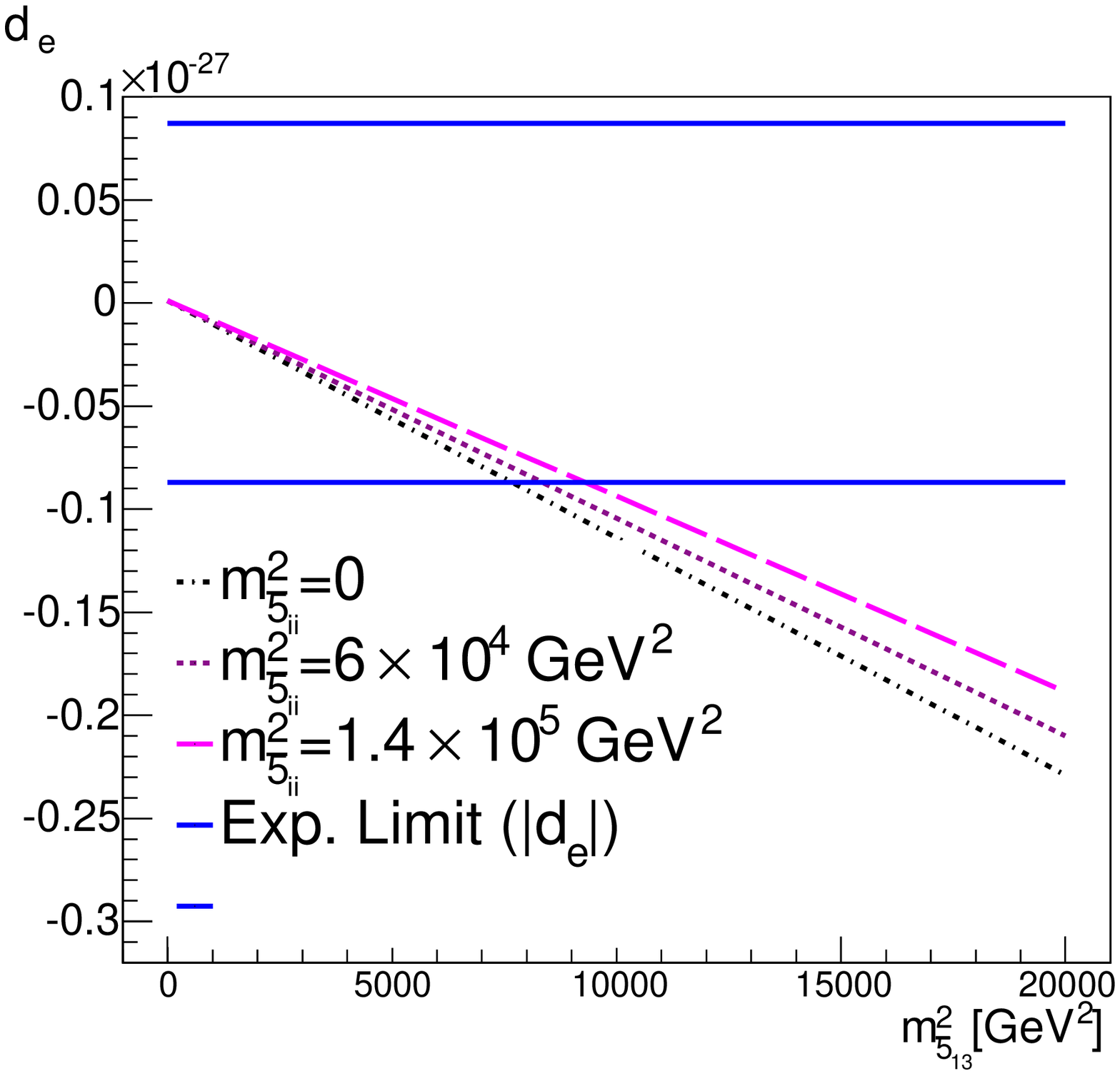}
\includegraphics[width=8.1cm, height=7cm]{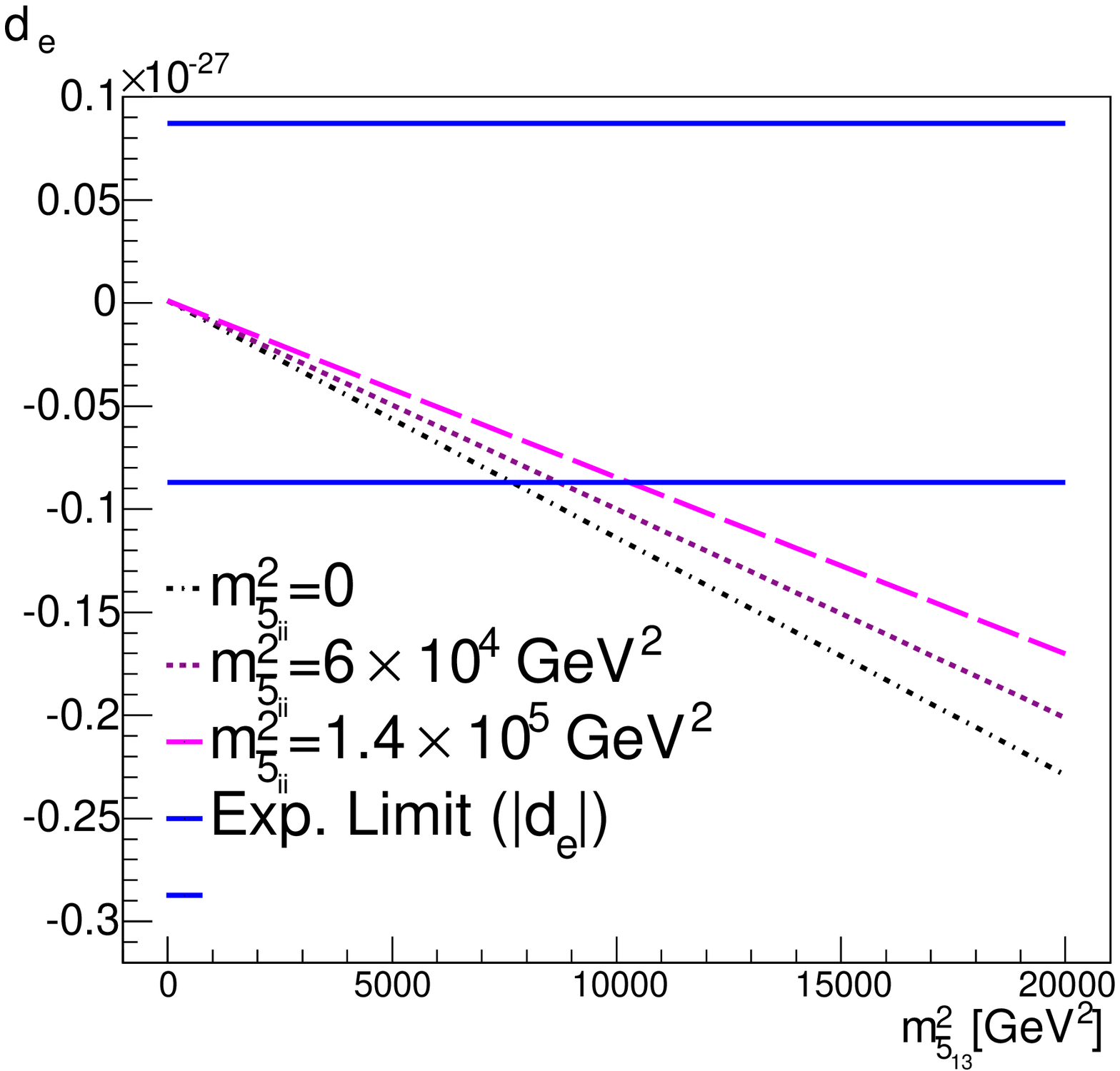}\\%NEW
\includegraphics[width=8.1cm, height=7cm]{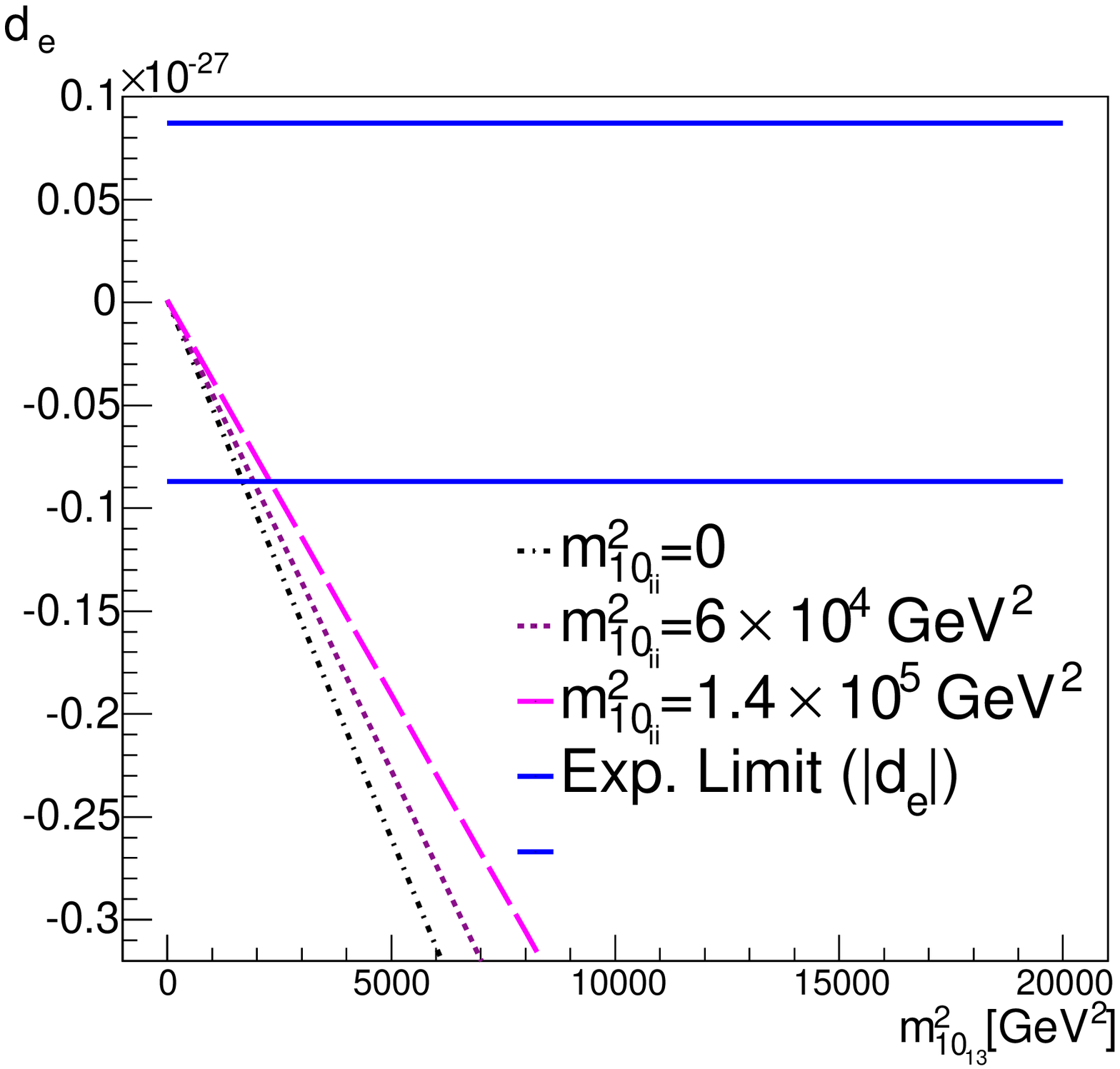}
\caption{\small{\it Top: we show $b = m^2_{\bar{5}_{13}}$ vs. $d_e$ for $a = m^2_{\bar{5}_{ii}}=0$, $6\times 10^5$ GeV$^2$  and $1.4\times 10^5$ GeV$^2$; in the first panel $m^2_{10}=0$ ($c=d=0$), while in the second panel
    $m^2_{10_{ii}}=m^2_{5_{ii}}$ ($c=a$, $d=0$). Bottom: here we show $d = m^2_{10_{13}}$ vs. $d_e$ for $m^2_{\bar{5}_{ii}}=m^2_{10_{ii}}$ ($c=a$, $b=0$).
} \label{fig:m2513vseEDM} }
\end{figure}

We see from Table \ref{Tbl:relobsII} that in A1 the value of the electron EDM, $d_e$, is already close to the experimental limit and from the diagram in
\Figref{fig:edm_flavor_cont} we see the potential importance for this observable when increasing the real part of
 $m^2_{\bar{5}_{13}}$. In A1, the leading contribution to $d_e$ comes from the terms in \eq{eq:leadingcont_d131} and once  $\rm{Re}\left[m^2_{\bar{5}_{13}}\right]$ is non-zero at $\Mi$, then at the GUT scale
it will be bigger than in the no-scale case and particularly  $\rm{Re}\left[(\delta^D_{31})_{RR}\right]$ will have a significant increase. 
As a consequence,  for $\rm{Re}\left[m^2_{\bar{5}_{13}}\right] \neq 0$ at the input scale, the most constraining observable is the electron EDM. \Figref{fig:m2513vseEDM} shows $d_e$ as a function of $b = m^2_{\bar{5}_{13}}$. The bounds on the electron EDM are shown by the two horizontal
solid lines straddling $d_e = 0$. As one can see, the flavour off-diagonal entry in the squark mass matrix is
bounded by $b < 10^4$ GeV$^2$ for our maximal value of $a$. 
Once again, allowing $c = a$ (with $d  = 0$) does not greatly affect this limit.
On the other hand, when we consider the EDM as a function of $d = m^2_{10_{13}}$ (with $b = 0$), 
we find that $d < 2000$ GeV$^2$ for the maximal value of $c = a$. 

\begin{figure}[ht!]
\centering
\includegraphics[width=8.1cm, height=7cm]{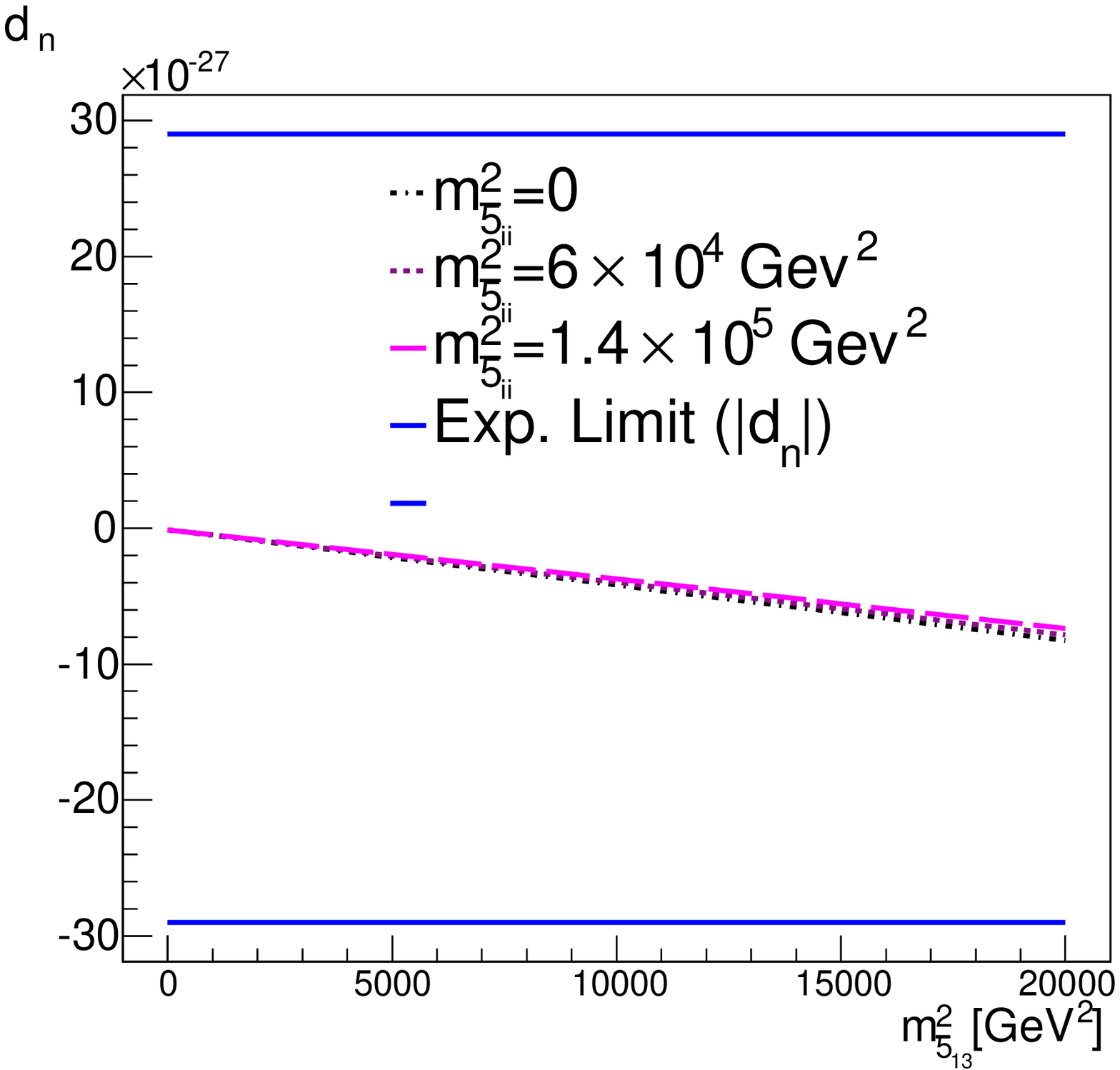}
\includegraphics[width=8.1cm, height=7cm]{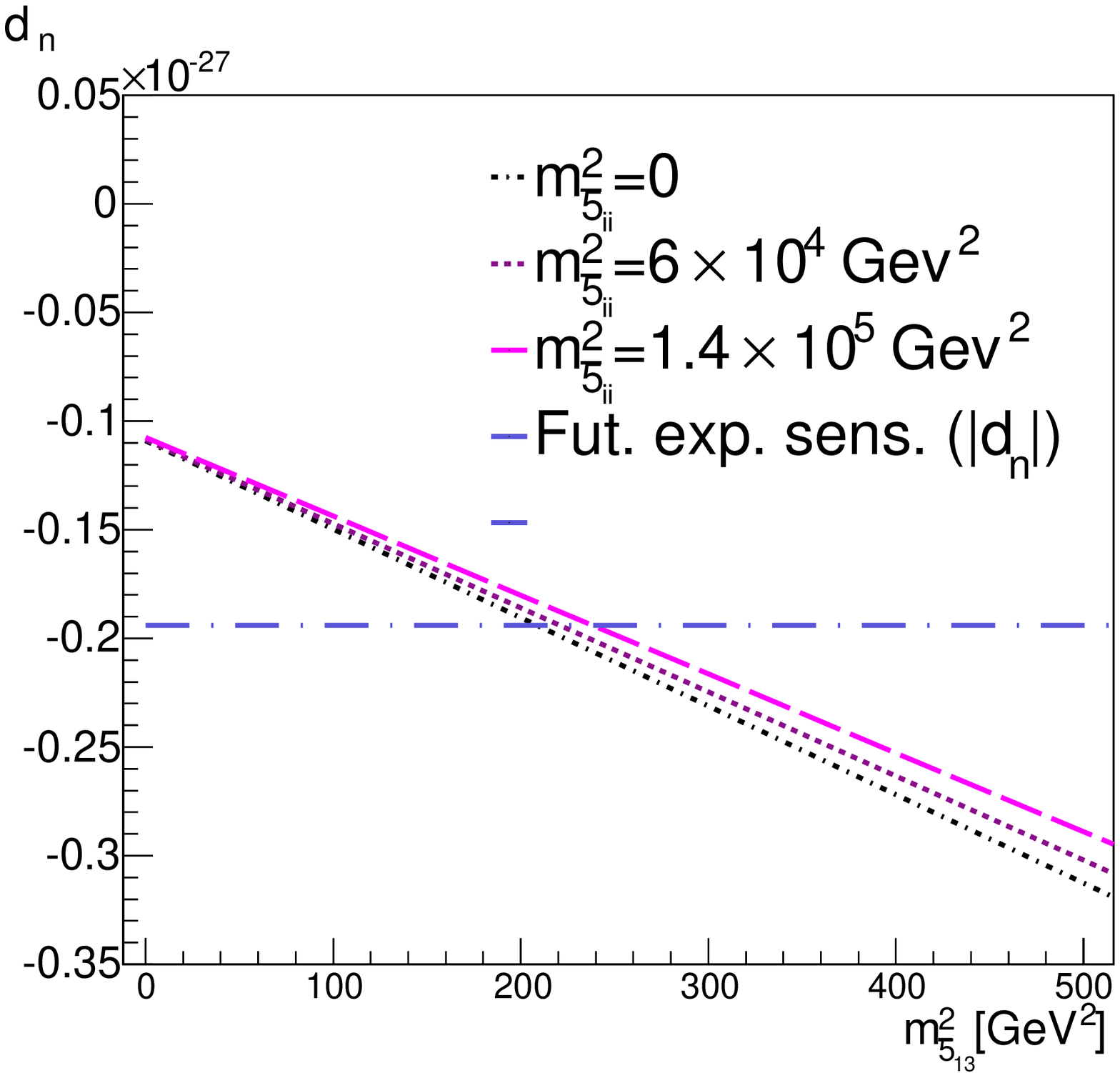}
\includegraphics[width=8.1cm, height=7cm]{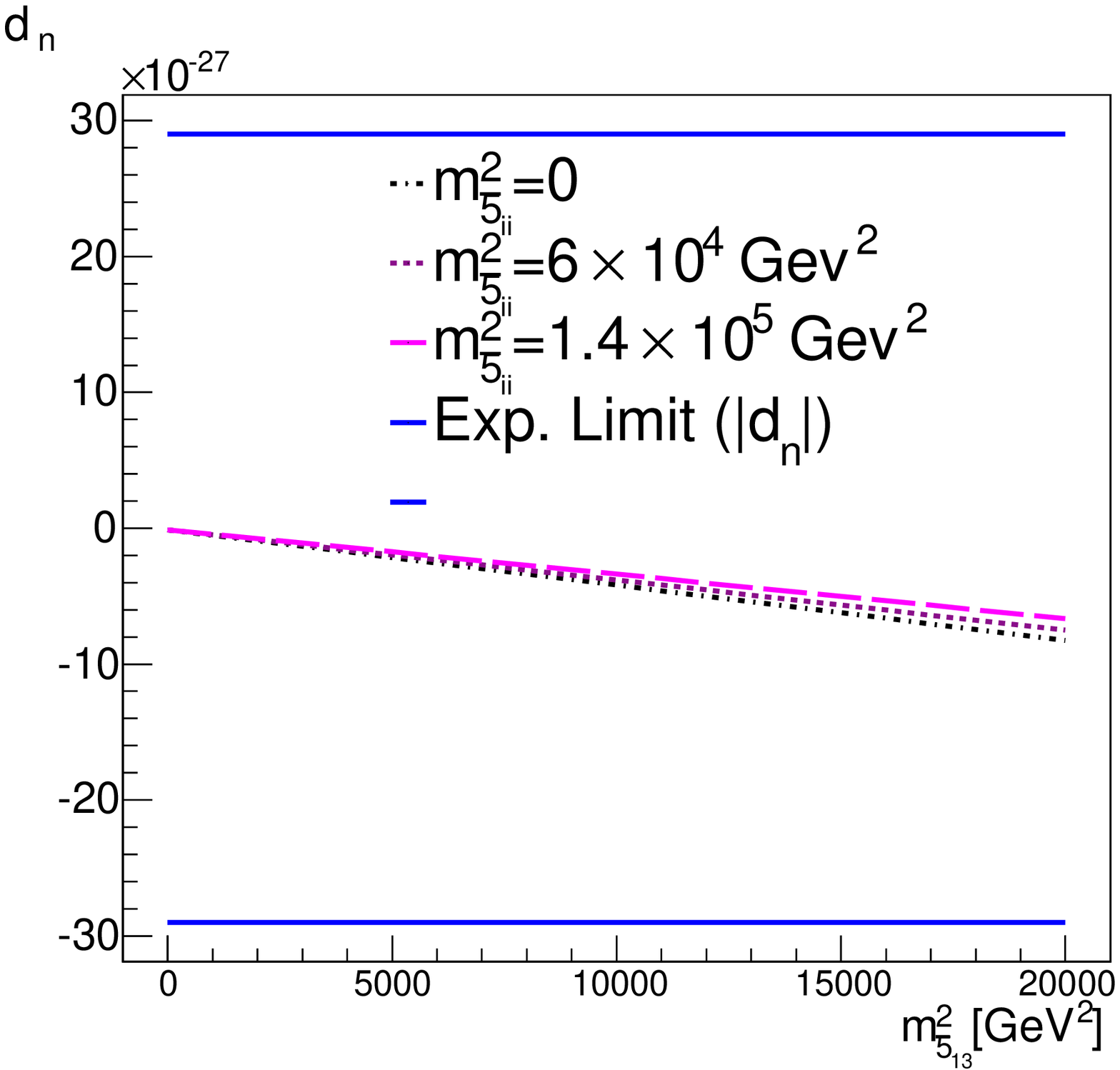}%NEW:sector 3
\includegraphics[width=8.1cm, height=7cm]{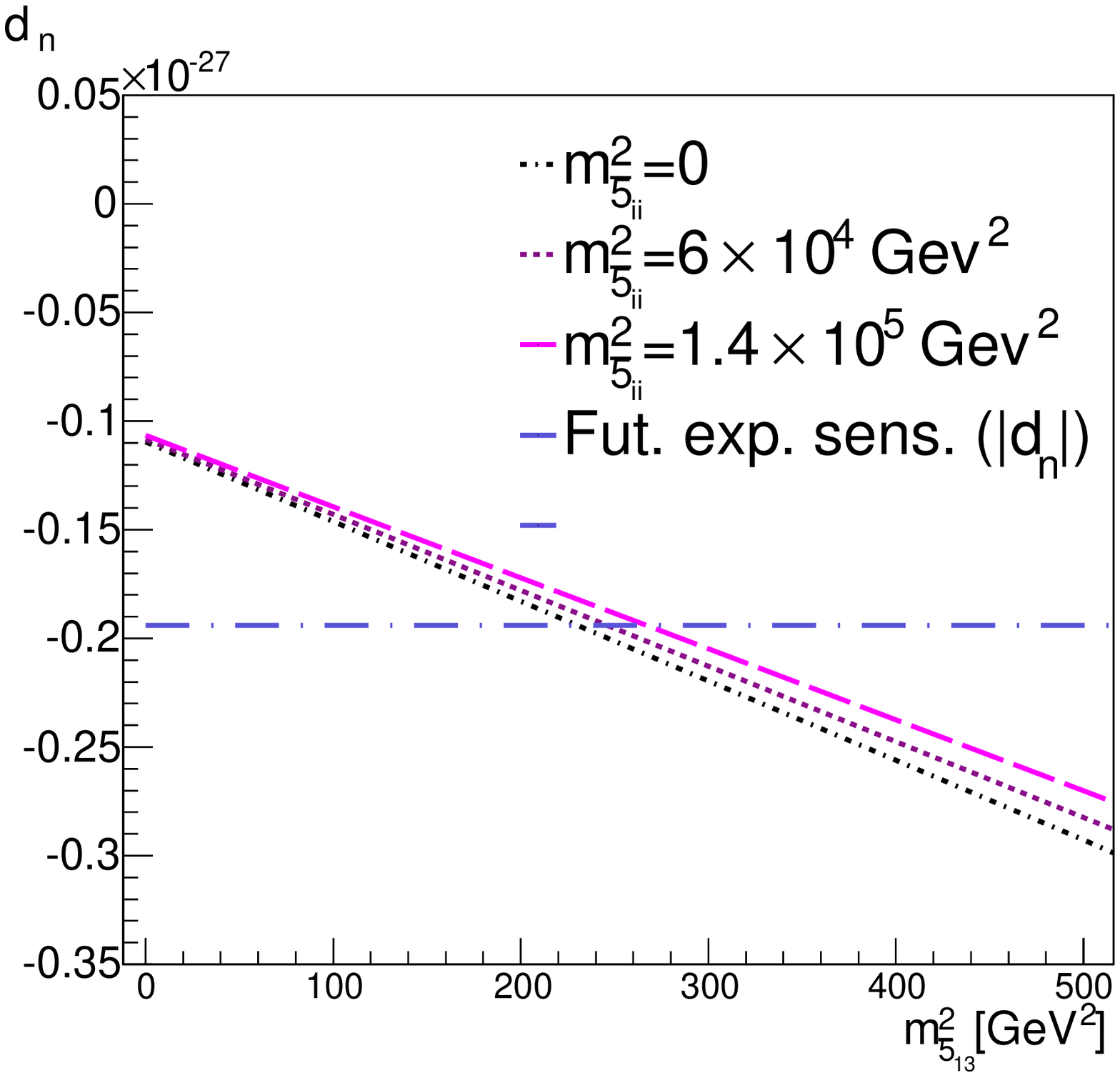}\\%:sector 3:
\includegraphics[width=8.1cm, height=7cm]{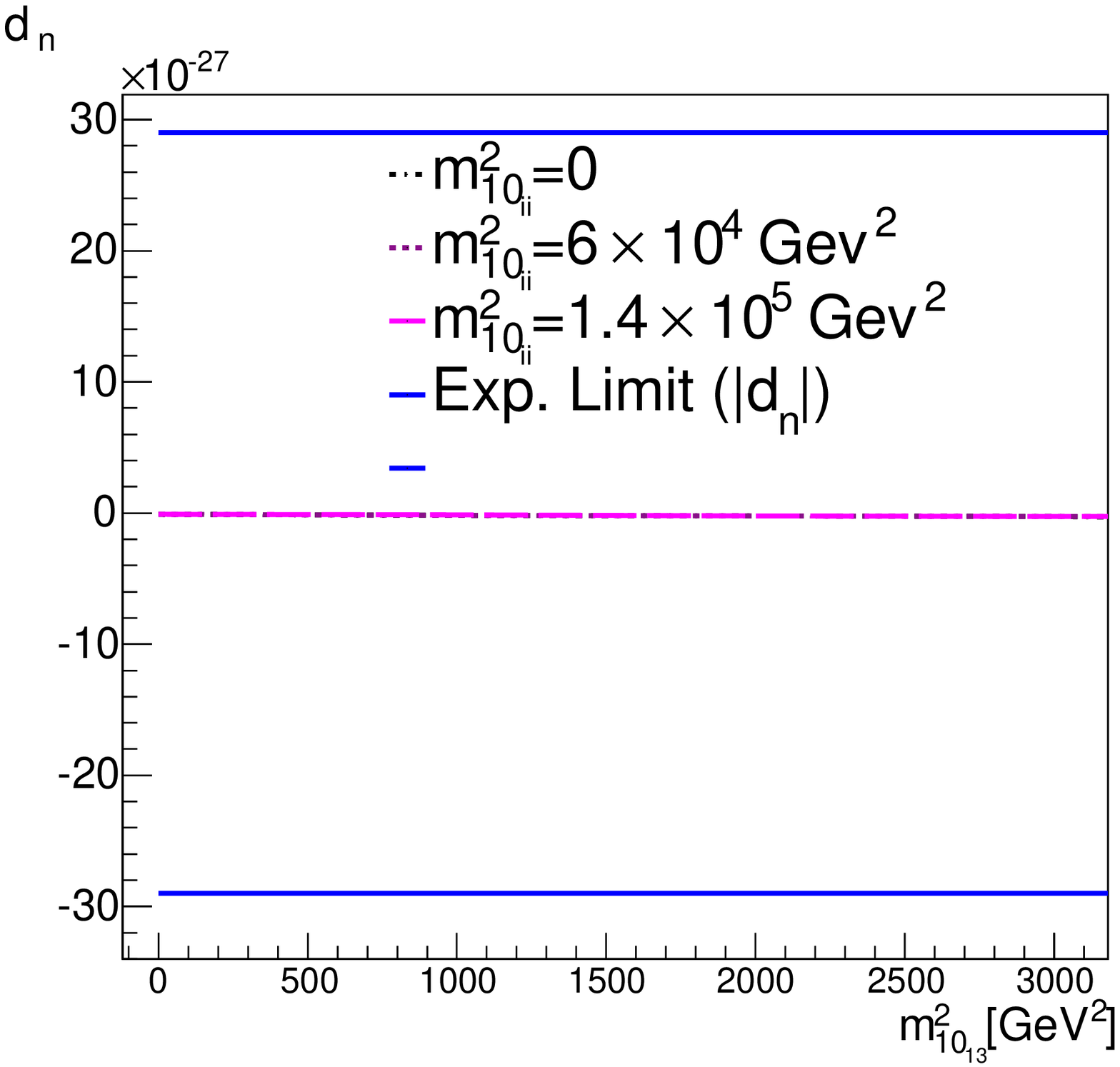}%sector 4%
\includegraphics[width=8.1cm, height=7cm]{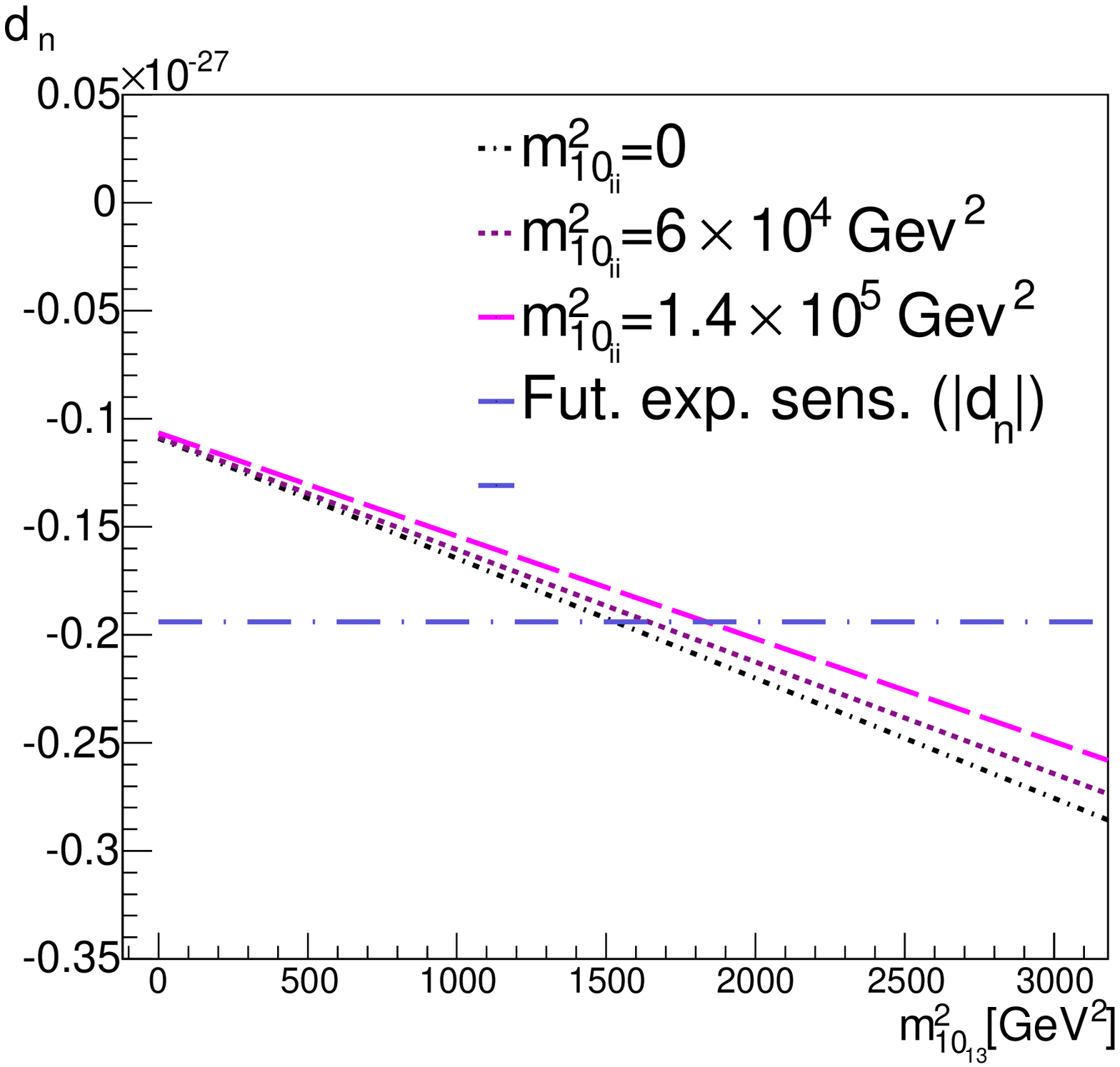} %sector 4
\caption{\small{\it Top: $b = m^2_{\bar{5}_{13}}$ vs. $d_n$ for $a = m^2_{\bar{5}_{ii}}=0$, $6\times 10^5$ GeV$^2$  and $1.4\times 10^5$ GeV$^2$. Here $m^2_{10}=0$ ($c=d=0$).   For definiteness, we have used the value of
    $1.94\times 10^{-28}$ \cite{Altarev:2009zz}, for the future sensitivity. Other values (all of $O(10^{-28})$) can be found in \cite{Hewett:2012ns}.
Middle: $m^2_{10_{ii}}=m^2_{5_{ii}}$ ($c=a$, $d=0$). Bottom: $d = m^2_{10_{13}}$ vs. $d_n$  for $m^2_{10_{ii}}=m^2_{\bar{5}_{ii}}$ ($c=a$, $b=0$).
} \label{fig:m2513vsNEDM} }
\end{figure}

 The neutron EDM is well below the present experimental limit, but at close to the future sensitivity of $O(10^{-28})$~e-cm~\cite{Hewett:2012ns}.
In Fig.~\ref{fig:m2513vsNEDM} we show the neutron EDM as a function of $m^2_{\bar{5}_{13}}$ for three values of $m^2_{\bar{5}_{ii}}=0$, $6\times 10^4$ GeV$^2$ and $1.4\times 10^5$ GeV$^2$.
The left panels shows the current (lack of) constraints, while the right panels show the anticipated 
future constraints. In the top row, 
$m^2_{10}=0$ ($c = d = 0$).  In the middle row, $m^2_{10_{ii}}=m^2_{5_{ii}}$ ($c=a$, $d=0$), and in the bottom row, we plot $d_n$ versus $d = m^2_{10_{13}}$ with $c = a$ and $b = 0$. As one can see,
there are no current constraints on either $b$ or $d$ for our allowed range in $a$ and $c$.
However, we expect that future constraints can place a limit of about $b \lesssim 230$ GeV$^2$ and
$d \lesssim 1540$ GeV$^2$.

\begin{figure}[!h]
\centering
\includegraphics[width=8.1cm, height=7cm]{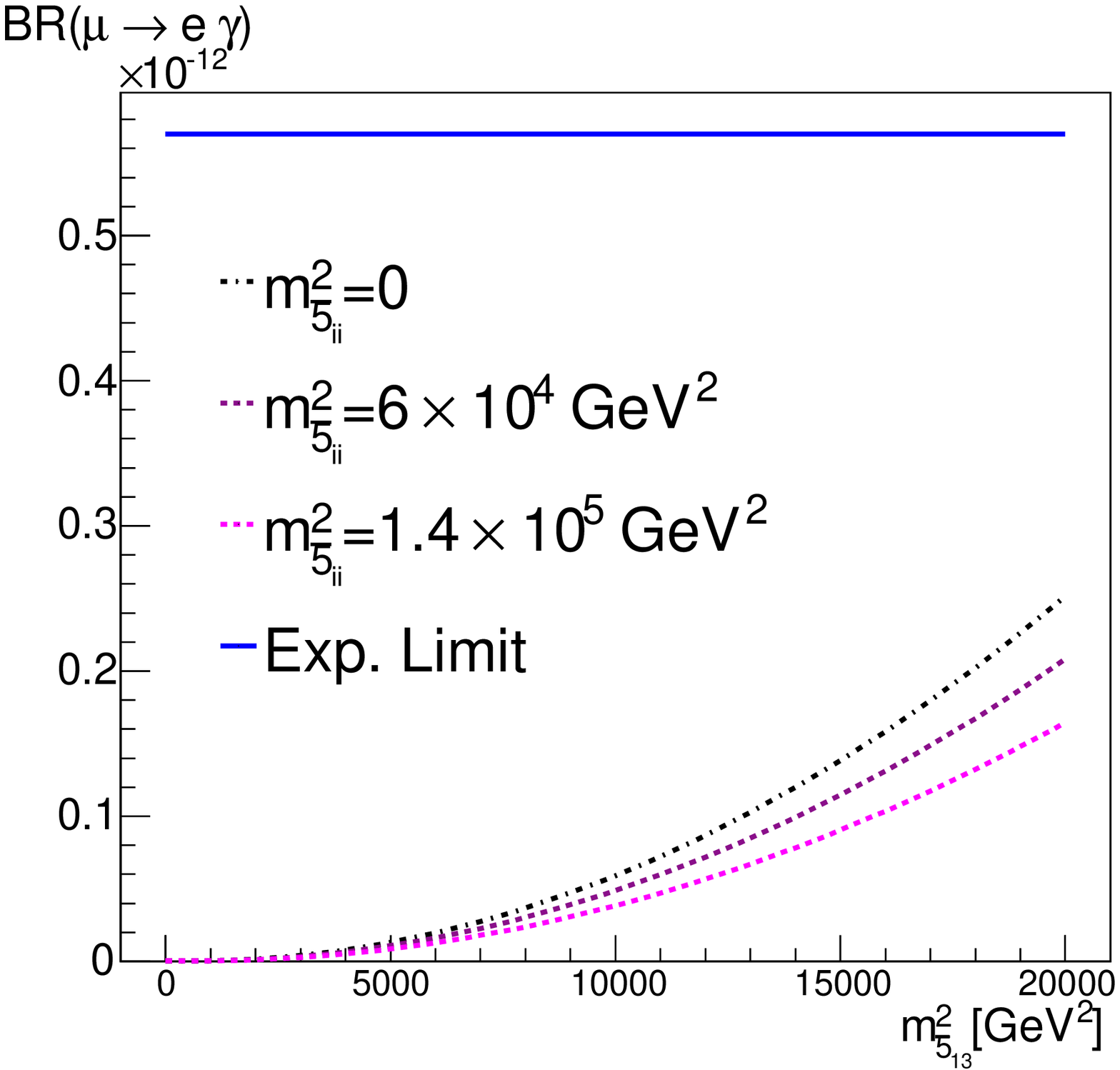}
\includegraphics[width=8.1cm, height=7cm]{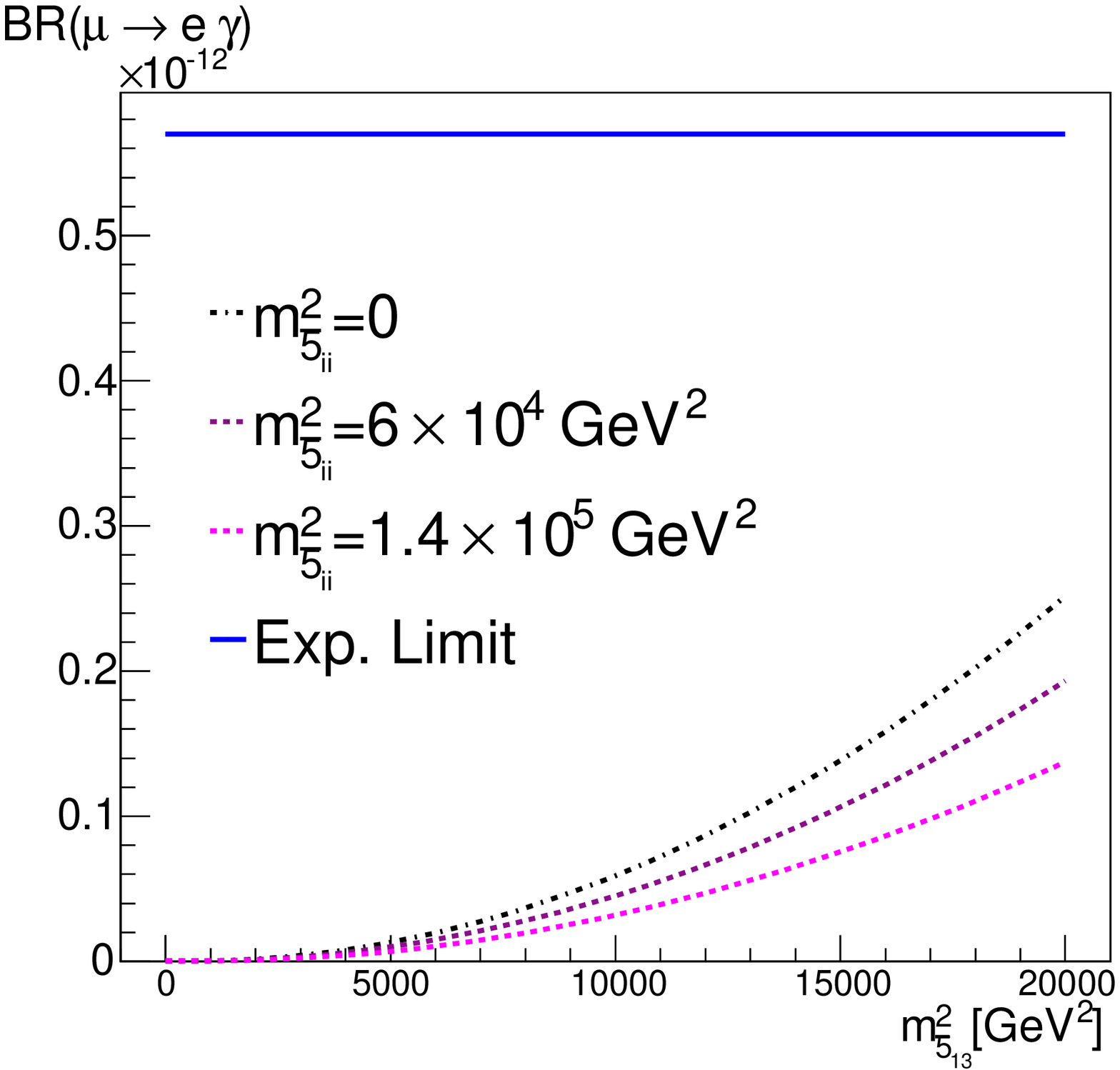}\\
\includegraphics[width=8.1cm, height=7cm]{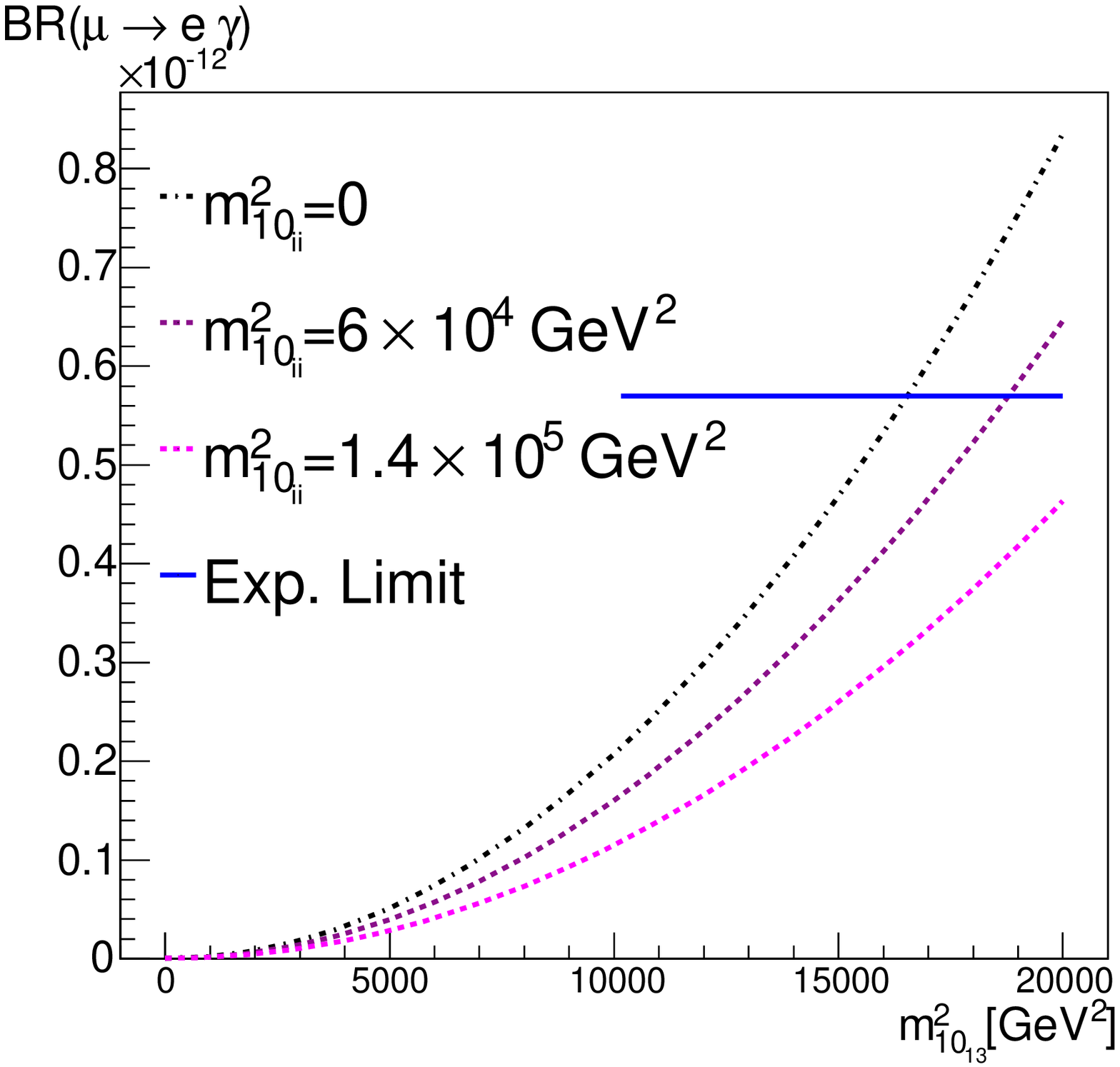}
\caption{\small{\it Top: $b = m^2_{\bar{5}_{13}}$ vs. $a = \Bmueg$ for $m^2_{\bar{5}_{ii}}=0$, $6\times 10^4$ GeV$^2$  and $1.4\times 10^5$ GeV$^2$. In the first panel $m^2_{10}=0$ ($c=d=0$), while in the second
 panel   $m^2_{10_{ii}}=m^2_{5_{ii}}$ ($c=a$, $d=0$). Bottom: $d = m^2_{10_{13}}$ vs. $\Bmueg$ for  $m^2_{10_{ii}}=m^2_{5_{ii}}$ ($c=a$, $b=0$). }. 
\label{fig:m2513vsBRmuegamma} }
\end{figure}

Other parameters that
are affected by switching on a non-zero off-diagonal parameter are $\epsilon_K$, $\Bmueg$ and $\Btaueg$,  
which remain (for the most part) within the experimental limits. Fig.~\ref{fig:m2513vsBRmuegamma} shows the
case for $\Bmueg$. This branching ratio is particularly sensitive to the increase in  $d = m^2_{10_{13}}$ if this is allowed to grow much faster than $m^2_{10_{12}}$ and $m^2_{\bar{5}_{12}}$, which is depicted in the
third panel in Fig.~\ref{fig:m2513vsBRmuegamma}.
When the real part of $m^2_{10_{13}}$  is non-zero at $\Mi$, while the real parts of $m^2_{10_{12}}$ and $m^2_{\bar{5}_{12}}$  are zero,   the values of $m^2_{E_{12}}$ and $m^2_{L_{12}}$ will evolve to values close to their no-scale values at $\Mw$, producing a
value for $\amuegL$ close to the no-scale case. 
When $m^2_{10_{13}}$ is allowed to increase to $O(10^4)$~GeV$^2$, the $\amuegR$ contribution
to $\Bmueg$, will dominate over $\amuegL$,  and the terms in $\amuegR$, \eq{eq:amuegLRLargeMu},  for $m=6$, which corresponds to the
lightest slepton, will drive $\Bmueg$ to levels above the experimental bound.

\begin{figure}[!h]
\centering
\includegraphics[width=8.1cm, height=7cm]{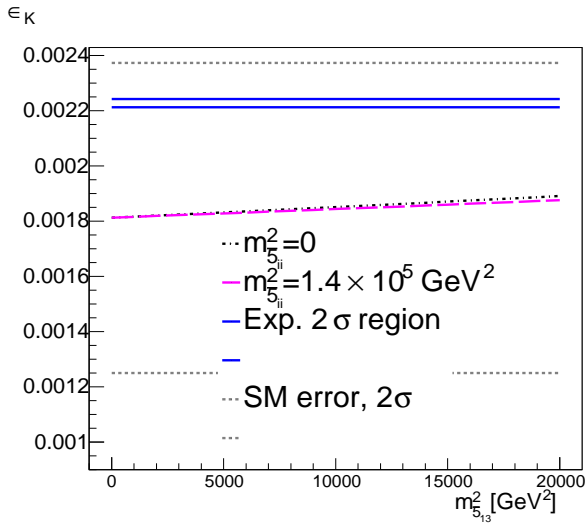}
\caption{\small{\it The 2 $\sigma$ experimental region of $|\epsilon_K|$ is $(2.213 - 2.243)\times 10^{-3}$, and the range of
the supersymmetric prediction together with the 2 $\sigma$ Standard Model error is   $(1.25 - 2.37)\times 10^{-3}$. 
} \label{fig:epskm2b513}}.
\end{figure}

In the case of $\epsilon_K$, the Standard Model uncertainty, 
which must be added to the supersymmetric value, must be taken into account.
In Fig.~\ref{fig:epskm2b513} we show $\epsilon_K$ as a function of $b = m^2_{\bar{5}_{13}}$
together with the area allowed by the NNLO SM error~\cite{Brod:2011ty} and the 2 $\sigma$ experimental region. This
clarifies that the value that we obtain for $\epsilon_K$ is in agreement with observations.

\subsubsection{The $(2,3)$ sector}

\begin{figure}[!h]
\centering
\includegraphics[width=8.1cm, height=7cm]{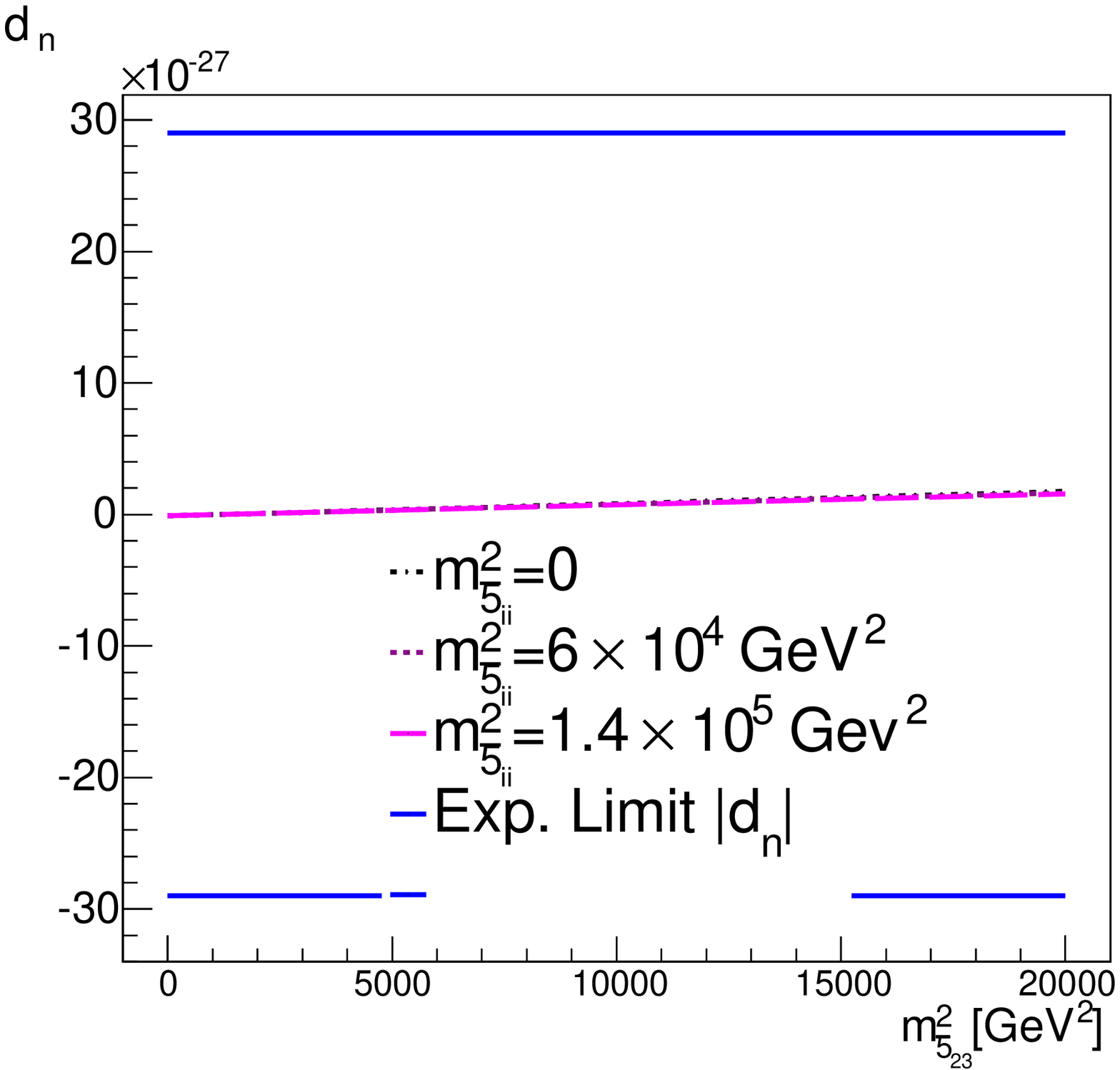}
\includegraphics[width=8.1cm, height=7cm]{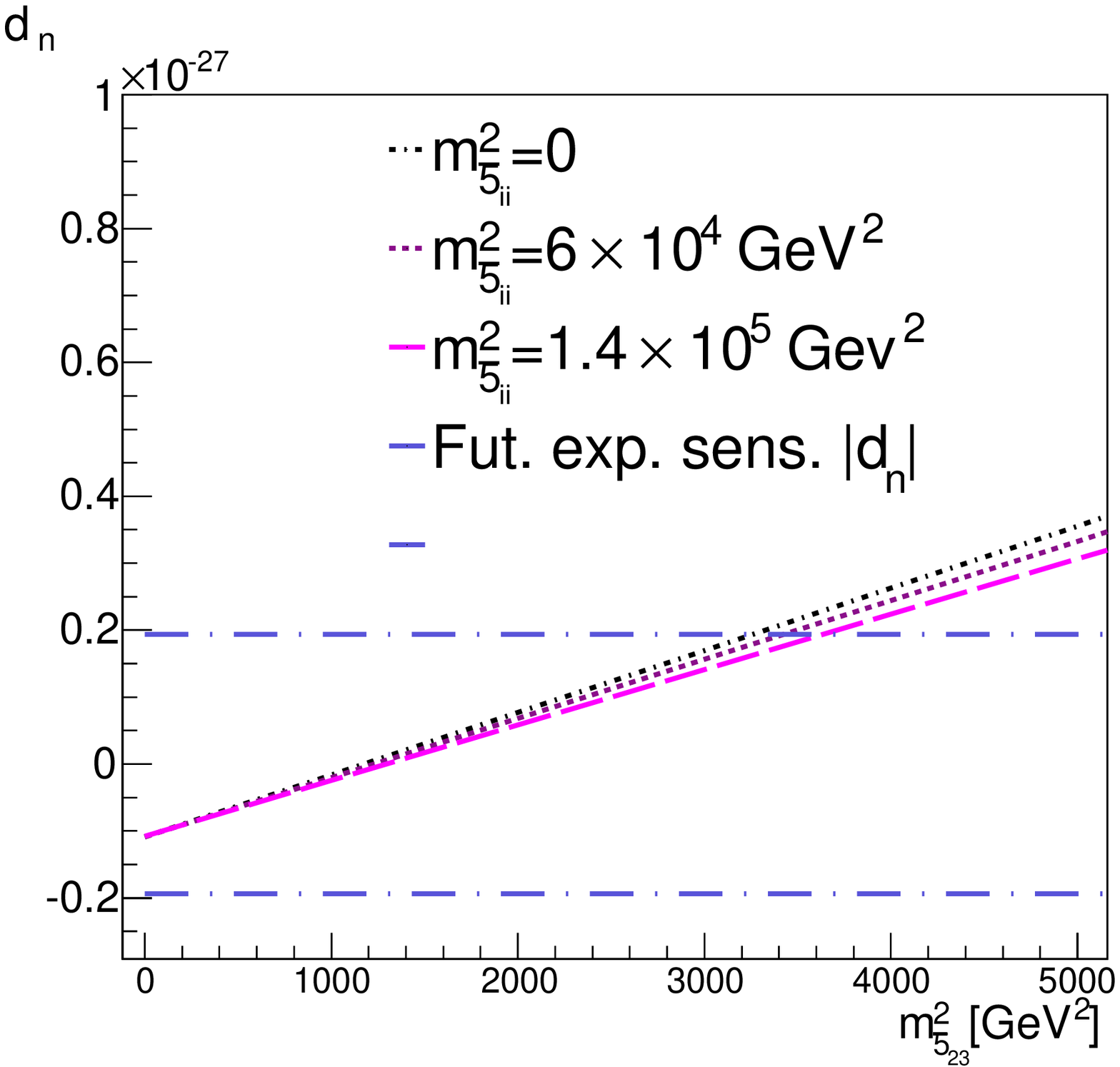}\\
\includegraphics[width=8.1cm, height=7cm]{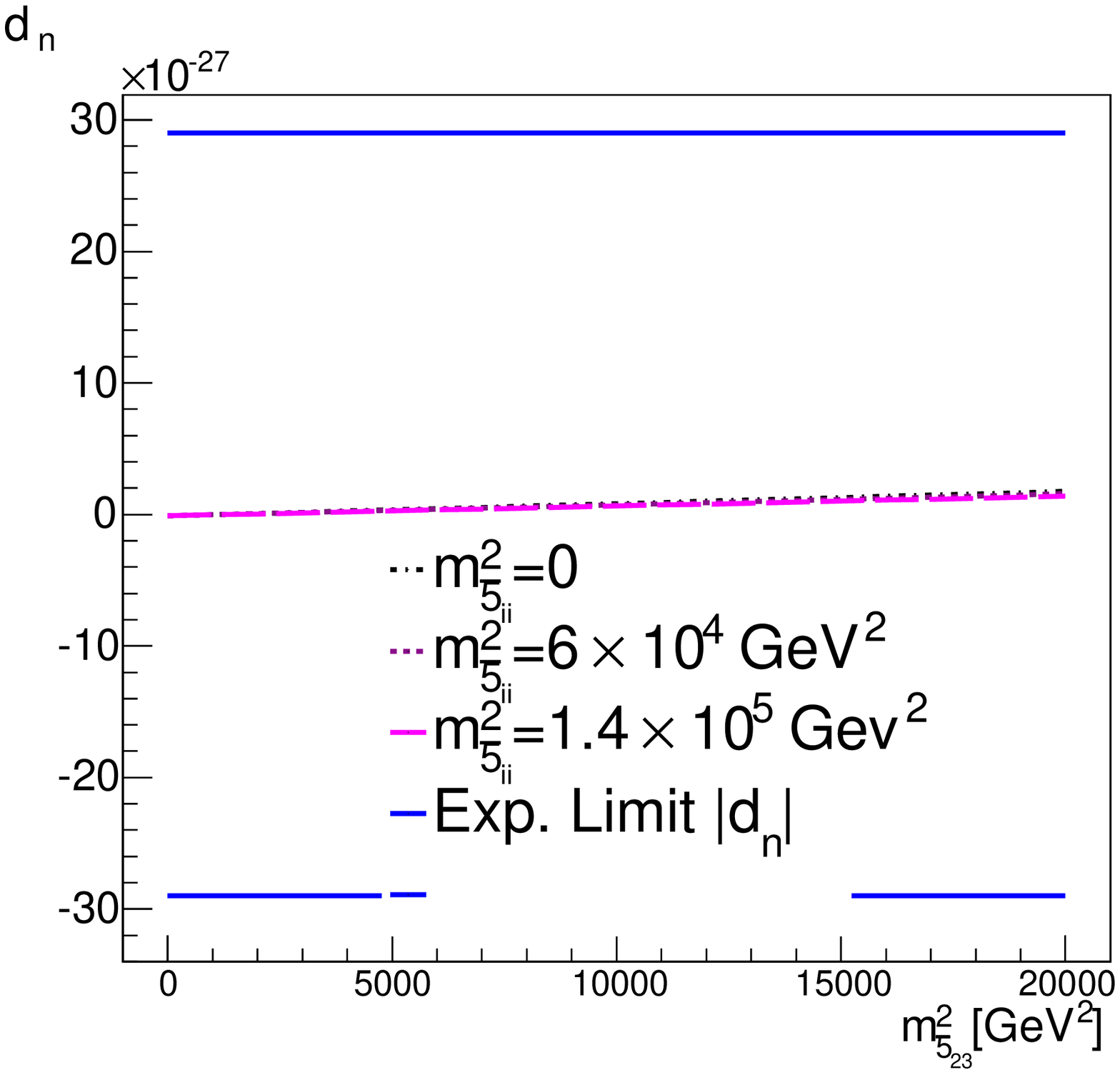}
\includegraphics[width=8.1cm, height=7cm]{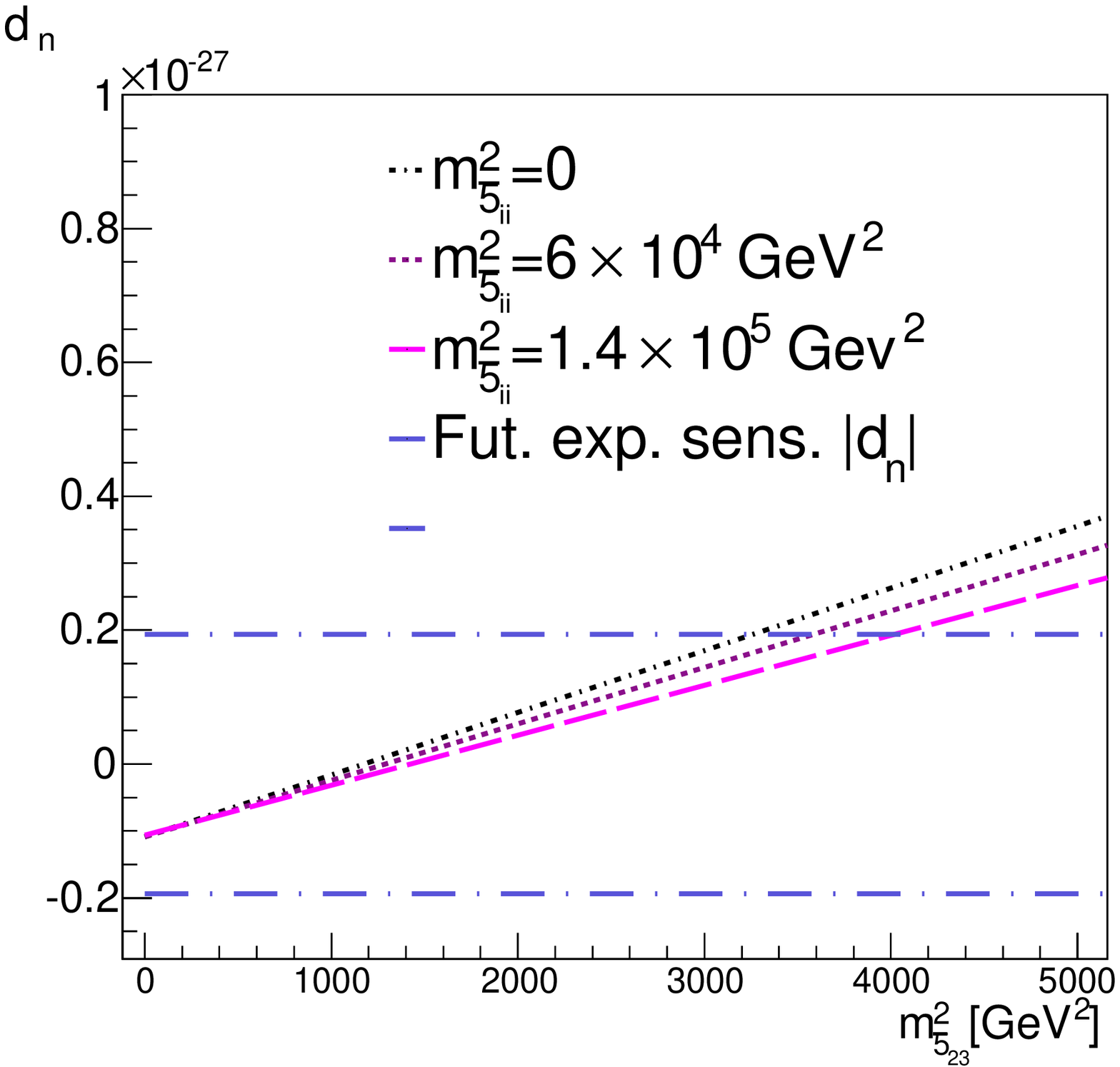}\\%Sector3
\includegraphics[width=8.1cm, height=7cm]{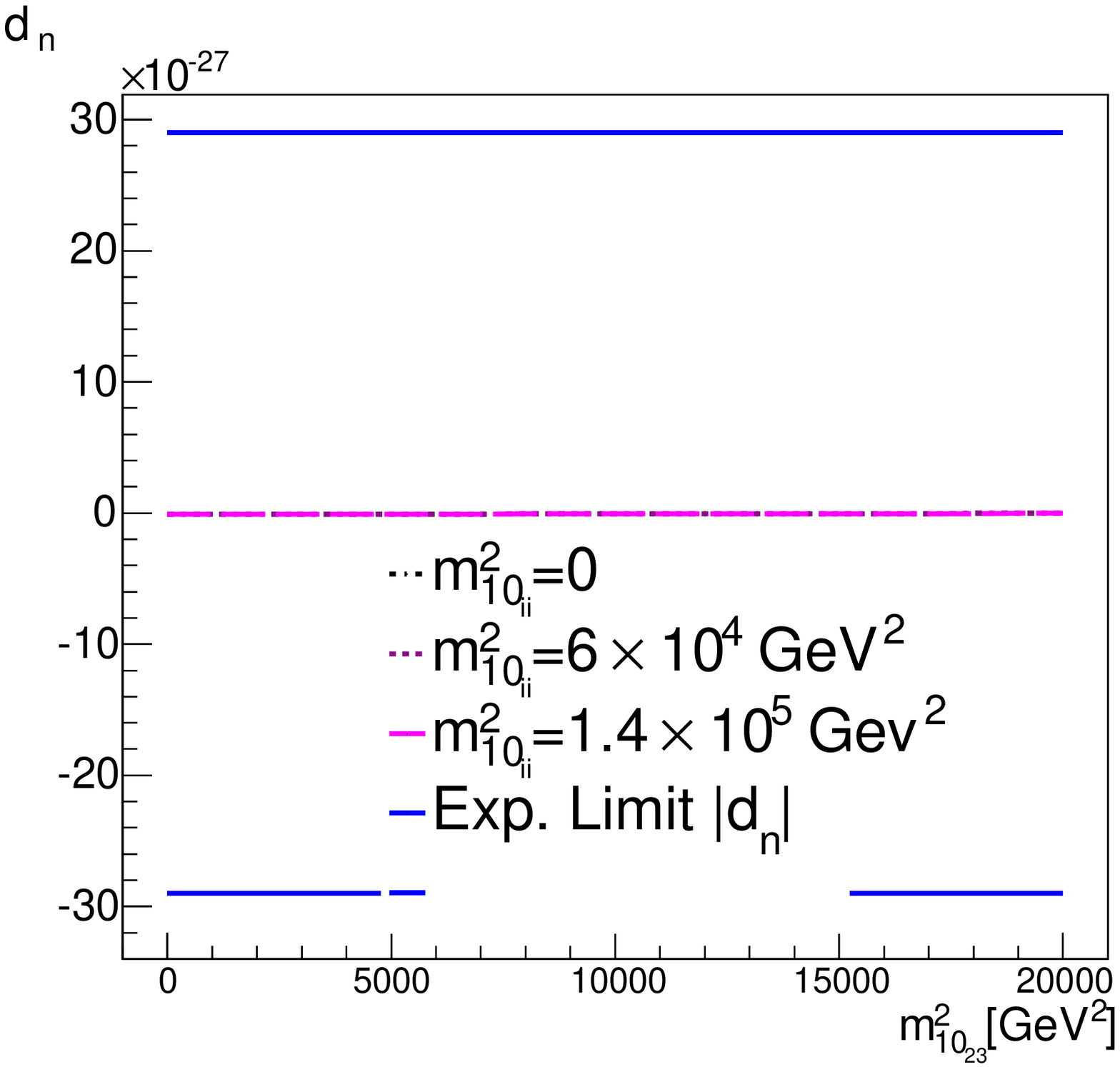}%Sector 4
\includegraphics[width=8.1cm, height=7cm]{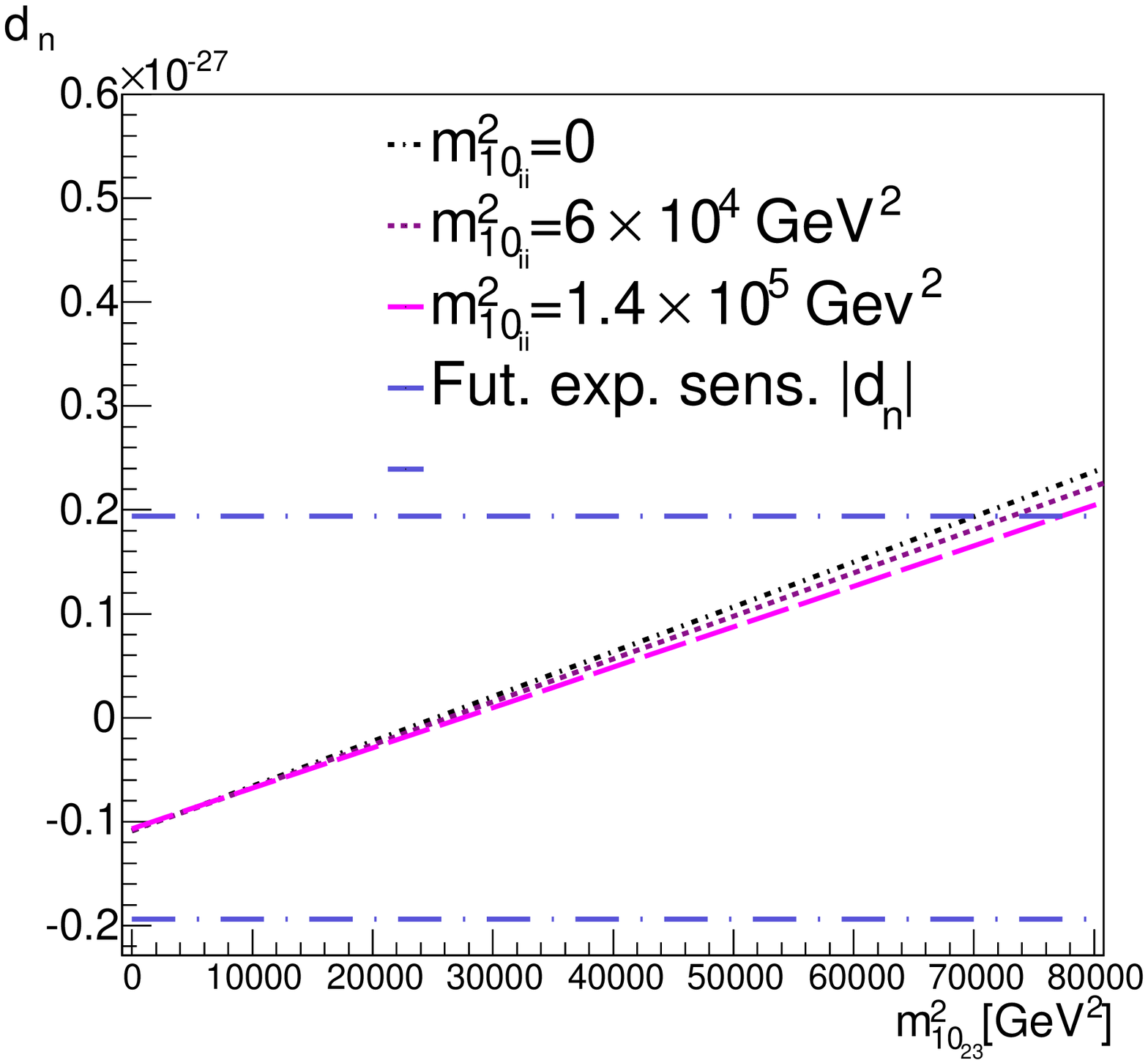}%sector 4
\caption{\small{\it Top: $m^2_{\bar{5}_{23}}$ vs. $d_n$ for $m^2_{\bar{5}_{ii}}=0$, $6\times 10^5$ GeV$^2$  and $1.4\times 10^5$ GeV$^2$. Here $m^2_{10_{ii}}=0$ ($c=d=0$).   For definiteness, we have used the value of
    $1.94\times 10^{-28}$ \cite{Altarev:2009zz}, for the future sensitivity. Other values (all of $O(10^{-28})$) can be found in \cite{Hewett:2012ns}.
Middle: the same as above, except for $m^2_{10_{ii}}=m^2_{5_{ii}}$ ($c=a$, $d=0$). Bottom: the case for $m^2_{10_{23}}$ vs. $d_n$  for $m^2_{10_{ii}}=m^2_{\bar{5}_{ii}}$ ($c=a$, $b=0$).
} \label{fig:m2523vsNEDM} }
\end{figure}

In this sector the most constraining parameter is the neutron EDM. Just as in the case of sector $(1,3)$, the neutron EDM is well below the present experimental limit and up to a value of 
$m^2_{\bar{5}_{23}}\approx 3.2\times 10^3$ GeV$^2$ (for $m^2_{\bar{5}_{ii}}=0$) also below the expected limit of the future 
sensitivity of  $O(10^{-28})$ ~e-cm.
 In  Fig.~\ref{fig:m2523vsNEDM} we show the neutron EDM as a function of $m^2_{\bar{5}_{23}}$ for three values of $m^2_{\bar{5}_{ii}}=0$, $6\times 10^4$ GeV$^2$ and $1.4\times 10^5$ GeV$^2$. 
 As in Fig.~\ref{fig:m2513vsNEDM}, we show current constraints from the neutron EDM in the 
 left panels and future limits on right. In the top row 
for the present experimental limit, while in the second for the future sensitivity.
 In the top row, 
$m^2_{10}=0$ ($c = d = 0$).  In the middle row, $m^2_{10_{ii}}=m^2_{5_{ii}}$ ($c=a$, $d=0$), and in the bottom row, we plot $d_n$ versus $d = m^2_{10_{23}}$ with $c = a$ and $b = 0$. As one can see,
there are no current constraints on either $b$ or $d$ for our allowed range in $a$ and $c$.
However, we expect that future constraints can place a limit of about $b \lesssim 3500$ GeV$^2$ and
$d \lesssim 8\times 10^4$ GeV$^2$.

Other parameters that
are affected by allowing for a non-zero off-diagonal parameter in the (23) sector are $\epsilon_K$, $\Bmueg$, $\Btaumug$ and $\Btaueg$,  which however remain within the experimental limits. Once again, for $\epsilon_K$ we also rely on taking into account the Standard Model uncertainty to obtain compatibility. We note also that
$\Bsg$ is sensitive to changes in this sector, though this is noticeable only for light ($\lesssim$ 1 TeV) supersymmetric 
spectra~\cite{Olive:2008vv}.

Table \ref{tbl:MaxSFV} summarize the constraints on the MaxSFV scenario
in the (12), (13) and (23) sectors. The numbers in the various entries of the Table are the maximum values in GeV$^2$
of $b$ (in the soft supersymmetry-breaking fiveplet mass-sequared matrix) where $b/a = 1$ is allowed
by present (and possible future) data, and the maximum values
of $b$ (in the soft supersymmetry-breaking tenplet mass-sequared matrix) where $d/c = 1$ is allowed
by present (and possible future) data. We see that the MaxSFV scenario in the fiveplet (12) sector is consistent with
the present data for $ m^2_{\bar{5}} \lesssim 170$~GeV$^2$, increasing to $\lesssim 520$~GeV$^2$ in the
tenplet (12) sector, $\lesssim 8000$~GeV$^2$ in the fiveplet (13) sector, and $\lesssim 1800$~GeV$^2$ in
the tenplet (13) sector. There are currently no constraints on MaxSFV in the (23) sector, though this may change with
future data.

\begin{table}[!h]
\begin{center}
\begin{tabular}{| l | l |  l  | l |}
%\hline
%\multicolumn{4}{c}{Comparison with of $|\rm{Re}[\deltaXY]|$ $f=L,E$ in Ans\"atze A1 and A4. }\\
\hline
\hline
Sector & (12) & (13) &(23) \\
 \hline
%$ m^2_{\bar{5}}$               & 200  & 10000  & - \\
$ m^2_{\bar{5}}$  - present             & 170  & 8000  & - \\
\hline
%$ m^2_{\bar{5}}$ - future              & -  & 200  & 3500 \\
$ m^2_{\bar{5}}$ - future              & -  & 230  & 3400 \\
\hline
%$ m^2_{10}$               & 600  & 2000  & - \\
$ m^2_{10}$  - present            & 520  & 1800  & - \\
\hline
%$ m^2_{10}$ - future             & -  & 1500  & 80000 \\
$ m^2_{10}$ - future             & -  & 1550  & 73000 \\
\hline
\end{tabular}
\end{center}
\caption{\it{Summary of constraints on $b = a$ and $d = c$ in the MaxSFV scenario.  All quantities are in GeV$^2$.
 \label{tbl:MaxSFV}}}
\end{table}

\section{Summary}

We have explored in this paper the phenomenological constraints on super-GUT models, in which
soft supersymmetry breaking inputs are postulated at some scale $\Mi$ intermediate between
$\Mg$ and the Planck scale, that are imposed by upper limits on flavor and CP violation.
For this purpose, we have chosen a benchmark supersymmetric model ({\bf B} in 
\eqref{eq:benchyuklambd} above) motivated by
no-scale supergravity that is consistent with other constraints from the LHC (e.g., $m_h$ and
the non-appearance of sparticles during Run~1) and cosmology (e.g., the density of cold dark matter
and inflation). Within this framework we have considered four possible scenarios for Yukawa couplings
that are compatible with CKM mixing and its extension to sparticles. Consideration of the runnings
of model parameters in two of these scenarios 
(A3 and A4 in Eqs. \ref{eq:Az3} and \ref{eq:Az4} above) were found to be generally
{\it incompatible} with the flavor-violation constraints, and not pursued further. However, the other two
scenarios (A1 and A2 in Eqs. \ref{eq:Az1} and \ref{eq:Az2} above) were found to be {\it compatible} with the flavour-violation
constraints. They have quite different predictions, and serve to illustrate the range of possibilities
for future flavor-violation measurements in no-scale super-GUTs.

We then considered possible deformations of the no-scale scenario, in which the soft supersymmetry-breaking
scalar masses $m_0$ are allowed to be non-zero but much smaller than the gaugino mass $m_{1/2}$ at $\Mi$.
In particular, we have investigated the maximal magnitudes of off-diagonal terms in the sfermion mass-squared
matrices $m_0^2$ for the SU(5) fiveplets and tenplets, and the possibility that these might be as large as the 
diagonal entries, a scenario we call MaxSFV. We find that the off-diagonal (12) entry in the fiveplet
mass-squared matrix could be as large as the diagonal entries if the latter are $\simeq 170$~GeV$^2$
and $\simeq 520$~GeV$^2$ in the $m^2_{\bar{5}}$ and $m^2_{10}$ mass-squared matrices, respectively.
The corresponding numbers in the (13) sector are $\simeq 8\times 10^3$~GeV$^2$
in the fiveplet mass-squared matrix and $\simeq 1.8\times 10^3$~GeV$^2$ for the tenplets. 
There are currently no useful bounds in the (23) sector.
The future sensitivity in $d_n$ would be sensitive to MaxSFV in the (13) sector of ${\cal O}(230)$~GeV$^2$ for the for the fiveplets
and of ${\cal O}(1.5\times 10^3)$~GeV$^2$ for the tenplets, and of ${\cal O}(3.4\times 10^3)$~GeV$^2$ in the
(23) sector for the fiveplets and ${\cal O}(7.3\times 10^4)$~GeV$^2$ for the tenplets.

Within these limits, {\it there would be no supersymmetric flavour problem} associated with sfermion masses
in the class of near-no-scale super-GUTs discussed in this paper.

\appendix 
\section{Three-family beta functions in SU(5)}

Many of the RGEs for the single-family case can be found in \cite{Polonsky:1994rz,Baer:2000gf,Ellis:2010jb,emo2}. 
Here we list the complete set of RGEs relevant for the 3-family case.
Whilst the beta function for $\lambda'$ in EQNO does not change with respect to the one-family case, that is
\small{
\bea
 \frac{\text{d}\lambda'}{\text{dt}} = 
\frac{\lambda'}{16 \pi ^2} \left(\frac{63 \lambda'^2}{20}+3 \lambda^2 \ - 30 \text{g}_{5}^2\right) \, ,
\nonumber
\eea
}
the other beta functions change as follows:
\small{
\bea
\label{eq:betafunc_su5_3g}
\frac{\text{d}\lambda}{\text{dt}}&=&
\frac{\lambda}{16 \pi ^2} \left(48 \text{Tr}[\text{h}_{10}^ {\dagger}\text{h}_{10}]+2 \
\text{Tr}[\text{h}_{\bar{5}}^ {\dagger}\text{h}_{\bar{5}}]+\frac{53 \lambda^2}{5}+\frac{21 \lambda'^2}{20}-\frac{98 \
\text{g}_{5}^2}{5}\right) \, ,
\nonumber\\
\frac{\text{dh}_{10}}{\text{dt}}&=&
\frac{1}{16 \pi ^2} \left[
\ \text{h}_{10}
\left(48 \text{Tr}[\text{h}_{10}^ {\dagger}\text{h}_{10}]+\frac{24 \
\lambda^2}{5}-\frac{96 \text{g}_{5}^2}{5}\right)+96 \
\text{h}_{10}\text{h}_{10}^ {\dagger}\text{h}_{10}
\right. \nonumber\\
&+&
\left. \text{h}_{10}\text{h}_{\bar{5}}^ {\dagger}\text{h}_{\bar{5}}+\text{h}_{\bar{5}}
\text{h}_{\bar{5}}^ {\dagger}\text{h}_{10}\right]  \, ,
\nonumber\\
 \frac{\text{dh}_{\bar{5}}}{\text{dt}}&=&
\frac{1 }{16 \pi ^2} 
\left[
\text{h}_{\bar{5}}
\left(2 \text{Tr}[\text{h}_{\bar{5}}^ {\dagger}\text{h}_{\bar{5}}]+\frac{24 \lambda^2}{5}-\frac{84 \
\text{g}_{5}^2}{5}\right)+48 \text{h}_{\bar{5}}\text{h}_{10}^ {\dagger}\text{h}_{10}
\right.  \nonumber\\
&+&
\left. 3 \
\text{h}_{\bar{5}} \text{h}_{\bar{5}}^ {\dagger}\text{h}_{\bar{5}}\right]  \, ; \nonumber
\eea
}
\small{
\bea
\frac{\text{dA}_\lambda}{\text{dt}} &=&
\frac{1}{8 \pi ^2}
 \left[48 \text{Tr}[\text{h}_{10}^ {\dagger}\text{a}_{10}]+2 \
\text{Tr}[\text{h}_{\bar{5}}^ {\dagger}\text{a}_{\bar 5}]+\frac{53 \lambda^2 \text{A}_{\lambda}}{5}+\frac{21 \
\text{A}_{\lambda'} \lambda'^2}{20} - \frac{98 \text{g}_{5}^2 \text{M}_{5}}{5}\right]  \, ,
\nonumber\\
\frac{\text{dA}_{\lambda'}}{\text{dt}} 
 &=&
\frac{1}{8 \pi ^2}
 \left[3 \lambda^2 \text{A}_{\lambda}+\frac{63 \text{A}_{\lambda'} \lambda'^2}{20}-30 \text{g}_{5}^2 \
\text{M}_{5}\right]  \, ,
\nonumber\\
 \frac{\text{da}_{10}}{\text{dt}} &=&
\frac{1}{16 \pi ^2}\left[
\frac{24}{5} \text{a}_{10} \left(10 \text{Tr}[\text{h}_{10}^\dagger\text{h}_{10}]+\lambda^2-8
   \text{g}_{5}^2\right) 
   \right. \nonumber\\   
  &+&
  \frac{48}{5} \text{h}_{10} \left(10
   \text{Tr}[\text{h}_{10}^\dagger \text{a}_{10}]+ |\lambda|^2 \text{A}_{\lambda} -4 \text{g}_{5}^2  \text{M}_{5}\right)\nonumber\\
 &+&
 144 \ \text{a}_{10} \text{h}_{10}^\dagger \text{h}_{10}
 + 144 \
   \text{h}_{10} \text{h}_{10}^\dagger \text{a}_{10}  \nonumber\\
&+&
\left.
 \text{a}_{10} \text{h}_{\bar{5}}^\dagger\text{h}_{\bar{5}}+\text{h}_{\bar{5}}  \text{h}_{\bar{5}}^\dagger \text{a}_{10}
+2   (\text{a}_{\bar 5} \text{h}_{\bar{5}}^\dagger \text{h}_{10}+\text{h}_{10} \text{h}_{\bar{5}}^\dagger \text{a}_{\bar 5})
\right]  \, ,
\\ 
\frac{\text{da}_{\bar 5}}{\text{dt}} &=&
\frac{1}{16 \pi ^2}
\left[
\text{h}_{\bar{5}} \left(4 \text{Tr}[\text{h}_{\bar{5}}^ {\dagger}\text{a}_{\bar 5}]+\frac{48 \lambda^2 \
\text{A}_{\lambda}}{5}- \frac{168 \text{g}_{5}^2 \text{M}_5 }{5}\right) \right. \nonumber \\
&+&
\text{a}_{\bar 5} \left(2 \
\text{Tr}[\text{h}_{\bar{5}}^ {\dagger}\text{h}_{\bar{5}}]+\frac{24 |\lambda|^2}{5}
-\frac{84 \
\text{g}_{5}^2}{5}\right)+96 \text{h}_{\bar{5}}\text{h}_{10}^ {\dagger}\text{a}_{10} \nonumber\\
&+&
\left. 48 \
\text{a}_{\bar 5}\text{h}_{10}^ {\dagger}\text{h}_{10}+6 \text{a}_{\bar 5}\text{h}_{\bar{5}}^ {\dagger}\text{h}_{\bar{5}}\ +
3 \text{h}_{\bar{5}}\text{h}_{\bar{5}}^ {\dagger}\text{a}_{\bar 5}\right]  \, ; \nonumber
\eea
}
%\newpage
\small{
\bea
\frac{\text{dm}_{24}^{2}}{\text{dt}}
 &=&
\frac{1}{8 \pi ^2}
 \left[\lambda^2 \
\left(\text{A}_{\lambda}^2+\text{m}_{24}^{2}+\text{m}_{5_H}^{2}+\text{m}_{\bar{5}_H}^{2}\right)+\frac{21}{20} \
\lambda'^2 \left(\text{A}_{\lambda'}^2+3 \text{m}_{24}^{2}\right)-20 \text{g}_{5}^2 \text{M}_{5}^2 {\mathbf{1}} \right]  \, ,
\nonumber\\
\frac{\text{dm}_{5_H}^{2}}{\text{dt}}
 &=&
\frac{1}{8 \pi ^2}
 \left[3 (16 \text{Tr}[\text{a}_{10}^{\dagger}\text{a}_{10}]+32 \
\text{Tr}[\text{h}_{10}\text{m}_{10}^{2}\text{h}_{10}^ {\dagger}]+16 \text{m}_{5_H}^{2} \
\text{Tr}[\text{h}_{10}^ {\dagger}\text{h}_{10}])\right. \nonumber\\
&+&
\left. \frac{24}{5} \lambda^2 \
\left(\text{A}_{\lambda}^2+\text{m}_{24}^{2}+\text{m}_{5_H}^{2}+\text{m}_{\bar{5}_H}^{2}\right)-\frac{48 \text{g}_{5}^2 \
\text{M}_{5}^2}{5} {\mathbf{1}}   \right]  \, ,
\nonumber\\
\frac{\text{dm}_{\bar{5}_H}^{2}}{\text{dt}}
 &=&
\frac{1}{8 \pi ^2}
 \left[2 \
(\text{Tr}[\text{a}_{\bar 5}^{\dagger}\text{a}_{\bar 5}]+\text{Tr}[\text{h}_{\bar{5}}\text{m}_{10}^{2}\text{h}_{\bar{5}}^ {\dagger}]\
+\text{Tr}[\text{h}_{\bar{5}}^ {\dagger}\text{m}_{\bar{5}}^{2}\text{h}_{\bar{5}} {\mathbf{1}} ]
\right. \nonumber\\
&+&
\left. 
\text{m}_{\bar{5}_H}^{2} \
\text{Tr}[\text{h}_{\bar{5}}^ {\dagger}\text{h}_{\bar{5}}])+\frac{24}{5} \lambda^2 \
\left(\text{A}^2_{\lambda}+\text{m}_{24}^{2}+\text{m}_{5_H}^{2}+\text{m}_{\bar{5}_H}^{2}\right)- \frac{48}{5} \text{g}_{5}^2 \
\text{M}_{5}^2   {\mathbf{1}} \right]  \, ,
\nonumber\\ 
\frac{\text{dm}_{\bar{5}}^{2}}{\text{dt}}
&=&
\frac{1}{16 \pi ^2}
\left[
2 (2 \text{a}_{\bar 5}^{\dagger}\text{a}_{\bar 5}+2 \text{m}_{\bar{5}_H}^{2} \
\text{h}_{\bar{5}}^ {\dagger}\text{h}_{\bar{5}}+\text{m}_{\bar{5}}^{2}\text{h}_{\bar{5}}^ {\dagger}\text{h}_{\bar{5}}+2 \
\text{h}_{\bar{5}}^ {\dagger}\text{m}_{10}^{2}\text{h}_{\bar{5}}+\text{h}_{\bar{5}}^ {\dagger}\text{h}_{\bar{5}}\text{m}_{\bar{5}}^{2})\right.
\nonumber\\
&-&
\left.\frac{96 \
\text{g}_{5}^2 \text{M}_{{5}}^2}{5} {\mathbf{1}}
\right]  \, , \nonumber\\
\frac{\text{dm}_{10}^{2}}{\text{dt}}
 &=&
\frac{1}{16 \pi ^2}
\left[
 -\frac{144}{5} \text{g}_{5}^2 \text{M}_{5}^2  {\mathbf{1}} +\text{m}_{10}^{2}\text{h}_{\bar{5}}^ {\dagger}\text{h}_{\bar{5}}+96 \
(\text{m}_{5_H}^{2} \text{h}_{10}^ {\dagger}\text{h}_{10}+\text{h}_{10}^ {\dagger}\text{m}_{10}^{2}\text{h}_{10})\right.\nonumber\\
&+&
48 \
(\text{m}_{10}^{2}\text{h}_{10}^ {\dagger}\text{h}_{10}+\text{h}_{10}^ {\dagger}\text{h}_{10}\text{m}_{10}^{2})+2 \
(48 \text{a}_{10}^{\dagger}\text{a}_{10}+\text{a}_{\bar 5}^{\dagger}\text{a}_{\bar 5}+\text{m}_{\bar{5}_H}^{2} \
\text{h}_{\bar{5}}^ {\dagger}\text{h}_{\bar{5}}+\text{h}_{\bar{5}}^ {\dagger}\text{m}_{\bar{5}}^{2}\text{h}_{\bar{5}})
\nonumber\\
&+&
\left.
\text{h}_{\bar{5}}^ {\dagger}\text{h}_{\bar{5}}\
\text{m}_{10}^{2} 
\right]  \, ; \label{eq:su5betafuncts}
\eea
}

\small{
\bea
\frac{\text{dB}_{24}}{\text{dt}}
 &=&
\frac{1}{8 \pi ^2}
 \left[2 \lambda^2 \text{A}_{\lambda}+\frac{21 \text{A}_{\lambda'} \lambda'^2}{10}
-20 \text{g}_{5}^2   \text{M}_{5} {\mathbf{1}}  \right]  \, ,
\nonumber\\
\frac{\text{dB}_{5}}{\text{dt}}
 &=&
\frac{1}{8 \pi ^2}
 \left[48 \text{Tr}[\text{h}_{10}^ {\dagger}\text{a}_{10}]+2 \
\text{Tr}[\text{h}_{\bar{5}}^ {\dagger}\text{a}_{\bar 5}]+\frac{48 \lambda^2 \text{A}_{\lambda}}{5}
-\frac{48 \ \text{g}_{5}^2 \text{M}_{5}}{5} {\mathbf{1}}   \right]  \, .
\eea
}

\section{Transformation rules} \label{sec:tr}

%new, we may need it
The physical mass eigenstates for each flavor are calculated in the SCKM basis by diagonalizing the $6\times 6$ matrices 
\bea
({\mathcal{M}}^{\skm}_{ f})^2_{ij}\!\!\!&=&\!\!\!\! \left[
\begin{array}{cc}
\hat{m}^2_{ f  LL} & \hat{m}^2_{ f  LR}\\
\hat{m}^2_{ f  LR} & \hat{m}^2_{ f  RR}
\end{array}
\right]_{ij} \equiv  (\widehat{\mathcal{M}}^{2}_{ f})_{ij} \nn\\
&=& \!\!\!\! \left[\begin{array}{cc}
\left(V^{f\dagger}_{L}m^2_{ Q}V^{f}_L\right)_{ij}+(\hat m_{f_i})^2 \delta_{ij} +D^f_{L} & 
-((V^{f\dagger}_L a_{f} V^{f}_R)_{ij} v_f+\mu^* \tan^p\beta \ \hat m_{f_i}\delta_{ij} )\\
 -( (V^{f}_L a^{\dagger}_f V^{f\dagger}_{R})_{ij} v_f+\mu \tan^p \beta\ \hat m_{f_i}\delta_{ij})&
 (V^{f\dagger}_{R} m^2_{ f_R} V^{f}_R)_{ij}+ (\hat{m}_{f_i})^2\delta_{ij} + D^f_{R}
\label{eq:eff_rotatedSCKMmassm}
\end{array}\right]\! \!,\nn\\
p\ &=&\  \left\{\begin{array}{c} 1,\ f=D, L\\ -1,\ f=U. \end{array}\right.,
\eea
where $\hat m_{f}$ are the diagonal quark mass matrices. We use the following notation for the matrices diagonalizing the soft mass-squared matrices:
\bea
\label{eq:diagmatKl}
K \ ({\mathcal{M}}^{\skm}_{ f})^2\  K^{\dagger} =
\rm{diag}(m_{\tilde\ell_1}^2,\hdots,m_{\tilde\ell_6}^2),
\eea
where $m_{\tilde\ell_i}^2$ are the mass eigenstates.

% end of addition

To help understand the impact of the RGEs on the final parameter values at $M_{EW}$, 
we show contributions to the beta functions in the SCKM basis taken from Eq. (4.34) of \cite{Martin:1993zk}: 
\bea
\beta^{(1)}_{\mds} &\supset  & \mds h_D h_D^\dagger,\quad  h_D \mQs h_D^\dagger,\quad  h_D h_D^\dagger \mds,\quad m^2 h_D h_D^\dagger, \nonumber \\
\beta^{(1)}_{\mQs} &\supset  & \mQs h_U^\dagger h_U,\quad \mQs h_D^\dagger h_D,\quad  h_U^\dagger h_U \mQs,\quad  h_D^\dagger h_D \mQs, \quad  h_U^\dagger\mus h_U, \quad   h_D^\dagger\mds h_D ,
\nn\\ 
& & a_D^\dagger a_D,\quad m^2 h_U^\dagger h_U , \quad m^2 h_D^\dagger h_D,
\label{eq:depbetfM2dM2Q}
\eea
where $m^2$ is a mass-squared term and not a matrix, for example $m^2=m^2_{H_d}$.
Using $dm^2=\int dt \beta^{(1)}_{m^2} / 16\pi^2$, we can calculate the
form of the mass-squared terms at the EW scale in the SCKM basis. % and
                                % compare the differences in the
                                % transformation of the Ansatazs A1
                                % and A2. 
In addition to the transformation given in Eq. (\ref{eq:SCKMtransf}),                                 
the general transformations for all the terms appearing in \eq{eq:depbetfM2dM2Q}, independent of the Ansatz, are as follows:
%\newpage
\small{
\begin{align}
%\begin{array}{ll}
%\vspace{6cm}
&{\it{Transformation ~of ~terms ~in~ }} \mQs & \it{Transformation ~of ~terms ~in~ } \mds  : \nn \\
&  \it{\hspace{2cm} to ~the ~SCKM ~basis:} & \nn\\
&\; & \; \nn\\
%&(\mQs)^{\skm}=V^{d\dagger}_L \mQs V^d_L & (\mds)^{\skm}=V^{d\dagger}_R \mds V^d_R \nn\\
%&\; & \; \nn\\
%&\\
%(\mQs)^{\skm}=V^{d\dagger}_L \mQs V^d_L & 
%(\mds)^{\skm}=V^{d\dagger}_R \mds V^d_R   \\
%&\\
%\vspace{0.5cm}
1. & ~ \mQs   h_D^\dagger h_D   \rightarrow   \mQsSK \hat{h}_D^2 &
1. ~ \mds   h_D h_D^\dagger \rightarrow  V^{d\dagger}_R \mds V^D_R \hat{h}^2_D \nn \\
2. &~ \mQs   h_U^\dagger h_U \rightarrow \mQsSK  \V^\dagger \hat{h}_U^2 \V &
2. ~ h_D \mQs h_D^\dagger  \rightarrow  \hat{h}_D \mQsSK \hat{h}_D \nn\\
3. &~ h_U^\dagger h_U  \mQs\rightarrow   \V^\dagger \hat{h}^2_U \V    \mQsSK & 
3. ~
h_D h_D^\dagger \mds \rightarrow \hat{h}_D \left(V^{\dagger}_{DR} \mds V_{DR}  \right) \nn\\
4. &~ h_D^\dagger h_D  \mQs\rightarrow   \hat{h}^2_D  \mQsSK &
4. ~ m^2 h_D h_D^\dagger\rightarrow m^2 \hat{h}_D \hat{h}^\dagger_D  \nn\\
5. &~ h_U^\dagger \mus h_U \rightarrow  
{\V}^\dagger \hat{h}_U \left( V^{u\dagger}_R  \mus V^{U}_R  \right) \hat{h}_U \V & \nn\\
6. &~ h_D^\dagger \mds h_D  \rightarrow   \hat{h}_D^\dagger \mdsSK \hat{h}_D  &\nn\\
7. &~ a_D^\dagger a_D  \rightarrow  \hat{a}_D^\dagger \hat{a}_D & \nn\\
8. &~ m^2 h_U^\dagger h_U    \rightarrow m^2 \V^\dagger \hat{h}_U \V & \nn\\
9. &~ m^2 h_D^\dagger h_D    \rightarrow \hat{h}^\dagger_D\hat{h}_D.
\label{eq:transfsckm} %\\
%\end{array}\nn\\
\end{align}
}

The transformations (\ref{eq:transfsckm}) to the SCKM basis are valid at all energy scales. 
With the exception of the fifth entry in (\ref{eq:transfsckm}) for the transformations of   $\mQs$, the rest of the terms have the same form.
(Actually, the initial value of $\mQsSK=V^{D\dagger}_L \mQs V^D_L$ also has the same form, since $V^D_L=\V$ for both cases.)
The fifth entry in (\ref{eq:transfsckm}) for the transformation of  $\mQs$, could potentially be different, 
because it is not given only in terms of $\V$ or squared quark masses,  but  $V^U_R$. 
However, since in both Ans\"atze we have the same form of $V^U_R$ and $\mus$, i.e., diagonal matrices, 
this term would not make a numerical difference in either Ansatz. 
Hence the only differences in $\mQsSK$ at $\Mg$
are the different values of the off-diagonal elements in $\mQs$ and $\mds$. 
There are more differences in the transformations of terms in $\mdsSK$,
since terms (1) and (3) of  (\ref{eq:transfsckm}) explicitly involve $V^D_R$, which is different in the two cases.

\begin{align}
&{\it{Transformation ~of ~terms ~in~ }} a_D & &\nn\\
1. &~a_D h_D^\dagger h_D  \rightarrow  \aDH \hat{h}^2_D  & = \ & \aDHE \hat{h}^2_D &\nn\\
2. &~a_D h_U^\dagger h_U \rightarrow    \aDH \hat{h}^2_U V^{U\dagger}_R V^D_R &= \ &  \aDHE \hat{h}^2_U V^{U\dagger}_R V^D_R &\nn\\
3. &~h_D h_U^\dagger a_D \rightarrow \hat{h}^2_D\aDH &=\ &    \hat{h}^2_D\aDHE & \nn\\
4. &~h_D h_U^\dagger a_U \rightarrow     \hat{h}_D \left(V^{D\dagger}_R V^U_R \hat{h}_U V^{U\dagger}_L a_U V^D_R \right).   &  &  &
\label{eq:transfsckmaD} 
\end{align}

\newcommand{\aUh}

\section*{Acknowledgements}
The work of JE was supported in part by the London Centre for Terauniverse Studies (LCTS), 
using funding from the European Research Council via the Advanced Investigator Grant 26732, 
and in part by the STFC Grant ST/L000326/1. The work of  K.A.O. was supported in part by DOE grant DE-SC0011842 at the 
University of Minnesota. We thank J. Rosiek for help with {\tt SUSY\_FLAVOR}.


\begin{thebibliography}{99}

\bibitem{EN}
J.~R.~Ellis and D.~V.~Nanopoulos,
  %``Flavor Changing Neutral Interactions in Broken Supersymmetric Theories,''
  Phys.\ Lett.\ B {\bf 110} (1982) 44.
  %%CITATION = PHLTA,B110,44;%%
  %417 citations counted in INSPIRE as of 23 Jun 2014
  
\bibitem{BG}
R.~Barbieri and R.~Gatto,
  %``Conservation Laws for Neutral Currents in Spontaneously Broken Supersymmetric Theories,''
  Phys.\ Lett.\ B {\bf 110} (1982) 211.
  %%CITATION = PHLTA,B110,211;%%
  %202 citations counted in INSPIRE as of 23 Jun 2014
  
  \bibitem{gaugino}
  Z.~Chacko, M.~A.~Luty, A.~E.~Nelson and E.~Ponton,
  %``Gaugino mediated supersymmetry breaking,''
  JHEP {\bf 0001}, 003 (2000)
  [hep-ph/9911323];
  %%CITATION = HEP-PH/9911323;%%
   M.~Schmaltz and W.~Skiba,
  %``Minimal gaugino mediation,''
  Phys.\ Rev.\  D {\bf 62}, 095005 (2000)
  [arXiv:hep-ph/0001172];
  %%CITATION = PHRVA,D62,095005;%%
M.~Schmaltz and W.~Skiba,
  %``The superpartner spectrum of gaugino mediation,''
  Phys.\ Rev.\  D {\bf 62}, 095004 (2000)
  [arXiv:hep-ph/0004210].
  %%CITATION = PHRVA,D62,095004;%%
  
  \bibitem{DG}
  S.~Dimopoulos and H.~Georgi,
  %``Softly Broken Supersymmetry and SU(5),''
  Nucl.\ Phys.\ B {\bf 193} (1981) 150.
  doi:10.1016/0550-3213(81)90522-8
  %%CITATION = doi:10.1016/0550-3213(81)90522-8;%%
  %2335 citations counted in INSPIRE as of 01 f思r. 2016
  
  \bibitem{Sakai}
  N.~Sakai,
  %``Naturalness in Supersymmetric Guts,''
  Z.\ Phys.\ C {\bf 11} (1981) 153.
  doi:10.1007/BF01573998
  %%CITATION = doi:10.1007/BF01573998;%%
  %1247 citations counted in INSPIRE as of 01 f思r. 2016
  
  \bibitem{cmssm}
M.~Drees and M.~M.~Nojiri,
Phys.\ Rev.\ D {\bf 47} (1993) 376 [arXiv:hep-ph/9207234];
%%CITATION = HEP-PH 9207234;%%
  G.~L.~Kane, C.~F.~Kolda, L.~Roszkowski and J.~D.~Wells,
  %``Study of constrained minimal supersymmetry,''
  Phys.\ Rev.\  D {\bf 49} (1994) 6173
  [arXiv:hep-ph/9312272];
  %%CITATION = PHRVA,D49,6173;%%
H.~Baer and M.~Brhlik,
Phys.\ Rev.\ D {\bf 53} (1996) 597 [arXiv:hep-ph/9508321];
%%CITATION = HEP-PH 9508321;%%
  Phys.\ Rev.\  D {\bf 57} (1998) 567
  [arXiv:hep-ph/9706509];
  %%CITATION = PHRVA,D57,567;%%
  J.~R.~Ellis, T.~Falk, K.~A.~Olive and M.~Schmitt,
Phys.\ Lett.\ B {\bf 388} (1996) 97
[arXiv:hep-ph/9607292];
%%CITATION = HEP-PH 9607292;%%
Phys.\ Lett.\ B {\bf 413} (1997) 355
[arXiv:hep-ph/9705444];
%%CITATION = HEP-PH 9705444;%%
J.~R.~Ellis, T.~Falk, G.~Ganis, K.~A.~Olive and M.~Schmitt,
Phys.\ Rev.\ D {\bf 58} (1998) 095002
[arXiv:hep-ph/9801445];
%%CITATION = HEP-PH 9801445;%%
V.~D.~Barger and C.~Kao,
Phys.\ Rev.\ D {\bf 57} (1998) 3131
[arXiv:hep-ph/9704403];
%%CITATION = HEP-PH 9704403;%%
J.~R.~Ellis, T.~Falk, G.~Ganis and K.~A.~Olive,
Phys.\ Rev.\ D {\bf 62} (2000) 075010
[arXiv:hep-ph/0004169];
  H.~Baer, M.~Brhlik, M.~A.~Diaz, J.~Ferrandis, P.~Mercadante, P.~Quintana and X.~Tata,
    Phys.\ Rev.\  D {\bf 63} (2001) 015007
  [arXiv:hep-ph/0005027];
J.~R.~Ellis, T.~Falk, G.~Ganis, K.~A.~Olive and M.~Srednicki,
Phys.\ Lett.\ B {\bf 510} (2001) 236
[arXiv:hep-ph/0102098];
%%CITATION = HEP-PH 0102098;%%
V.~D.~Barger and C.~Kao,
Phys.\ Lett.\ B {\bf 518} (2001) 117
[arXiv:hep-ph/0106189];
%%CITATION = HEP-PH 0106189;%%
L.~Roszkowski, R.~Ruiz de Austri and T.~Nihei,
JHEP {\bf 0108} (2001) 024
[arXiv:hep-ph/0106334];
%%CITATION = HEP-PH 0106334;%%
A.~Djouadi, M.~Drees and J.~L.~Kneur,
JHEP {\bf 0108} (2001) 055
[arXiv:hep-ph/0107316];
%%CITATION = HEP-PH 0107316;%%
U.~Chattopadhyay, A.~Corsetti and P.~Nath,
Phys.\ Rev.\ D {\bf 66} (2002) 035003
[arXiv:hep-ph/0201001];
%%CITATION = HEP-PH 0201001;%%
J.~R.~Ellis, K.~A.~Olive and Y.~Santoso,
New Jour.\ Phys.\  {\bf 4} (2002) 32
[arXiv:hep-ph/0202110];
%%CITATION = HEP-PH 0202110;%%
H.~Baer, C.~Balazs, A.~Belyaev, J.~K.~Mizukoshi, X.~Tata and Y.~Wang,
JHEP {\bf 0207} (2002) 050
[arXiv:hep-ph/0205325];
%%CITATION = HEP-PH 0205325;%%
R.~Arnowitt and B.~Dutta,
arXiv:hep-ph/0211417.
%%CITATION = HEP-PH 0211417;%%

\bibitem{bfs}
R.~Barbieri, S.~Ferrara and C.~A.~Savoy,
  %``Gauge Models with Spontaneously Broken Local Supersymmetry,''
  Phys.\ Lett.\ B {\bf 119}, 343 (1982).
  %%CITATION = PHLTA,B119,343;%%

\bibitem{gauge}
M.~Dine, A.~E.~Nelson and Y.~Shirman,
  %``Low-energy dynamical supersymmetry breaking simplified,''
  Phys.\ Rev.\ D {\bf 51}, 1362 (1995)
  [hep-ph/9408384].
  %%CITATION = HEP-PH/9408384;%%

\bibitem{EK}
S.~A.~R.~Ellis and G.~L.~Kane,
  %``Lepton Flavour Violation via the K\"ahler Potential in Compactified M-Theory,''
  arXiv:1505.04191 [hep-ph].
  %%CITATION = ARXIV:1505.04191;%%
  
%\cite{Ciuchini:2007ha}
\bibitem{Ciuchini:2007ha} 
  M.~Ciuchini, A.~Masiero, P.~Paradisi, L.~Silvestrini, S.~K.~Vempati and O.~Vives,
  %``Soft SUSY breaking grand unification: Leptons versus quarks on the flavor playground,''
  Nucl.\ Phys.\ B {\bf 783}, 112 (2007)
  doi:10.1016/j.nuclphysb.2007.05.032
  [hep-ph/0702144 [HEP-PH]].
  %%CITATION = doi:10.1016/j.nuclphysb.2007.05.032;%%
  %113 citations counted in INSPIRE as of 13 Sep 2016

\bibitem{nosc1}
  E.~Cremmer, S.~Ferrara, C.~Kounnas and D.~V.~Nanopoulos,
  %``Naturally Vanishing Cosmological Constant In N=1 Supergravity,''
  Phys.\ Lett.\  B {\bf 133}, 61 (1983);
  %%CITATION = PHLTA,B133,61;%%
 J.~R.~Ellis, C.~Kounnas and D.~V.~Nanopoulos,
  %``No Scale Supersymmetric Guts,''
  Nucl.\ Phys.\  B {\bf 247}, 373 (1984).
  %%CITATION = NUPHA,B247,373;%%

\bibitem{GMSB}
  G.~F.~Giudice, R.~Rattazzi,
  %``Theories with gauge mediated supersymmetry breaking,''
  Phys.\ Rept.\  {\bf 322 } (1999)  419-499.
  [hep-ph/9801271].


\bibitem{eno5}
J.~R.~Ellis, D.~V.~Nanopoulos and K.~A.~Olive,
  %``Lower limits on soft supersymmetry breaking scalar masses,''
  Phys.\ Lett.\ B {\bf 525} (2002) 308
  [hep-ph/0109288].
  %%CITATION = HEP-PH/0109288;%%
  %30 citations counted in INSPIRE as of 24 Jun 2014
    
 \bibitem{emo2}
  J.~Ellis, A.~Mustafayev and K.~A.~Olive,
  %``Resurrecting No-Scale Supergravity Phenomenology,''
  Eur.\ Phys.\ J.\ C {\bf 69}, 219 (2010)
  [arXiv:1004.5399 [hep-ph]].
  %%CITATION = ARXIV:1004.5399;%%
  
\bibitem{ENO2013}
J.~Ellis, D.~V.~Nanopoulos and K.~A.~Olive,
  %``A No-Scale Framework for Sub-Planckian Physics,''
  Phys.\ Rev.\ D {\bf 89} (2014) 043502
  [arXiv:1310.4770 [hep-ph]].
  %%CITATION = ARXIV:1310.4770;%%
  %15 citations counted in INSPIRE as of 23 Jun 2014
  
  \bibitem{sg}
  L.~Calibbi, Y.~Mambrini and S.~K.~Vempati,
  %``SUSY-GUTs, SUSY-Seesaw and the Neutralino Dark Matter,''
  JHEP {\bf 0709}, 081 (2007)
  [arXiv:0704.3518 [hep-ph]]; 
  L.~Calibbi, A.~Faccia, A.~Masiero and S.~K.~Vempati,
  %``Lepton flavour violation from SUSY-GUTs: Where do we stand for MEG, PRISM /
  %PRIME and a super flavour factory,''
  Phys.\ Rev.\  D {\bf 74}, 116002 (2006)
  [arXiv:hep-ph/0605139];
  E.~Carquin, J.~Ellis, M.~E.~Gomez, S.~Lola and J.~Rodriguez-Quintero,
  %``Search for Tau Flavour Violation at the LHC,''
  JHEP {\bf 0905} (2009) 026
  [arXiv:0812.4243 [hep-ph]];
  %%CITATION = JHEPA,0905,026;%%
J.~Ellis, A.~Mustafayev and K.~A.~Olive,
  %``Constrained Supersymmetric Flipped SU(5) GUT Phenomenology,''
  Eur.\ Phys.\ J.\ C {\bf 71}, 1689 (2011)
  [arXiv:1103.5140 [hep-ph]].
  %%CITATION = ARXIV:1103.5140;%%

  \bibitem{Ellis:2010jb}
  J.~Ellis, A.~Mustafayev and K.~A.~Olive,
  %``What if Supersymmetry Breaking Unifies beyond the GUT Scale?,''
  Eur.\ Phys.\ J.\ C {\bf 69}, 201 (2010)
  [arXiv:1003.3677 [hep-ph]].
  %%CITATION = ARXIV:1003.3677;%%
  
  \bibitem{dmmo}
  E.~Dudas, Y.~Mambrini, A.~Mustafayev and K.~A.~Olive,
  %``Relating the CMSSM and SUGRA Models with GUT Scale and Super-GUT Scale Supersymmetry Breaking,''
  Eur.\ Phys.\ J.\ C {\bf 72}, 2138 (2012)
  [Erratum-ibid.\ C {\bf 73}, 2430 (2013)]
  [arXiv:1205.5988 [hep-ph]].
  %%CITATION = ARXIV:1205.5988;%%
  


  \bibitem{Polonsky:1994rz}
   N.~Polonsky and A.~Pomarol,
  %``GUT effects in the soft supersymmetry breaking terms,''
  Phys.\ Rev.\ Lett.\  {\bf 73}, 2292 (1994)
  [arXiv:hep-ph/9406224],
  and 
%  N.~Polonsky and A.~Pomarol,
%  ``Nonuniversal GUT corrections to the soft terms and their implications 
%    in supergravity models,''
  Phys.\ Rev.\ D {\bf 51} (1995) 6532
  [arXiv:hep-ph/9410231].
  %%CITATION = HEP-PH 9410231;%%

 \bibitem{vcmssm}
  J.~R.~Ellis, K.~A.~Olive, Y.~Santoso and V.~C.~Spanos,
  %``Phenomenological constraints on patterns of supersymmetry breaking,''
  Phys.\ Lett.\ B {\bf 573}, 162 (2003)
  [hep-ph/0305212].
  %%CITATION = HEP-PH/0305212;%%

\bibitem{Borzumati}
 F.~Borzumati and T.~Yamashita,
  %``Minimal supersymmetric SU(5) model with nonrenormalizable operators: Seesaw mechanism and violation of flavour and CP,''
  Prog.\ Theor.\ Phys.\  {\bf 124}, 761 (2010)
  [arXiv:0903.2793 [hep-ph]].
  %%CITATION = ARXIV:0903.2793;%%
  
\bibitem{EG}
  J.~R.~Ellis and M.~K.~Gaillard,
  %``Fermion Masses and Higgs Representations in SU(5),''
  Phys.\ Lett.\ B {\bf 88} (1979) 315.
  doi:10.1016/0370-2693(79)90476-3
  %%CITATION = doi:10.1016/0370-2693(79)90476-3;%%
  %201 citations counted in INSPIRE as of 01 Feb 2016
  


  \bibitem{rpp} 
   K.~A.~Olive {\it et al.}  [Particle Data Group Collaboration],
  %``Review of Particle Physics,''
  Chin.\ Phys.\ C {\bf 38}, 090001 (2014).
  %%CITATION = CHPHD,C38,090001;%%
  
  \bibitem{Chetyrkin:2000yt} 
  K.~G.~Chetyrkin, J.~H.~Kuhn and M.~Steinhauser,
  %``RunDec: A Mathematica package for running and decoupling of the strong coupling and quark masses,''
  Comput.\ Phys.\ Commun.\  {\bf 133}, 43 (2000)
  [hep-ph/0004189].
  %%CITATION = HEP-PH/0004189;%%
  
  \bibitem{Xing:2007fb} 
  Z.~-z.~Xing, H.~Zhang and S.~Zhou,
  %``Updated Values of Running Quark and Lepton Masses,''
  Phys.\ Rev.\ D {\bf 77}, 113016 (2008)
  [arXiv:0712.1419 [hep-ph]].
  %%CITATION = ARXIV:0712.1419;%%
  
      \bibitem{ssard} Information about this code is available from K. A. Olive: it contains important contributions 
from T. Falk,  G. Ganis, A. Mustafayev,
J. McDonald, F. Luo, K. A. Olive, P. Sandick, Y. Santoso, V. Spanos, and M. Srednicki. 

\bibitem{Olive:2008vv} 
  K.~A.~Olive and L.~Velasco-Sevilla,
  %``Constraints on Supersymmetric Flavour Models from b ---> s gamma,''
  JHEP {\bf 0805}, 052 (2008)
  [arXiv:0801.0428 [hep-ph]].
  %%CITATION = ARXIV:0801.0428;%%

\bibitem{lhc}
G.~Aad {\it et al.}  [ATLAS Collaboration],
  %``Search for squarks and gluinos with the ATLAS detector in final states with jets and missing transverse momentum using $\sqrt{s}=8$ TeV proton--proton collision data,''
  JHEP {\bf 1409} (2014) 176
  [arXiv:1405.7875 [hep-ex]];
  %%CITATION = ARXIV:1405.7875;%%
 JHEP {\bf 1510}, 054 (2015)
  doi:10.1007/JHEP10(2015)054
  [arXiv:1507.05525 [hep-ex]];
  %%CITATION = ARXIV:1507.05525;%%S.~Chatrchyan {\it et al.}  [CMS Collaboration],
  %``Search for new physics in the multijet and missing transverse momentum final state in proton-proton collisions at $\sqrt{s}$= 8 TeV,''
  JHEP {\bf 1406} (2014) 055
  [arXiv:1402.4770 [hep-ex]].
  %%CITATION = ARXIV:1402.4770;%%
  
    \bibitem{planck}
  P.~A.~R.~Ade {\it et al.}  [Planck Collaboration],
  %``Planck 2015 results. XIII. Cosmological parameters,''
  arXiv:1502.01589 [astro-ph.CO].
  %%CITATION = ARXIV:1502.01589;%%
  %37 citations counted in INSPIRE as of 02 mar 2015

  
   
 \bibitem{fh}
   S.~Heinemeyer, W.~Hollik and G.~Weiglein,
  Eur.\ Phys.\ J.\ C {\bf 9} (1999) 343
  [arXiv:hep-ph/9812472];
  %%CITATION = HEP-PH 9812472;%%
  S.~Heinemeyer, W.~Hollik and G.~Weiglein,
  Comput.\ Phys.\ Commun.\  {\bf 124} (2000) 76
  [arXiv:hep-ph/9812320];
  %%CITATION = CPHCB,124,76;%%
   M.~Frank {\it et al.}, 
  JHEP {\bf 0702} (2007) 047
  [arXiv:hep-ph/0611326];
  %%CITATION = JHEPA,0702,047;%%
  T.~Hahn, S.~Heinemeyer, W.~Hollik, H.~Rzehak and G.~Weiglein,
  Comput.\ Phys.\ Commun.\  {\bf 180} (2009) 1426.
  %%CITATION = CPHCB,180,1426;%%
  see {\tt http://www.feynhiggs.de}.

 \bibitem{LHCmh}
G.~Aad {\it et al.} [ATLAS and CMS Collaborations],
  %``Combined Measurement of the Higgs Boson Mass in $pp$ Collisions at $\sqrt{s}=7$ and 8 TeV with the ATLAS and CMS Experiments,''
  Phys.\ Rev.\ Lett.\  {\bf 114} (2015) 191803
  doi:10.1103/PhysRevLett.114.191803
  [arXiv:1503.07589 [hep-ex]].
  %%CITATION = doi:10.1103/PhysRevLett.114.191803;%%
  %283 citations counted in INSPIRE as of 01 f思r. 2016
  

%\cite{DeBruyn:2012wj}
\bibitem{DeBruyn:2012wj} 
  K.~De Bruyn, R.~Fleischer, R.~Knegjens, P.~Koppenburg, M.~Merk and N.~Tuning,
  %``Branching Ratio Measurements of $B_s$ Decays,''
  Phys.\ Rev.\ D {\bf 86}, 014027 (2012)
  doi:10.1103/PhysRevD.86.014027
  [arXiv:1204.1735 [hep-ph]].
  %%CITATION = doi:10.1103/PhysRevD.86.014027;%%

%\cite{Arbey:2012ax}
\bibitem{Arbey:2012ax} 
  A.~Arbey, M.~Battaglia, F.~Mahmoudi and D.~Martínez Santos,
  %``Supersymmetry confronts $B_s → μ^+μ^-$ : Present and future status,''
  Phys.\ Rev.\ D {\bf 87}, 035026 (2013)
  doi:10.1103/PhysRevD.87.035026
  [arXiv:1212.4887 [hep-ph]].
  %%CITATION = doi:10.1103/PhysRevD.87.035026;%%

%%%%%%%%%%%%%%%%%%%

\bibitem{susyflavor}
%\cite{Rosiek:2014sia}
%\bibitem{Rosiek:2014sia} 
  J.~Rosiek,
  %``SUSY FLAVOR v2.5: a computational tool for FCNC and CP-violating processes in the MSSM,''
  Comput.\ Phys.\ Commun.\  {\bf 188}, 208 (2014)
  %doi:10.1016/j.cpc.2014.10.003
  [arXiv:1410.0606 [hep-ph]];
  %%CITATION = doi:10.1016/j.cpc.2014.10.003;%%
  %\cite{Crivellin:2012jv}\bibitem{Crivellin:2012jv} 
  A.~Crivellin, J.~Rosiek, P.~H.~Chankowski, A.~Dedes, S.~Jaeger and P.~Tanedo,
  %``{SUSY}{\_}{FLAVOR} v2: A Computational tool for {FCNC} and {CP}-violating processes in the {MSSM},''
  Comput.\ Phys.\ Commun.\  {\bf 184}, 1004 (2013)
  %doi:10.1016/j.cpc.2012.11.007
  [arXiv:1203.5023 [hep-ph]];
  %%CITATION = doi:10.1016/j.cpc.2012.11.007;%%
  %\cite{Rosiek:2010ug}\bibitem{Rosiek:2010ug} 
  J.~Rosiek, P.~Chankowski, A.~Dedes, S.~Jager and P.~Tanedo,
  %``{SUSY}{\_}{FLAVOR}: A Computational Tool for {FCNC} and {CP}-violating Processes in the {MSSM},''
  Comput.\ Phys.\ Commun.\  {\bf 181}, 2180 (2010)
 % doi:10.1016/j.cpc.2010.07.047
  [arXiv:1003.4260 [hep-ph]].
  %%CITATION = doi:10.1016/j.cpc.2010.07.047;%%
  

\bibitem{bmm}
 R.Aaij {\it et al.}  [LHCb and CMS Collaborations],
LHCb-CONF-2013-012, CMS PAS BPH-13-007;
V.~Khachatryan {\it et al.}  [CMS and LHCb Collaborations],
  %``Observation of the rare $B^0_s\to\mu^+\mu^-$ decay from the combined analysis of CMS and LHCb data,''
  Nature {\bf 522}, 68 (2015)
  [arXiv:1411.4413 [hep-ex]].
  %%CITATION = ARXIV:1411.4413;%%

%%%%%%%%%%%%%%%%%%%%%%%%%%%%%%
  %\cite{Jager:2008fc}
\bibitem{Jager:2008fc} 
  S.~J\"ager,
  %``Supersymmetry beyond minimal flavour violation,''
  Eur.\ Phys.\ J.\ C {\bf 59}, 497 (2009)
  [arXiv:0808.2044 [hep-ph]].
  %%CITATION = ARXIV:0808.2044;%%
  %7 citations counted in INSPIRE as of 29 Oct 2014
  
%\cite{Altmannshofer:2009ne}
\bibitem{Altmannshofer:2009ne}
  W.~Altmannshofer, A.~J.~Buras, S.~Gori, P.~Paradisi and D.~M.~Straub,
  %``Anatomy and Phenomenology of FCNC and CPV Effects in SUSY Theories,''
  Nucl.\ Phys.\ B {\bf 830} (2010) 17
  [arXiv:0909.1333 [hep-ph]].
  %%CITATION = ARXIV:0909.1333;%%
  %178 citations counted in INSPIRE as of 23 Oct 2014
  
  \bibitem{Arana-Catania:2013nha}
M.~Arana-Catania, S.~Heinemeyer, and M.~Herrero, 
%``{New Constraints on General  Slepton Flavor Mixing}'',
  %\href{http://dx.doi.org/10.1103/PhysRevD.88.015026}
  {{\em Phys.Rev.} {\bf D88}
  (2013) 015026},
%\href{http://arxiv.org/abs/1304.2783}{{\tt arXiv:1304.2783 [hep-ph]}}.
[arXiv:1304.2783].
%%CITATION = ARXIV:1304.2783;%%.


\bibitem{AbdusSalam_VS:2016}
S. ~AbdusSalam and L. ~Velasco-Sevilla. In preparation. Bounds extracted from a comprehensive scan are compatible with the conditions presented here. 

%\cite{Pospelov:2005pr}
\bibitem{Pospelov:2005pr} 
  M.~Pospelov and A.~Ritz,
  %``Electric dipole moments as probes of new physics,''
  Annals Phys.\  {\bf 318}, 119 (2005)
  doi:10.1016/j.aop.2005.04.002
  [hep-ph/0504231].
  %%CITATION = doi:10.1016/j.aop.2005.04.002;%%



 
%\cite{Hewett:2012ns}
\bibitem{Hewett:2012ns} 
  J.~L.~Hewett {\it et al.},
  %``Fundamental Physics at the Intensity Frontier,''
  doi:10.2172/1042577
  arXiv:1205.2671 [hep-ex].
  %%CITATION = doi:10.2172/1042577;%%
  %211 citations counted in INSPIRE as of 20 Mar 2016 /

  %\cite{Aoyama:2012wj}
\bibitem{Aoyama:2012wj} 
  T.~Aoyama, M.~Hayakawa, T.~Kinoshita and M.~Nio,
  %``Tenth-Order QED Contribution to the Electron g-2 and an Improved Value of the Fine Structure Constant,''
  Phys.\ Rev.\ Lett.\  {\bf 109}, 111807 (2012)
  [arXiv:1205.5368 [hep-ph]].
  %%CITATION = ARXIV:1205.5368;%%
  %75 citations counted in INSPIRE as of 22 Aug 2014
  
  \bibitem{Adam:2013mnn}
J.~Adam {\it et al.} [MEG Collaboration],
  %``New constraint on the existence of the $\mu^+ \to e^+\gamma$ decay,''
  Phys.\ Rev.\ Lett.\  {\bf 110}, 201801 (2013)
  doi:10.1103/PhysRevLett.110.201801
  [arXiv:1303.0754 [hep-ex]].
  %%CITATION = doi:10.1103/PhysRevLett.110.201801;%%
  
\bibitem{Aubert:2009ag}
BaBar Collaboration, B.~Aubert {\em et al.}, 
%``{Searches for Lepton Flavor  Violation in the Decays $\tau\rightarrow e \gamma$ and t$\tau\rightarrow \mu  \gamma$}'',
  % \href{http://dx.doi.org/10.1103/PhysRevLett.104.021802}
  {{\em
  Phys.Rev.Lett.} {\bf 104} (2010)  021802},
%\href{http://arxiv.org/abs/0908.2381}{{\tt arXiv:0908.2381 [hep-ex]}}.
[arXiv:0908.238]
%%CITATION = ARXIV:0908.2381;%%.



    \bibitem{Amhis:2012bh}
Heavy Flavor Averaging Group, Y.~Amhis {\em et al.}, 
%``{Averages of B-Hadron,   C-Hadron, and tau-lepton properties as of early 2012}'',
%\href{http://arxiv.org/abs/1207.1158}{{\tt arXiv:1207.1158 [hep-ex]}}.
arXiv:1207.1158 [hep-ex].
%%CITATION = ARXIV:1207.1158;%%.


%\cite{Gabrielli:1995bd}
\bibitem{Gabrielli:1995bd} 
  E.~Gabrielli, A.~Masiero and L.~Silvestrini,
  %``Flavor changing neutral currents and CP violating processes in generalized supersymmetric theories,''
  Phys.\ Lett.\ B {\bf 374}, 80 (1996)
  doi:10.1016/0370-2693(96)00158-X
  [hep-ph/9509379].
  %%CITATION = doi:10.1016/0370-2693(96)00158-X;%%
  %109 citations counted in INSPIRE as of 03 Mar 2016

%\cite{Kersten:2012ed}
\bibitem{Kersten:2012ed} 
  J.~Kersten and L.~Velasco-Sevilla,
  %``Flavour constraints on scenarios with two or three heavy squark generations,''
  Eur.\ Phys.\ J.\ C {\bf 73}, no. 4, 2405 (2013)
  doi:10.1140/epjc/s10052-013-2405-y
  [arXiv:1207.3016 [hep-ph]].
  %%CITATION = doi:10.1140/epjc/s10052-013-2405-y;%%
  %11 citations counted in INSPIRE as of 21 Apr 2016

%\cite{Brod:2011ty}
\bibitem{Brod:2011ty} 
  J.~Brod and M.~Gorbahn,
  %``Next-to-Next-to-Leading-Order Charm-Quark Contribution to the CP Violation Parameter epsilon_K and Delta M_K,''
  Phys.\ Rev.\ Lett.\  {\bf 108}, 121801 (2012)
  doi:10.1103/PhysRevLett.108.121801
  [arXiv:1108.2036 [hep-ph]].
  %%CITATION = doi:10.1103/PhysRevLett.108.121801;%%
  %95 citations counted in INSPIRE as of 03 Mar 2016

%\cite{Kersten:2014xaa}
\bibitem{Kersten:2014xaa} 
  J.~Kersten, J.~h.~Park, D.~St\"ockinger and L.~Velasco-Sevilla,
  %``Understanding the correlation between $(g-2)_\mu$ and $\mu \rightarrow e \gamma$ in the MSSM,''
  JHEP {\bf 1408}, 118 (2014)
  doi:10.1007/JHEP08(2014)118
  [arXiv:1405.2972 [hep-ph]].
  %%CITATION = doi:10.1007/JHEP08(2014)118;%%
  
  %\cite{Mertens:2011ts}
\bibitem{Mertens:2011ts} 
  P.~Mertens and C.~Smith,
  %``The s ---> d gamma decay in and beyond the Standard Model,''
  JHEP {\bf 1108}, 069 (2011)
  doi:10.1007/JHEP08(2011)069
  [arXiv:1103.5992 [hep-ph]].
  %%CITATION = doi:10.1007/JHEP08(2011)069;%%
  %21 citations counted in INSPIRE as of 21 Apr 2016


%\cite{Dedes:2008iw}
\bibitem{Dedes:2008iw} 
  A.~Dedes, J.~Rosiek and P.~Tanedo,
  %``Complete One-Loop MSSM Predictions for B --> lepton lepton' at the Tevatron and LHC,''
  Phys.\ Rev.\ D {\bf 79}, 055006 (2009)
  doi:10.1103/PhysRevD.79.055006
  [arXiv:0812.4320 [hep-ph]].
  %%CITATION = doi:10.1103/PhysRevD.79.055006;%%
  %29 citations counted in INSPIRE as of 21 Apr 2016





%\bibitem{NEDMFSens}
%\cite{Altarev:2009zz}
\bibitem{Altarev:2009zz} 
  I.~Altarev {\it et al.},
  %``Towards a new measurement of the neutron electric dipole moment,''
  Nucl.\ Instrum.\ Meth.\ A {\bf 611}, 133 (2009).
  doi:10.1016/j.nima.2009.07.046.
  %%CITATION = doi:10.1016/j.nima.2009.07.046;%%
  %23 citations counted in INSPIRE as of 01 Mar 2016

   \bibitem{Baer:2000gf}
  H.~Baer, M.~A.~Diaz, P.~Quintana and X.~Tata,
  %``Impact of physical principles at very high energy scales on the
  %superparticle mass spectrum,''
  JHEP {\bf 0004}, 016 (2000)
  [arXiv:hep-ph/0002245].
  



%\cite{Martin:1993zk}
\bibitem{Martin:1993zk}
  S.~P.~Martin and M.~T.~Vaughn,
  %``Two loop renormalization group equations for soft supersymmetry breaking couplings,''
  Phys.\ Rev.\ D {\bf 50}, 2282 (1994)
  [Erratum-ibid.\ D {\bf 78}, 039903 (2008)]
  [hep-ph/9311340].
  %%CITATION = HEP-PH/9311340;%%
  %621 citations counted in INSPIRE as of 28 Oct 2014
  
  
 
  
\end{thebibliography}
\end{document}